\documentclass[dvipsnames]{lmcs}
\pdfoutput=1

\usepackage{lastpage}
\lmcsdoi{16}{1}{29}
\lmcsheading{}{\pageref{LastPage}}{}{}%
{May~23,~2018}{Mar.~02,~2020}{}

\ACMCCS{[{\bf Theory of computation}]: Logic---Constraint and logic programming, [{\bf Theory of computation}]: Logic---Automated reasoning,
[{\bf Software and its engineering}]: Software notations and tools---Context specific languages--- Domain specific languages, [{\bf Security and privacy}]: Cryptography---Symmetric cryptography and hash functions, [{\bf Security and privacy}]: Cryptography---Cryptanalysis and other attacks}


\theoremstyle{plain}

\usepackage[english]{babel}

\addto\extrasenglish{%

}

\usepackage{multirow}
\usepackage{appendix}
\usepackage{subfig}
\usepackage{bold-extra}

\begin{document}

\title[]{Translation of Algorithmic Descriptions of Discrete\\
Functions to SAT with Applications to Cryptanalysis Problems}

\author[A.~Semenov]{Alexander Semenov}	
\address{Russia, Irkutsk, ISDCT SB RAS}	
\email{biclop.rambler@yandex.ru}  

\author[I.~Otpuschennikov]{Ilya Otpuschennikov}	
\address{Russia, Irkutsk, ISDCT SB RAS}	
\email{otilya@yandex.ru}  

\author[I.~Gribanova]{Irina Gribanova}	
\address{Russia, Irkutsk, ISDCT SB RAS}	
\email{the42dimension@gmail.com}  

\author[O.~Zaikin]{Oleg Zaikin}	
\address{Russia, Irkutsk, ISDCT SB RAS}	
\email{zaikin.icc@gmail.com}  

\author[S.~Kochemazov]{Stepan Kochemazov}	
\address{Russia, Irkutsk, ISDCT SB RAS}	
\email{veinamond@gmail.com}  

\begin{abstract}
  \noindent
  In the present paper, we propose a technology for translating
  algorithmic descriptions of discrete functions to SAT. The
  proposed technology is aimed at applications in algebraic cryptanalysis.
  We describe how cryptanalysis problems are reduced to SAT in
  such a way that it should be perceived as natural by the
  cryptographic community. In~the theoretical part of
  the paper we justify the main principles of general reduction
  to SAT for discrete functions from a class containing the majority
  of functions employed in cryptography. Then, we describe the \textsc{Transalg} software tool developed based on these principles
  with SAT-based cryptanalysis specifics in mind. We demonstrate the
  results of applications of \textsc{Transalg} to construction
  of a number of attacks on various cryptographic functions.
  Some of the corresponding attacks are state of the art.
  We compare the functional capabilities of the proposed
  tool with that of other domain-specific software tools which can be used to reduce cryptanalysis problems to SAT, and also with the CBMC system widely employed in symbolic verification. The paper also presents vast experimental data, obtained using the SAT solvers that took first places at the SAT competitions in the recent several years.
\end{abstract}

\maketitle

\section*{Introduction}\label{sec:intro}

The state-of-the-art algorithms for solving the Boolean satisfiability
problem (SAT) are successfully used in many practical areas including
symbolic verification, scheduling and planning, bioinformatics,
network science, etc. In the recent years, a growth of interest to 
applications of SAT in cryptanalysis is observed. The corresponding results are covered in, e.g., \cite{Massacci:1999:UWR:1624218.1624261,
DBLP:journals/jar/MassacciM00,
LewisM03,
DBLP:conf/SAT/MironovZ06,
DBLP:conf/SAT/DeKV07,
CourtoisB2007,
ErkokPLPV09,
Bard:2009:AC:1618541,
DBLP:conf/SAT/SoosNC09,
DBLP:conf/pact/SemenovZBP11,
Courtois-tatra,
Nossum2012,
Courtois-cryptologia,
Courtois-eprint,
Semenov2016,
Stoffelen:2016:OSI:3081631.3081642,
10.1007/978-3-319-66263-3_16,
NejatiLGCG17,SemeonovZOKI2018}.

SAT-based cryptanalysis implies two stages: on the first stage 
a SAT encoding of a considered cryptanalysis problem is constructed.
On the second stage the obtained SAT instance is solved using some 
SAT solving algorithm. The success on the second stage is not guaranteed
because SAT is an NP-hard problem, and also due to the fact that the hardness of cryptanalysis problems is usually preserved during their translation to SAT form \cite{Cook97findinghard}. Despite an impressive progress in the development of applied algorithms for solving SAT, the majority of cryptanalysis problems remain too hard even for cutting edge SAT solvers.

Meanwhile, the first stage is guaranteed to be effective for most cryptanalysis problems, at least in theory. It follows from the fact 
that cryptographic transformations are usually designed to be performed
very fast. To reduce some cryptanalysis problem to SAT, one has to perform
the so-called propositional encoding of a corresponding function. In 
practice, this task is quite nontrivial because modern cryptographic
algorithms are constructed from a large number of basic primitives.
Often, a researcher has to carry out a substantial
amount of manual work to make a SAT encoding for a considered cipher
due to some of its constructive features or because of specific 
requirements of an implemented attack. 

There are two main classes of tools that can be used to automatically reduce a cryptanalysis problem to SAT. 
First, it is possible to use one of the several  available domain-specific systems. 
Among them we would like to mention \textsc{SAW} \cite{Carter2013} that can operate with the \textsc{Cryptol} language \cite{LewisM03,ErkokPLPV09,DBLP:conf/fmcad/ErkokCW09} designed 
for specifying cryptographic algorithms. Another system that allows
one to reduce cryptanalysis problems to SAT (albeit with some additional steps) 
is the \textsc{URSA} system \cite{journals/lmcs/predrag}.  Finally, it is possible to use the  \textsc{Grain-of-Salt} tool \cite{DBLP:conf/tools/Soos10} to construct SAT encodings of cryptographic functions from a 
limited class, formed by keystream generators based on feedback shift registers. 
Another approach to reducing cryptanalysis problems to SAT 
consists in using generic systems for symbolic
verification in the form of Bounded Model Checking  
\cite{Biere:1999:SMC:646483.691738,Biere:1999:SMC:309847.309942,ckl2004}.
For example, one can employ the \textsc{CBMC} system \cite{ckl2004,DBLP:series/faia/Kroening09}
or \textsc{LLBMC} system \cite{LLBMC}.

Both generic systems and domain-specific systems have their pros and cons. On the one hand, generic systems usually support widely employed programming 
languages, such as C, and therefore it is easy to adapt an
existing implementation of a cryptographic function to such a tool.
They also have a wide spectrum of applications, are  well supported
and have a good documentation. 
On the other hand, while domain-specific tools may lack in
convenience of use, their languages 
are often purposefully enriched (compared to generic
languages) by instructions and data types that improve their ability
to deal with cryptographic functions. For example, such instructions 
and data types may allow them to work
with bits directly, or to implement specific cryptographic attacks,
such as guess-and-determine attacks \cite{Bard:2009:AC:1618541},
attacks based on differential paths (e.g., the ones from 
\cite{DBLP:conf/eurocrypt/WangLFCY05,DBLP:conf/eurocrypt/WangY05}), etc. 

In the present paper, we introduce a new software tool designed to encode algorithms that specify cryptographic functions to SAT. It is named \textsc{Transalg} (from TRANSlation of ALGorithms). \textsc{Transalg} uses a domain-specific language called TA language. The TA language is formed from a subset of 
the C language. It is  designed to be much simpler and to avoid platform-dependent or undefined behavior. The language is
extended by several specific data types and instructions
that often allow \textsc{Transalg} to tackle common cryptanalysis tasks  
better than the competition. In particular, the TA language has 
a specific \texttt{bit} data type to represent a single bit of data and supports \texttt{bit} arrays of arbitrary size. This allows \textsc{Transalg} to reduce the redundancy of constructed propositional encodings
and better preserve the structure of an original problem. 
The \textsc{Transalg} basic concept implies that a cryptographic
function is interpreted starting from its input and ending with its 
output. Thus, the language has specific directives to declare 
variables and arrays as input and output ones. As a result, the 
first variables in the constructed SAT encoding always 
correspond to the function's input, and the last ones to its output.
\textsc{Transalg} eliminates all the auxiliary variables that do not depend on input
and do not influence 
the output, and uses minimization (in the form of \textsc{Espresso} logic minimization tool  \cite{Brayton:1984:LMA:577427})
during the construction of propositional encodings to reduce their size. 
The TA language also uses
several specific constructions that make it easier for  \textsc{Transalg} users to employ the constructed SAT encodings for implementing 
cryptographic attacks in the SAT context.
In the present paper, we describe the \textsc{Transalg} tool and its theoretical
foundations in detail, compare it with the competition and show its capabilities in application
to cryptanalysis of several cryptographic systems which are currently
used or have been used in recent past.

Let us give an outline of the paper. In \autoref{sec:sat}, we briefly
touch the basics of the Boolean satisfiability problem. In \autoref{sec:theory}, we give the theoretical foundations of SAT-based cryptanalysis. Here we discuss several features of the procedures for translating 
programs defining discrete functions to SAT. As we show below, they are
particularly important in the context of cryptographic applications.
Also in the same section we discuss the main theoretical results that form
the basis of the software tool that performs effective
reductions of inversion problems of discrete functions to SAT. This software tool named \textsc{Transalg} is described in \autoref{sec:transalg}. In \autoref{sec:compar}, we 
compare the functionality of \textsc{Transalg} with that of other software tools
which can be used to encode cryptographic problems to SAT:
\textsc{CBMC}; \textsc{SAW}; \textsc{URSA}; \textsc{Grain-of-Salt}. 
In \autoref{sec:applications}, we describe SAT-based attacks on several relevant cryptographic functions. The corresponding SAT encodings were constructed using \textsc{Transalg}.  It should be noted that some of the described attacks are currently the best known. 
\autoref{sec:related}
contains a brief review of related works.
  
The present paper is an extended version of the report \cite{DBLP:conf/ecai/OtpuschennikovS16} presented at the ECAI 2016 conference. The sources of \textsc{Transalg} are available at 
\cite{transalg-repository}. The examples of \textsc{Transalg} programs for various
cryptographic primitives can be found at \cite{satencodings-repository}.
All the instances considered in \autoref{sec:compar} are also available online at
\cite{instances-repository}.

\section{The Boolean satisfiability problem and algorithms for its solving used~in~cryptanalysis}\label{sec:sat}

The Boolean satisfiability problem (SAT) is a decision problem, in which for an arbitrary Boolean formula $F$ it is necessary to decide whether there exists such truth assignment for its variables
on which formula $F$ takes the value of $True$.
Hereinafter, let us 
denote values $True$ and $False$ by $1$ and $0$, respectively.
It can be shown that SAT for an arbitrary Boolean formula $F$ can be 
effectively (in polynomial time in the size of 
description of $F$) reduced to SAT for a formula in a Conjunctive Normal Form (CNF). Further we will consider SAT exactly in this sense. Also, below we view SAT not only as a decision problem but also as the corresponding search problem: if CNF $C$ is
satisfiable to find any truth assignment that satisfies $C$.

The decision variant of SAT is NP-complete \cite{DBLP:conf/stoc/Cook71}, while the 
search variant is NP-hard \cite{Garey}. Nevertheless, during the last three decades several highly effective practical SAT solving algorithms were developed, driven by numerous applications in various areas of science. The detailed information on SAT and the algorithms for its solving can be found in  the book \cite{DBLP:series/faia/2009-185}. 

In the present paper, we consider the applications of SAT solving algorithms 
to cryptanalysis problems, in particular to the problem of finding a preimage
of a cryptographic function given its image. 
It can be viewed as a problem of finding solutions
of a system of 
algebraic equations which interconnect ciphering algorithm's steps. Lately,
the direction of research in which a cryptanalysis problem is viewed 
in the general context of the problem of solving  algebraic equations is often referred to as \textit{algebraic cryptanalysis}
\cite{Bard:2009:AC:1618541}. As it will be shown below, from an algebraic system
or even from an algorithm defining a cryptographic function one can effectively
transition to SAT. A number of valuable results in algebraic cryptanalysis were obtained thanks to the use of SAT solvers: \cite{DBLP:conf/SAT/MironovZ06,Courtois-tatra,CourtoisB2007,Courtois-cryptologia,
DBLP:conf/SAT/DeKV07,Bard:2009:AC:1618541,
DBLP:conf/SAT/SoosNC09,DBLP:conf/pact/SemenovZBP11} and several others. 
The particular area of algebraic cryptanalysis which employs SAT solvers 
is known as \textit{SAT-based cryptanalysis}.

We are not aware of theoretical results that would demonstrate advantage of some algorithms for solving SAT-based cryptanalysis instances over others.
However, based on a large
number of papers (both cited above and below) one can conclude that CDCL 
SAT solvers \cite{MSLM09} suit best for solving such problems. 
The construction of first effective CDCL SAT solvers was the result of 
a deep modernization of the well-known DPLL algorithm
\cite{Davis:1962,Davis:1960:CPQ:321033.321034}, 
which was undertaken in  \cite{Marques-Silva:1997,DBLP:journals/tc/Marques-SilvaS99,Moskewicz:2001,Zhang:2001:ECD:603095.603153,DBLP:conf/sat/EenS03}.
After this, CDCL-based SAT solvers became de facto
algorithmic tools for solving computational problems in a number of areas,
first and foremost in symbolic verification \cite{Biere:1999:SMC:646483.691738,Biere:1999:SMC:309847.309942,Biere2003BoundedMC,Prasad2005,JPMS2008}. The computational potential of CDCL 
in application to cryptanalysis problems was realized approximately in the middle of 2000-s. 
As it was noted above, to the present day a lot of papers have
been published in which CDCL SAT solvers were applied
to cryptanalysis problems. A short review of the
most prominent  results in this direction will be given in \autoref{sec:related}.

\section{Theoretical foundations of SAT-based cryptanalysis}\label{sec:theory}

As mentioned above, SAT-based cryptanalysis is an area of algebraic cryptanalysis (see \cite{Bard:2009:AC:1618541}) in which SAT solvers are used to solve equations that connect the input of a cryptographic algorithm with its output.
A cryptanalysis problem usually involves searching for a preimage of a known image of a considered function (in this case, the term \textit{preimage attack} is also used). In some cases, it is necessary to find several inputs whose outputs satisfy some additional constraints. Such constraints are used, for example, in the problem of finding collisions of cryptographic hash functions. 
Hereinafter, we refer to all these problems using the general term \textit{inversion problems}.

In this section, we provide theoretical foundations of SAT-based cryptanalysis. In~particular, we look into the construction of a Boolean formula that encodes a considered inversion problem. Below we follow the methodology of symbolic execution and in particular, Bounded Model Checking. 

\textit{Symbolic Execution} \cite{King:1976:SEP:360248.360252} is a technique that associates
with a program for a computer or for an abstract machine some symbolic expressions,
usually Boolean formulas. 
\textit{Bounded Model Checking} involves applying automated reasoning and combinatorial algorithms to a Boolean expression associated with a finite state system to prove some properties of such system \cite{Biere2003BoundedMC}. From our point of view, this approach best fits the problem of constructing SAT encodings for inversion of discrete functions 
in general, and for cryptographic functions especially.

Hereinafter, denote by $\{0,1\}^n$ the set of all possible binary words of length $n$. By $\{0,1\}^{+}$ we denote the set of all binary words of length $n=1,2,\ldots$.
Let us consider functions of the kind
\begin{equation}
\label{eq1}
f:\{0,1\}^{+} \rightarrow \{0,1\}^{+},
\end{equation}
i.e. functions that map arbitrary binary words into binary words. Additionally,
we assume that each function of the kind \eqref{eq1} is defined everywhere on $\{0,1\}^{+}$ (i.e. is total) and is specified by a Turing 
machine program $A(f)$, the complexity of which is bounded by a polynomial in
the length of an input word. A program $A(f)$ specifies an infinite family of 
functions of the kind
\begin{equation}
\label{eq2}
f_n:\{0,1\}^n \rightarrow \{0,1\}^{+}, n = 1,2,\ldots
\end{equation}
It is clear that for an arbitrary $n=1,2,\ldots$ it follows that $Dom\,f_n =\{0,1\}^n$. 
Hereinafter, to functions \eqref{eq1} and \eqref{eq2} we refer as \textit{discrete
functions}.

\begin{defi}
For a discrete function $f$ of the kind \eqref{eq1} the problem of its inversion
consists in the following. Given $A(f)$, for an arbitrary $n=1,2,\ldots$ and 
arbitrary $y\in Range\,f_n$ to find such $x\in\{0,1\}^n$ that $f_n(x)=y$.
\end{defi}

It is quite easy to give examples of cryptanalysis problems that can be naturally formulated in the context of  inversion problems of corresponding functions. The main terms related to cryptography that we use below can be found, e.g., in \cite{Menezes:1996:HAC:548089}.

In our first example, suppose that given a secret key $x\in \{0,1\}^n$, $f_n$ 
generates a pseudorandom sequence (generally speaking, of an arbitrary length) that is later used to cipher some plaintext via bit-wise XOR. Such a sequence is
called a \textit{keystream}. Knowing some fragment of plaintext lets us know the 
corresponding fragment of keystream, i.e. some word $y$ for which we can consider
the problem of finding such $x\in \{0,1\}^n$ that $f_n (x)=y$. Regarding
cryptographic keystream generators, this corresponds to the so called 
\textit{known plaintext attack}.

Let us give another example. Total functions of the kind $f:\{0,1\}^{+}
\rightarrow\{0,1\}^c$, where $c$ is some constant, are called \textit{hash
functions}. If $n$ is the length of an input message, and $n>c$, then there exist
such $x_1$, $x_2$, $x_1\neq x_2$ that $f_n(x_1)=f_n(x_2)$. Such a pair $x_1$, 
$x_2$ is called a \textit{collision} of a hash function $f$. A cryptographic hash 
function is considered compromised if one is able to find collisions of that 
function in reasonable time.

For an arbitrary function of the kind \eqref{eq1} there exists an effective 
in theory procedure for reducing the problem of its inversion to SAT. Essentially,
it follows from the Cook--Levin theorem, and to prove it one can use any known
technique, e.g., from \cite{DBLP:conf/stoc/Cook71} or \cite{Goldreich:2008:CCC:1373317}. 
Below we briefly review the main techniques used to prove statements of 
such a kind. They play a crucial role in understanding basic
principles that serve as a foundation of our software tool for 
translating algorithmic descriptions of discrete functions to SAT.

So, let us fix $n$ and consider an arbitrary function of the kind \eqref{eq2}.
Since the runtime of a program defining the corresponding function is finite for
any $x\in\{0,1\}^n$, we can consider this function in the following form:
\begin{equation}
\label{eq3}
f_n:\{0,1\}^n\rightarrow \{0,1\}^m.
\end{equation}
It can be specified by a Boolean circuit $S(f_n)$ over an
arbitrary complete basis. Hereinafter, we use the basis $\{\wedge,\neg\}$. On the current stage, assume that we are given a circuit $S(f_n)$. Note that it has $n$ inputs and $m$ outputs.
Let us fix some order on the sets of inputs and outputs of $S(f_n)$.
With each input of $S(f_n)$ we associate a Boolean variable. Denote the 
obtained ordered set of variables by $X=\{x_1,\ldots,x_n\}$. We will say that $X$ encodes
the input of function $f_n$. Similarly, let us encode the output of $f_n$ via
the Boolean variables forming the ordered set $Y=\{y_1,\ldots,y_m\}$. 
For a circuit $S(f_n)$  in linear time in the number of 
nodes in $S(f_n)$ we can construct a CNF denoted by $C(f_n)$. 
The corresponding algorithm traverses each inner node of 
a circuit exactly once. With each gate $G \in \{\wedge,\neg\}$ it associates
an auxiliary variable $v(G)$ from the set $V:V\cap X=\emptyset$.
For an arbitrary $v(G)$ a CNF $C(G)$ is then constructed which 
uses at most $3$ Boolean variables. The exact representation of
$C(G)$ depends on the gate $G$. The result of this process is 
the CNF: 
\begin{equation}
\label{eq5}
C(f_n) = \bigwedge_{G\in S(f_n)} C(G).
\end{equation}
The described technique of constructing a CNF for a circuit 
$S(f_n)$ is known as \textit{Tseitin transformations} \cite{Tseitin70}. 

\begin{defi}
\label{template}
To CNF $C(f_n)$ of the kind \eqref{eq5} we further refer as \emph{template CNF}
encoding the algorithm that specifies function $f_n$, or in short \emph{template CNF
for $f_n$}. 
\end{defi}

Let $u$ be an arbitrary Boolean variable. Below we will use the following notation:
by $l_\lambda(u)$, $\lambda\in \{0,1\}$ denote literal $\neg u$ if $\lambda=0$,
and literal $u$ if $\lambda=1$. Let $C(f_n)$ be a template CNF for $f_n$. Now let 
$x=(\alpha_1,\ldots,\alpha_n)$ and $y=(\gamma_1,\ldots,\gamma_m)$ be arbitrary truth
assignments from $\{0,1\}^n$ and $\{0,1\}^m$, respectively. In other words, let us consider
$x=(\alpha_1,\ldots,\alpha_n)$ as an assignment of variables from $X=\{x_1,\ldots,x_n\}$,
and $y=(\gamma_1,\ldots,\gamma_m)$ as an assignment of variables from  $Y=\{y_1,\ldots,y_m\}$.
Consider the following two CNFs:
\begin{equation*}
\begin{array}{lcc }
C(x,f_n) & = &l_{\alpha_1} (x_1)\wedge\ldots\wedge l_{\alpha_n} (x_n )\wedge C(f_n ),\\
C(f_n,y) & = &C(f_n )\wedge l_{\gamma_1} (y_1)\wedge\ldots\wedge l_{\gamma_m} (y_m ).
\end{array}
\end{equation*}
For many practical applications of SAT and, in particular, to describe many
cryptographic attacks studied in algebraic cryptanalysis (see, e.g., 
\cite{SemeonovZOKI2018}) the following fact plays a very important role.

\begin{rem}
The application of only the \textit{Unit Propagation} rule \cite{dowling-jlogp84,MSLM09}
to CNF $C(x,f_n)$ for a particular $x=(\alpha_1,\ldots,\alpha_n)$ results
in the derivation of values for all remaining variables, including 
that of variables from $Y$: $y_1=\gamma_1,\ldots,y_m=\gamma_m$, such that 
$f_n(x)=y$, $y=(\gamma_1,\ldots,\gamma_m)$.
\end{rem}

This property was several times used in other papers (see, e.g., 
\cite{Jarvisalo2009,DBLP:journals/jar/JarvisaloBH12,DBLP:conf/pact/SemenovZ15}).
Its proof in a very similar formulation can be found in 
\cite{DBLP:conf/ijcai/BessiereKNW09}. Essentially, it
follows from the fact that the set $X$ in $C(f_n)$ is
a Strong Unit Propagation Backdoor Set 
(SUPBS) \cite{Williams:2003:BTC:1630659.1630827}.

The following statement is a variant of the Cook--Levin
theorem in the context of the problem of inverting functions of the
kind \eqref{eq3}. The basic steps of its proof are standard and 
can be found, for example, in \cite{Goldreich:2008:CCC:1373317}. However, from our
point of view there are several technical issues that should be
clarified for better understanding of how the software tool described below 
works with data. That is why we present the short proof of this 
statement detailing only the features that play an important role 
in the context of this study.

\begin{thm}
\label{Thm1}
Let $f$ be an arbitrary function of the kind \eqref{eq1}. Then there exists an algorithm $A'$ such that given as an input a program $A(f)$, a
number $n$ (in unary form), and a word $y\in\{0,1\}^{+}$, in polynomial time it constructs a CNF $C(f_n,y)$
with the following properties:
\begin{enumerate}
\item For $y\notin Range\,f_n$ the CNF $C(f_n,y)$ is unsatisfiable.
\item For $y\in Range\,f_n$ the CNF $C(f_n,y)$ is satisfiable and from any of its 
satisfying assignments one can extract such a word $x\in\{0,1\}^n$ that $f_n(x)=y$.
\end{enumerate}
\end{thm}
\noindent\textit{Sketch proof.}
Assume that the program $A(f)$ is executed on the Turing
machine, described in \cite{Garey}, which works only with binary data. By algorithm
$A'$ we mean the informal procedure that constructs a circuit $S(f_n)$ based on the text of the program $A(f)$ and a number~$n$. 

Note that the transition function in the machine from
\cite{Garey} looks as follows:
\begin{equation}
\label{eq6}
\delta:(q,\theta)\rightarrow(q',\theta',s),
\end{equation}
where $\theta\in\{0,1,\emptyset\}$ is an input symbol, $q\in Q$ is an arbitrary state, and $s$ is a variable that defines the direction in which the head of the Turing machine is going 
to shift, i.e. $s\in\{-1,0,+1\}$. The function \eqref{eq6} describes the execution of one elementary command.

Let us consider the execution of program $A(f)$ as a sequence of time moments, such 
that during the transition from one time moment to another exactly one elementary command
is executed. The moment $t=0$ corresponds to a starting configuration. With each moment
$t$ we associate the set of Boolean variables $X_t$, $X_i\cap X_j =\emptyset$ if 
$i\neq j$. With the transition from $t$ to $t+1$ we associate a formula
\begin{equation}
\label{eq7}
\Psi_{t\rightarrow t+1}= \bigvee_{(q,\theta)}\Phi^{(q,\theta)}_{t\rightarrow t+1}.
\end{equation}
An arbitrary formula $\Phi^{(q,\theta)}_{t\rightarrow t+1}$ is a formula of the kind
\begin{equation*}
\phi_t(q,\theta)\Rightarrow \phi_{t+1}(q',\theta'),
\end{equation*}
(here by $\Rightarrow$ we denote logical implication), which is constructed in the 
following manner. The formula $\phi_t(q,\theta)$ is a conjunction of literals over the
set $X_t$ that encodes a particular pair $(q,\theta)$ and also the state of the head of the Turing machine at
the moment $t$. The formula $\phi_{t+1}(q',\theta')$ is the conjunction of literals over
the set of Boolean variables $X_{t+1}$ that encodes a pair $(q',\theta')$ and the 
state of the head corresponding to the triple $(q',\theta',s)$. The correspondence 
between $(q,\theta)$ and $(q',\theta',s)$ is defined by the transition function 
\eqref{eq6}. In \eqref{eq7} the disjunction is performed over all possible pairs 
$(q,\theta)$ in the program $A(f)$.

It is very important to note that the cardinality of $Q$ does not depend on $n$. Therefore,
the size of the formula $\Psi_{t\rightarrow t+1}$ is a constant that does not depend
on $n$. Omitting some details, let us note that from the above a formula 
$\Psi_{t\rightarrow t+1}$ defines a function
\begin{equation*}
F_{t\rightarrow t+1}:\{0,1\}^{|X_t|}\rightarrow \{0,1\}^{|X_{t+1}|},
\end{equation*}
which can also be specified using Boolean circuit $S(F_{t\rightarrow t+1})$ over the 
basis $\{\wedge,\neg\}$, which has $|X_t|$ inputs and $|X_{t+1}|$ outputs. The sets 
$X_t$ and $X_{t+1}$ are the sets of input and output variables of a circuit 
$S(F_{t\rightarrow t+1})$, respectively, and $X_{t+1}$ is the set of input variables
of circuit $S(F_{t+1\rightarrow t+2})$. 

Let $t(n)$ be the upper bound on the runtime of the program $A(f_n)$ over all inputs from
$\{0,1\}^n$. By combining the circuits $S(F_{t_0\rightarrow t_1}),\ldots,S(
F_{t(n)-1\rightarrow t(n)})$ according to the above we construct the circuit for 
which it is easy to see that it specifies the function $f_n$ of the kind \eqref{eq3}. This circuit is $S(f_n)$.

Let us construct the template CNF $C(f_n)$ for the circuit $S(f_n)$.
Let $y$ be an arbitrary
assignment from $\{0,1\}^m$. Consider a CNF $C(f_n,y)$. 
Now we use the property mentioned above and conclude that 
$X$ is a SUPBS in $C(f_n,y)$. Thus it follows that the points (1) and (2) from the Theorem formulation are valid. \qed

We would like to give some additional comments regarding Theorem \ref{Thm1}.
Using the properties of the Tseitin transformations, it is easy to show
that the reduction presented in Theorem \ref{Thm1} is parsimonious 
\cite{Garey}, i.e. the number of assignments that satisfy CNF $C(f_n,y)$, $y\in Range\,f_n$, 
is equal to the number of preimages of $y$.
For further purposes it will be enough for us to find at least one preimage of $y \in Range\,f_n$.
It should be noted that the values of variables from $X$ in an arbitrary satisfying assignment 
of $C(f_n,y)$, $y\in Range\,f_n$, specify a preimage of $y$. This follows directly from the fact that
$X$ is SUPBS for $C(f_n,y)$.
Indeed, let $\alpha$ be a satisfying assignment of $C(f_n,y)$ 
and $x=(\alpha_1,...,\alpha_n)$ be an assignment of variables from $X$ extracted from $\alpha$.
Suppose that $x$ is \textit{not} a preimage of $y$. 
Then, since $X$ is SUPBS, the application of UP to CNF 
$$
l_{\alpha_1}(x_1) \wedge \ldots \wedge l_{\alpha_n}(x_n) \wedge C(f_n,y)
$$
should lead to a conflict. Thus, we have a contradiction with the fact that $x$ 
is a part of the assignment that satisfies $C(f_n,y)$.

Based on Theorem \ref{Thm1}, let us formulate the general concept of SAT-based cryptanalysis.
Assume that we have a function $f$ of the kind \eqref{eq1}, and  consider a problem
of finding a preimage of a particular $y\in Range\,f_n$ for a fixed $n$. Then, using Theorem \ref{Thm1} we construct
CNF $C(f_n,y)$. From any satisfying assignment of $C(f_n,y)$ it is easy to extract such $x\in\{0,1\}^n$ that $f_n(x)=y$.

As a concluding remark we would like to note that an input word 
from $\{0,1\}^n$, employed by the procedure used in the proof of 
Theorem \ref{Thm1} to transition from a program $A(f)$ to a template CNF $C(f_n)$,  is not constrained in any way. In fact, this procedure takes Boolean variables $x_1,\ldots,x_n$ as an input and outputs $C(f_n)$, thus essentially 
performing symbolic execution.

\section{Transalg: software tool for encoding algorithmic descriptions of~discrete~functions~to~SAT}\label{sec:transalg}

In the present section, we describe the \textsc{Transalg} software tool 
that in essence implements the translation procedure
for transforming algorithmic descriptions of functions of the kind \eqref{eq3} to
SAT, which was outlined in Theorem \ref{Thm1}. The only conceptual
difference is that instead of the Turing machine \textsc{Transalg}
uses an abstract machine with random access to memory cells.

\textsc{Transalg} takes as an input an algorithm that specifies 
a discrete function in a special TA language. Then it 
uses this description to construct a symbolic representation of 
the algorithm (in the sense of symbolic execution). We refer to 
the obtained representation as to \textit{propositional encoding}.
The propositional encoding is first built as a set of Boolean formulas, and then can be transformed to the DIMACS CNF format or
the AIGER format \cite{Biere-FMV-TR-07-1}. 

In \textsc{Transalg} the process of computing a value of a discrete 
function $f_n$ is represented as a sequence of elementary operations 
with memory cells of an abstract machine. Each memory cell contains 
one bit of information. Any elementary operation $o$ over data in 
memory cells is essentially a Boolean function of arity $k$, $k\geq 1$, 
where $k$ is some constant that does not depend on the size of input 
(strictly speaking, it is possible to consider $k\in\{1,2\}$). 
For example, if $o$ has the arity of $2$ then the result of applying $o$ to data in memory cells $c_1$ and $c_2$ is one bit that is 
written to memory cell $c_3$. However, during the construction of 
a propositional encoding  \textsc{Transalg}  does not use the real 
data. Instead, it links with the cells $c_1$, $c_2$, $c_3$ 
the Boolean variables $v_1$, $v_2$ and $w$, respectively. Then it 
associates with operation $o$ the Boolean formula
\begin{equation}
\label{eq3-1}
w\equiv \phi_o(v_1,v_2),
\end{equation}
where $\phi_o(v_1,v_2)$ specifies a function $o$. 
For an arbitrary formula of the kind $\phi_o(v_1,v_2)$, we represent 
the corresponding function as a Boolean circuit. \textsc{Transalg}
can work with different basis functions, but in the most simple case
it can construct a circuit over $\{\wedge,\neg\}$.  

\subsection{TA language}

To describe discrete functions, \textsc{Transalg} uses a domain specific 
language called TA language. The TA language has a C-like syntax and block structure.
An arbitrary block (composite operator) is essentially a list of instructions, and has 
its own (local) scope. In the TA language, one can use nested blocks with no limit on 
depth. During the analysis of a program, \textsc{Transalg} constructs a scope tree with the global
scope at its root. Every identifier in a TA program belongs to some scope. Variables 
and arrays declared outside of any block and also all functions belong to the global 
scope and therefore can be accessed in any point of a program.

A TA program is a list of functions. The \texttt{main} function is the entry point and, thus, 
must exist in every program. The TA language supports basic constructions used in 
procedural languages (variable declarations, assignment operators, conditional 
operators, loops, function calls, etc.), various integer operations and bit operations 
including bit shifting and comparison. 

Similar to most symbolic execution systems, \textsc{Transalg} supports loops with fixed length
and processes them via unwinding. It also supports conditional operators with any depth of nesting. 
On the level of ideas the corresponding solutions do not differ from those employed in symbolic 
verification systems, such as \textsc{CBMC} (see, e.g., \cite{DBLP:series/faia/Kroening09}). 
Briefly, the processing of a conditional operator is based on the following considerations. 
Each conditional operator of the kind \texttt{if then else} is associated with two arrays 
$R_1$ and $R_2$ in the memory of an abstract computing machine. The contents of these arrays 
represent two alternatives for data that will be in the memory of the machine after executing 
the conditional operator. With the cells of arrays $R_1$ and $R_2$ we first associate the 
encoding variables. Each encoding variable encodes the Boolean value that is the result of 
execution of this conditional operator. 

The main data type in the TA language is the 
\texttt{bit} type, which can be used to specify arrays of bits of an arbitrary
finite length.
\textsc{Transalg} uses this data type to establish links between 
variables used in a TA program and Boolean variables included into a corresponding
propositional encoding. 
It is important to note that \textsc{Transalg} does not establish such links for 
variables of other types, in particular \texttt{int} and \texttt{void}, 
which are used as service variables, e.g., as loop counters or to specify 
functions that do not return any value.
We will refer to variables that appear in a TA program as
\textit{program variables}. All variables included in a propositional encoding are 
called \textit{encoding variables}. Given a TA program $A$ that specifies $f_n$, 
\textsc{Transalg} constructs a propositional encoding of $f_n$. Below
we will refer to this process as to the \textit{translation} of the TA program $A$. 

Declarations of global \texttt{bit} variables can have the \texttt{\_\_in} or the 
\texttt{\_\_out} attribute. The \texttt{\_\_in} attribute marks variables that 
correspond to the algorithm's input. The \texttt{\_\_out} attribute marks variables 
that correspond to the algorithm's output. Local \texttt{bit} variables cannot be declared
with these attributes. Note, that the TA language strictly fixes the order in
which it introduces Boolean variables at all steps of a considered algorithm.
It means that the variables encoding the algorithm's input are always numbered from $1$ to $n$,
where $n$ is the length of input in bits. The variables corresponding to output 
are always represented by the last $m$ variables in an encoding, where $m$ is the 
length of output in bits.  Thus if necessary it is possible to exactly and
explicitly associate input and output of an algorithm with the corresponding 
Boolean variables in a CNF (e.g., when manually invoking a SAT solver and
processing its output).

\subsection{Techniques aimed at reducing the redundancy of propositional encodings}

In the process of symbolic execution of algorithms it is often the case
that redundant variables and constraints are introduced. By calling them redundant we mean that they do not provide any additional information and can be safely removed. \textsc{Transalg} uses several techniques that often make it possible to significantly reduce the redundancy of a resulting SAT encoding. 

The first technique exploits the fact that many algorithms can be 
represented in form of sequences of procedures which are simple and very 
similar to each other. Therefore, during the symbolic execution it is possible 
that the same Boolean formulas will be generated multiple times. Taking this 
fact into account, for each new formula \textsc{Transalg} first checks whether 
it is already present in the database. If the answer is ``no'', then the newly 
constructed formula is added to the database and associated with a new encoding 
variable. Otherwise, on the following steps the variable associated with the 
existing formula from the database is used. The approach is close to that 
introduced in \cite{10.1007/3-540-46419-0_28}.

Another technique is related to the \texttt{\_\_in} and \texttt{\_\_out}
attributes. Upon the generation of the resulting encoding, \textsc{Transalg} analyzes all 
functional dependencies of encoding variables on one another in order to define 
the minimally required set of variables that influence the construction of 
an output from an input. All the remaining variables and formulas defined using
them are safely removed from an encoding without influencing its correctness.
After this the variables are renumbered to exclude gaps.

The third technique aimed at reducing the number of auxiliary variables in a resulting 
propositional encoding works as follows.
\textsc{Transalg} can use Boolean functions with arity $k>2$ in the role of 
elementary operations over data in memory cells of its abstract machine. 
Therefore, as a result of each elementary step, a new encoding variable $v$ is introduced and the
following Boolean formula is constructed:
\begin{equation}
\label{eq8}
v\equiv \phi(\tilde{v}_1,\ldots,\tilde{v}_k).
\end{equation}
Here, $\tilde{v}_1,\ldots,\tilde{v}_k$ are some encoding variables introduced 
at the previous steps. In other words, it is possible to represent $f_n$ over any complete basis 
with arbitrarily complex basis functions. To transform \eqref{eq8} into CNF,  \textsc{Transalg} uses the well-known \textsc{Espresso} library \cite{Brayton:1984:LMA:577427}.
The arity of a function, which is given as an input to \textsc{Espresso}, is often a very serious limitation: for functions with 
more than 20 inputs the performance of \textsc{Espresso} is beginning to have a significant 
impact on the time of SAT encoding construction. 
To counter this issue, \textsc{Transalg} implements the ability to split formulas of this kind into several disjoint parts. 
Each such part is associated with a separate encoding variable. 
In the TA language these variables are declared using a special \texttt{\_\_mem} attribute.
The described technique gives the user more manual control: for example, using \texttt{\_\_mem}, one can change the ratio between the number of variables and the number of clauses in the resulting SAT encoding.

\subsection{Cryptographic-specific features of \textsc{Transalg}}
\textsc{Transalg} also has several features that are specific for cryptographic algorithms and
the use cases typical for algebraic cryptanalysis.
In fact, one of the main features of \textsc{Transalg} that make it 
better for cryptographic problems is the \texttt{bit} type.
The availability of this type is particularly useful when working with keystream generators that
have shift registers of sizes which are not multiple of 8 (e.g., 19, 22, 23 bits). Note
that in general purpose programming languages, such as C, memory is allocated in
blocks of bits of a fixed size that is dependent both on the size of supported data
types (8, 16, 32, 64 bits) and a particular compiler's implementation. Thus, a C program for a keystream generator with a register of size, say, 19 bits,  inevitably processes excess data. For example, to represent such a shift register (of size 19) it would use a 32-bit integer variable. Consequently, this problem remains relevant for generic systems that employ the C language in that they require additional procedures to remove redundant Boolean variables from an encoding. Another important feature of \textsc{Transalg} is that it allows us to work with \texttt{bit} arrays simultaneously as arrays and as integer variables (represented in binary form). In particular, it can perform basic arithmetic operations (addition, subtraction, multiplication) with \texttt{bit} arrays without any additional data type transformations.

Cryptographic algorithms often use various bit shifting operators and also copy bits
from one cell to another without changing their value. During the symbolic execution 
of such operators, there may appear elementary steps producing the formulas of the 
kind $v\equiv \tilde{v}$. However, we do not really need such formulas in a propositional encoding since it is evident that without the loss of correctness 
we can replace an arbitrary formula of the kind $v'\equiv \phi(v,\ldots)$ by a 
formula $v'\equiv \phi(\tilde{v},\ldots)$. 
In such cases \textsc{Transalg} does not introduce new encoding variables.

When implementing cryptographic attacks in SAT form, it is sometimes
desirable to manually track (and manipulate) the values of Boolean
variables corresponding to specific program variables. 
The TA language allows us to add a special directive that outputs 
the numbers of encoding variables corresponding to specific program
variables into the header of a propositional encoding. For this purpose
it uses the \texttt{core\_vars(<program variable>)}
instruction. Here \texttt{program variable} can be a \texttt{bit} variable or a 
\texttt{bit} array. During the translation of a TA program, \textsc{Transalg}
will put into the DIMACS file header the numbers of encoding
variables associated with specified program variable at the moment
when the \texttt{core\_vars()} instruction is executed. Usually,
the obtained variable numbers are used either to parameterize a
SAT solver or in special heuristics, such as in  \cite{DBLP:conf/SAT/DeKV07}.

It is often demanded by attacks to impose specific constraints 
on the values of variables at certain steps of an algorithm. 
Since \textsc{Transalg} monitors the values of program variables 
inside TA program at any step of computing, it also allows us to 
impose any constraints on such variables.
For this purpose the TA language uses the \texttt{assert(<expression>)} instruction. This instruction assumes that \texttt{expression} takes the value of $True$. The Boolean formula corresponding to \texttt{expression} is added to the resulting propositional encoding.

Note, that the functionality related to \texttt{core\_vars} and
\texttt{assert} instructions can only be used if \textsc{Transalg}
outputs the propositional encoding in the DIMACS CNF format. 

\subsection{Example of a TA program and its translation}
Let us consider the following example. Its goal is to demonstrate how new encoding variables and constraints involving them are introduced in the course of interpretation of a simple TA program.
Note, that in this example we do not consider all technical nuances of 
the propositional encoding procedures implemented in \textsc{Transalg}. In particular,
here we completely omit the post-processing stage, at which the tool removes all the variables
and constraints that are redundant (because they do not influence the 
output in any way). The variable reindexing is 
also performed during the post-processing.

\begin{exa}
\label{ex1}
Consider an encoding of a linear feedback shift register (LFSR) \cite{Menezes:1996:HAC:548089} with
\textsc{Transalg}. In \autoref{fig:rslos-ta}, we show the TA program for the LFSR with feedback 
polynomial $P(z)=z^{19}+z^{18}+z^{17}+z^{14}+1$ over $GF(2)$ (here $z$ is a formal 
variable).
\begin{figure}[htbp]
\centering
		\includegraphics[width=1\textwidth]{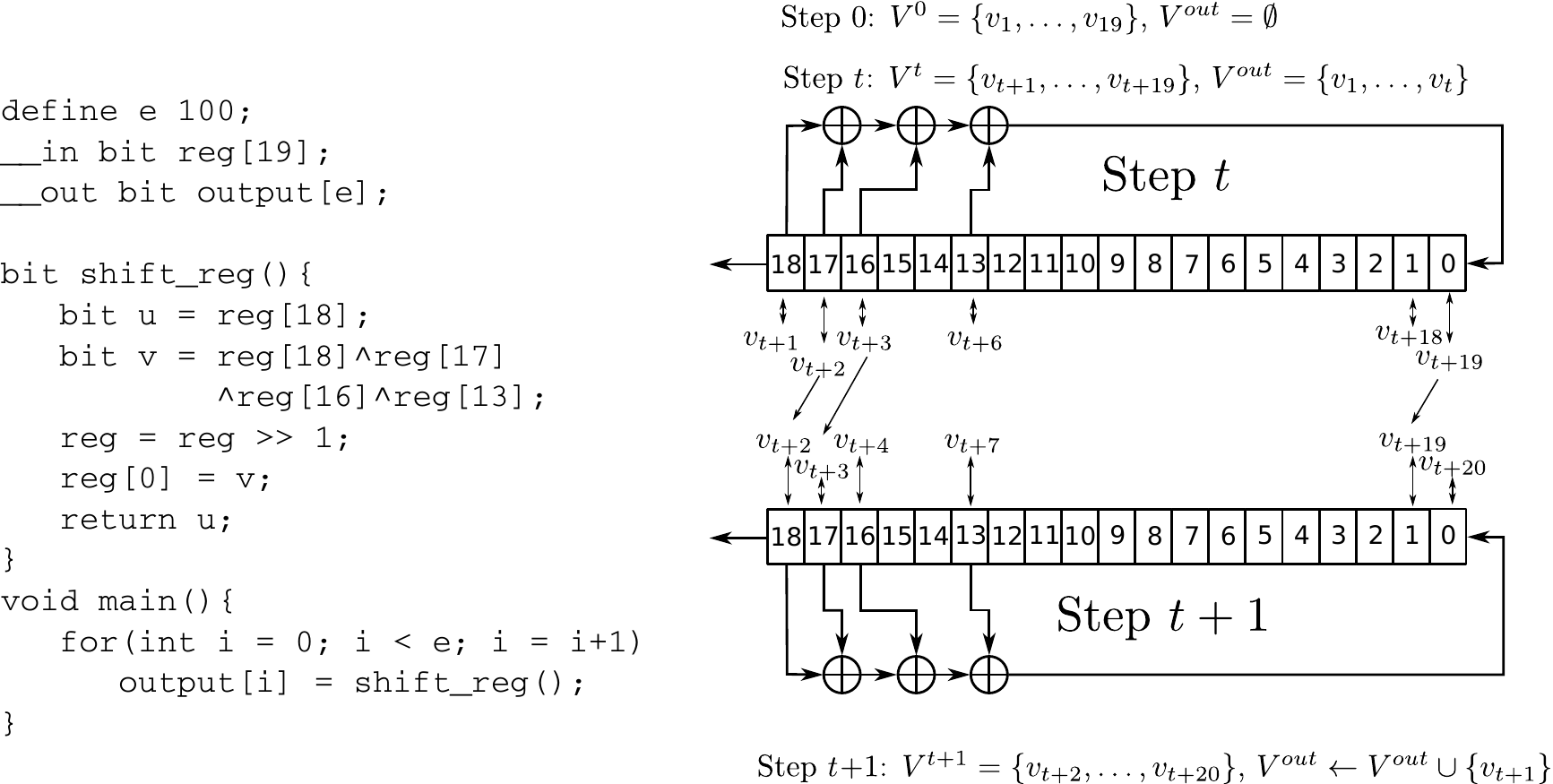}
	\caption{TA program for LFSR}
	\label{fig:rslos-ta}
\end{figure}
Let us view the process of executing this TA program as a sequence of data 
modifications in a memory of an abstract computing machine at moments $\{1,\ldots,e\}$.
At every moment $t\in \{0,1,\ldots,e\}$, \textsc{Transalg} links a set $V^t$ of encoding 
variables with program variables of the \texttt{bit} type. Denote  $V=\bigcup_{t=0}^{e}V^t$.
Let us separately denote by $V^{in}$ and $V^{out}$ the sets formed by encoding variables that correspond to input and output data, respectively. Note that during the translation 
of transition from moment $t$ to moment $t+1$ it is not necessary to create new 
encoding variables for every cell of the register. If we copy data from one register
cell to another, then we can use the same encoding variable to represent the corresponding 
data value at moments $t$ and $t+1$. Therefore, at each moment \textsc{Transalg} creates
only one new encoding variable and links it with program variable \texttt{reg[0]}. All 
the other program variables get linked with encoding variables created at previous moments.

In accordance with the above, the set of encoding variables corresponding to the initial values 
of the register is $V^{in}=V^0=\{v_1,\ldots,v_{19}\}$. After each shift we encode values 
of register's cells with sets 
\begin{equation*}
V^1=\{v_2,v_3,\ldots,v_{20}\},V^2=\{v_3,v_4,\ldots,v_{21}\},\ldots,V^e=\{v_{e+1},v_{e+2},
\ldots,v_{e+19}\}.
\end{equation*}
Note that $V^{out}=\{v_1,\ldots,v_e\}$. Thus the set of encoding variables for
this program is $V=\{v_1,v_2,\ldots,v_{e+19}\}$, and the corresponding variables are connected
between each other by the following Boolean formulas (here $\oplus$ stands for addition modulo 2):
\begin{equation*}
\begin{array}{lcc}
v_{20} &\equiv &v_1\oplus v_2\oplus v_3\oplus v_6\\
&\ldots\\
v_{e+19}&\equiv &v_e\oplus v_{e+1}\oplus v_{e+2}\oplus v_{e+5}.
\end{array}
\end{equation*}
\end{exa}
Note that in the example the register's size is not a multiple of $8$. As it was
noted above, in the tools that employ general purpose programming languages, such as C, the description of such an algorithm would require either using a 32-bit integer variable to
store register's state or using nineteen 8-bit variables each representing a single bit.
Either way there would be quite a lot of unused bits in a constructed SAT encoding, which
would need to be taken care of.


\section{Comparison of tools for constructing SAT encodings of cryptanalysis problems}\label{sec:compar}

In the present section, we briefly describe major software tools that can translate cryptographic algorithms to SAT, compare their functional capabilities with that of \textsc{Transalg}, and study the performance of state-of-the-art SAT solvers on several classes of SAT encodings obtained by all discussed tools.

\subsection{Tools for translating cryptographic algorithms to SAT}
We are aware of several domain-specific software tools (besides \textsc{Transalg}) that can be used to encode  cryptographic algorithms to SAT. Below we provide their brief description.

The \textsc{Grain-of-Salt} tool \cite{DBLP:conf/tools/Soos10} is designed to produce SAT encodings only for cryptographic keystream generators based on the shift registers.  It uses a special declarative language to describe each of keystream registers (by means of its feedback polynomials)
and the general configuration of a generator.
Unfortunately, \textsc{Grain-of-Salt} does not support  many 
standard cryptographic operations  
and therefore can be applied only to a limited spectrum of cryptographic functions.

\textsc{URSA} (a system for Uniform Reduction to SAT \cite{journals/lmcs/predrag}) is a  propositional encoding tool that is applicable to a wide class of combinatorial problems, varying from CSP (Constraint Satisfaction Problem) to cryptography. To describe these problems, \textsc{URSA} employs a declarative domain-specific language. The constructed SAT instances can be solved using two embedded solvers: \textsc{ARGOSAT} \cite{Maric09} and \textsc{CLASP} \cite{Gebser07}. 

\textsc{Cryptol} is a domain-specific Haskell-like programming language for specifying  cryptographic algorithms \cite{LewisM03,ErkokPLPV09,DBLP:conf/fmcad/ErkokCW09}. 
Software Analysis Workbench (\textsc{SAW}) \cite{Carter2013} allows for producing SAT and SMT encodings of cryptographic problems described in \textsc{Cryptol}. Further we refer to \textsc{SAW} that takes as an input a program written in \textsc{Cryptol} as \textsc{SAW+Cryptol}.

Another major class of tools that can translate cryptographic algorithms to SAT 
is formed by the systems designed for software verification. Probably the most well known
and powerful representative of this class is the \textsc{CBMC} tool (Bounded Model Checking for ANSI C \cite{ckl2004}).
\textsc{CBMC} works with programs written in the C language. Note, that 
the primary application domain of \textsc{CBMC} is software verification.
Therefore, for each program there should be a hypothesis that needs
verifying or falsifying. Putting cryptanalysis attacks in this context
requires some paradigm adjustment, but in the case of \textsc{CBMC} it 
can be done quite easily.

\subsection{Computational comparison of tools}
To compare the effectiveness of propositional encodings produced by the aforementioned tools, we chose several cryptographic keystream generators. Here they are, ordered by the resistance to SAT-based cryptanalysis (from the weakest to the strongest): Geffe \cite{Geffe}, Wolfram \cite{Wolfram}, Bivium \cite{ DBLP:conf/isw/Canniere06} and Grain \cite{Hell:2007:GSC:1358393.1358401}. The Geffe generator is a simple generator, which is not resistant to many cryptographic attacks including the correlation attack proposed in \cite{Siegenthaler84}. We considered the strengthened Geffe generator (to which we further refer as S\_Geffe), which is a particular case of the threshold generator \cite{Bruer}. It turned out that S\_Geffe with a secret key length up to 160 bits is not resistant to SAT-based cryptanalysis (when implementing the known plaintext attack). We considered the variant of the S\_Geffe generator that uses three LFSRs defined by the following primitive polynomials:
\begin{equation*}
\begin{array}{l}
z^{52} + z^{49} + 1;\\
z^{53} + z^{52} + z^{38} + z^{37} + 1;\\
z^{55} + z^{31} + 1.
\end{array}
\end{equation*}
Thus, the considered generator has a secret key of 160 bits.

Unlike many other generators, the Wolfram generator does not use shift registers. It is based on a one-dimensional cellular automaton \cite{vonN51}. This generator is not resistant to the attack proposed in 
\cite{MeierS91} if its secret key length is less than 500 bits. Meanwhile, the cryptanalysis of the Wolfram 
generator with the secret key length of more than 150 bits is already a hard problem for state-of-the-art SAT solvers. Therefore, below we consider the  version of the generator with 128-bit secret key. The Bivium generator \cite{DBLP:conf/isw/Canniere06} is a 
popular object for SAT-based and algebraic cryptanalysis. Nevertheless, SAT-based cryptanalysis of Bivium is quite a hard problem that, as the estimations show \cite{Semenov2016}, can be solved in reasonable time in a powerful distributed computing environment. Finally, we considered the Grain generator \cite{Hell:2007:GSC:1358393.1358401}, in particular, its v1 version. There are no known attacks on this version that are better than the brute force attack.

For each mentioned generator a SAT-based variant of the known plaintext attack was studied. It means that the following problem was considered: to invert a function of the form $g: \{0,1\}^n \rightarrow \{0,1\}^m$, where $n$ is the amount of bits of registers' state that produces the analyzed keystream, and $m$ is the length of the analyzed keystream. Using \textsc{SAW+Cryptol}, \textsc{URSA}, \textsc{CBMC}, and \textsc{Transalg}, we built propositional encodings of the following functions:
\begin{equation*} 
\begin{array}{lll}
g^{\mathit{S\_Geffe}}&:& \{0,1\}^{160} \rightarrow \{0,1\}^{250}, \\
g^{\mathit{Wolfram}}&:& \{0,1\}^{128} \rightarrow \{0,1\}^{128}, \\
g^{\mathit{Bivium}}&:& \{0,1\}^{177} \rightarrow \{0,1\}^{200}, \\
g^{\mathit{Grain}}&:& \{0,1\}^{160} \rightarrow \{0,1\}^{160}.
\end{array}    
\end{equation*}
It should be noted that since \textsc{Grain-of-Salt} operates only with shift registers, it was not possible to construct SAT encodings of the Wolfram generator via this tool. 
In \autoref{encodings_table} the obtained encodings are compared by the amount of variables, clauses, and literals.

\begin{table}[htbp]
\begin{center}
{\caption{The characteristics of SAT encodings.}
\label{encodings_table}}	
\begin{tabular}{|l|r|r|r|r|r|}
			\hline
			& \textsc{Grain-of-Salt} & \textsc{URSA} & \textsc{SAW+Cryptol} & \textsc{Transalg} & \textsc{CBMC} \\
            \hline
            \multicolumn{6}{|c|}{S\_Geffe} \\
			\hline
			Vars & 1~910 & 2~394 & 1~883 & 1~000 & 2~668 \\
			\hline
			Clauses & 8~224 & 8~436 & 6 891 & 6~474 & 9~514 \\
			\hline
			Literals & 28~976 & 23~308 & 19~793 & 25~226 & 26~536 \\
			\hline
			\hline
            \multicolumn{6}{|c|}{Wolfram} \\
			\hline
			Vars & - & 24~704 & 24~620 & 12~544 & 32~904 \\
			\hline
			Clauses & - & 86~144 & 85~784 & 74~112 & 114~830 \\
			\hline
			Literals & - & 233~600 & 232~811 & 246~400 & 311~460 \\
			\hline
			\hline
            \multicolumn{6}{|c|}{Bivium} \\
			\hline
			Vars & 842 & 1~637 & 1~432 & 1~172 & 1~985 \\
			\hline
			Clauses & 6~635 & 5~975 & 5~308 & 7~405 & 7~044 \\
			\hline
			Literals & 29~455 & 16~995 & 15~060 & 29~745 & 19~866 \\
			\hline
			\hline
			\multicolumn{6}{|c|}{Grain} \\
			\hline
			Vars & 4~546 & 9~279 & 4~246 & 1~945 & 10~088 \\
			\hline
			Clauses & 74~269 & 37~317 & 16~522 & 34~165 & 40~596 \\
			\hline
			Literals & 461~069 & 105~925 & 46~402 & 190~388 & 115~178 \\
			\hline
		\end{tabular}
		\end{center}
\end{table}

At the second stage, we used the constructed encodings for implementing the known plaintext attack on the described generators. As it was mentioned above, the inversion problems for $g^{\mathit{S\_Geffe}}: \{0,1\}^{160} \rightarrow \{0,1\}^{250}$ and $g^{\mathit{Wolfram}}: \{0,1\}^{128} \rightarrow \{0,1\}^{128}$ are quite easy even for sequential SAT solvers. Meanwhile, the inversion problem for $g^{\mathit{Bivium}}: \{0,1\}^{177} \rightarrow \{0,1\}^{200}$ is hard even for the best parallel SAT solvers. The inversion problem for $ g^{\mathit{Grain}}: \{0,1\}^{160} \rightarrow \{0,1\}^{160}$ is extremely hard and cannot be solved in reasonable time in any modern distributed computing system that we are aware of. 
That is why we studied weakened variants of the last two problems. In particular, SAT solvers were given some information about unknown registers' state. In the case of Bivium, the last 30 bits of the second register were assumed to be known. In the case of Grain, 102 (out of 160) bits were known: the whole 80-bit linear register and the last 22 bits of the 80-bit nonlinear register. The constructed SAT instances are denoted by Bivium30 and Grain102, respectively. 

For each considered generator, a set of 100 SAT instances was constructed 
by generating random values of corresponding registers states, which were used to produce keystreams. It means that for each generator we randomly constructed $100$ variants of registers' initial states,
then used the implementation of the generator to produce the corresponding
$100$ fragments of keystream. After this, we constructed SAT 
instances (by  each of the considered tools) and SMT instances (by
\textsc{SAW+Cryptol} and \textsc{CBMC}) for all generated pairs 
of initial states and keystream fragments. The goal of using the 
pre-generated pairs of initial register's states and keystream
fragments is to perform the testing in equal conditions: indeed,
it is often the case that for some keystream fragments the inversion
problems are easier to tackle due 
to specific features of a corresponding algorithm.
Thus it makes sense to compare 
over the exact same cryptanalysis problems encoded to SAT in different forms.
All constructed SAT and SMT instances are available online \cite{instances-repository}. 

Let us give a few more comments on the matter. It turned out that the \textsc{Grain-of-Salt}
tool neither allows for constructing a template CNF nor has any instructions to assign the Boolean 
variables corresponding to keystream bits to some pre-specified values. It basically
generates and uses a random keystream fragment every time it is run. Thus, we had to 
resort to extracting the randomly generated keystream fragment (and initial state)
from the \textsc{Grain-of-Salt} encoding and using it to assign the necessary bits 
in the encodings constructed by all the other tools. 
To construct such encodings via \textsc{Transalg} and \textsc{CBMC} we used template CNFs.
In \textsc{SAW+Cryptol} and \textsc{URSA} each encoding was constructed using individual program to which pre-specified data was added with special instructions.

In the experiments, we employed the SAT solvers that took prizes on the last SAT Competitions and also several SAT solvers which have shown good results on SAT-based cryptanalysis problems: 
\textsc{MapleLcmDistChronoBt} \cite{ChronoBT},
\textsc{Maple\_LCM\_Dist} \cite{maplelcm2017},
\textsc{MapleCOMSPS\_LRB\_VSIDS\_2} \cite{maplecomsps2017},
\textsc{MapleSAT} \cite{DBLP:conf/sat/LiangGPC16},
\textsc{Glucose} \cite{syrup2017},
\textsc{Maple\_LCM\_Scavel} \cite{Scavel2018},
\textsc{COMiniSatPS Pulsar} \cite{pulsar2017},
\textsc{Cryptominisat5} \cite{DBLP:conf/SAT/SoosNC09},
\textsc{Lingeling} \cite{lingeling2017},
\textsc{MiniSat} \cite{DBLP:conf/sat/EenS03}, and 
\textsc{Rokk} \cite{yasumoto-rokk}.

We also used the SMT solvers that took prizes on the last SMT Competitions: \textsc{CVC4} \cite{CVC4}, \textsc{Z3} \cite{Z3}, and \textsc{Yices} \cite{YICES}. In all experiments described below, we employed the HPC-cluster ``Academician V.M. Matrosov'' \cite{Matrosov-web} as a computing platform. Each node of this cluster is equipped with two 18-core Intel Xeon E5-2695 CPUs. Each solver was run in the sequential mode (on 1 CPU core) with the time limit of 5000 seconds (following the rules of SAT competitions).

\begin{table}[htbp]
\begin{center}
{\caption{Solving cryptanalysis problems for S\_Geffe, Wolfram, Bivium, and Grain. For each considered pair (keystream generator, tool) only results obtained by the best solver are shown. Time is shown in seconds.}
\label{solving_table}}	
		\begin{tabular}{|l|r|r|r|r|r||r|r|}	
			\hline
			& \textsc{TrAlg} & \textsc{GoS} & \textsc{URSA} & \textsc{Cr-SAT} & \textsc{Cb-SAT} & \textsc{Cr-SMT} & \textsc{Cb-SMT} \\
			\hline
            \hline
			\multicolumn{8}{|c|}{S\_Geffe} \\
			\hline
			Solver & \textsc{MiniSat} & \textsc{MiniSat} & \textsc{MiniSat} & \textsc{MiniSat} & \textsc{\textbf{MiniSat}} & \textsc{yices} & \textsc{yices} \\
			\hline
			Solved & 100/100 & 100/100 & 100/100 & 100/100 & \textbf{100/100} & 100/100 & 100/100 \\
			\hline
			Avg. time & 4.05 & 4.61 & 4.64 & 3.98 & \textbf{3.26} & 11.17 & 13.61 \\
			\hline
            \hline
			\multicolumn{8}{|c|}{\textsc{Wolfram}} \\
			\hline
			Solver & \textbf{\textsc{MaCom}} & - & \textsc{MaCom} & \textsc{rokk} & \textsc{MaCom} & \textsc{yices} & \textsc{z3} \\
			\hline
			Solved & \textbf{100/100} & - & 100/100 & 100/100 & 100/100 & 76/100 & 78/100 \\
			\hline
			Avg. time & \textbf{442} & - & 931 & 614 & 536 & 1631 & 1844 \\
			\hline
            \hline
			\multicolumn{8}{|c|}{Bivium30} \\
			\hline
			Solver & \textsc{MaLcm} & \textsc{MaLcm} & \textsc{rokk} & \textbf{\textsc{MaChr}} & \textsc{MaChr} & \textsc{yices} & - \\
			\hline
			Solved & 100/100 & 100/100 & 100/100 & \textbf{100/100} & 100/100 & 13/100 & 0/100 \\
			\hline
			Avg. time & 725 & 781 & 788 & \textbf{695} & 995 & 1759 & - \\
			\hline
			\hline
			\multicolumn{8}{|c|}{\textsc{Grain102}} \\
			\hline
			Solver & \textbf{\textsc{rokk}} & \textsc{rokk} & \textsc{rokk} & \textsc{rokk} & \textsc{rokk} & \textsc{yices} & - \\
			\hline
			Solved & \textbf{97/100} & 83/100 & 85/100 & 85/100 & 95/100 & 3/100 & 0/100 \\
			\hline
			Avg. time & \textbf{1407} & 2290 & 2038 & 2364 & 1737 & 3090 & - \\
			\hline
		\end{tabular}
		\end{center}		
\end{table}

In \autoref{solving_table} for each considered pair (generator, tool) only the results obtained by the best solver are shown. In this table we use the following abbreviations:
\begin{itemize}
\item \textsc{MaChr} for \textsc{MapleLcmDistChronoBt};
\item \textsc{MaCom} for \textsc{MapleCOMSPS\_LRB\_VSIDS\_2}; 
\item \textsc{MaLcm} for \textsc{Maple\_LCM\_Dist}; 
\item \textsc{Cr-SAT} for \textsc{SAW+Cryptol\_SAT}; 
\item \textsc{Cr-SMT} for \textsc{SAW+Cryptol\_SMT}; 
\item \textsc{Cb-SAT} for \textsc{CBMC\_SAT}; 
\item \textsc{Cb-SMT} for \textsc{CBMC\_SMT}; 
\item \textsc{TrAlg} for \textsc{Transalg}.
\end{itemize}
\noindent
For each generator the best result is marked with bold. The corresponding cactus plots are shown in \autoref{fig:best}. The detailed data for all considered solvers can be found in Appendix~\ref{appA}. All cactus plots in the present paper were made by the \textsc{mkplot} script \cite{web:mkplot}.

\begin{figure}[ht]
\subfloat[][S\_Geffe]{
    \includegraphics[width=0.5\textwidth]{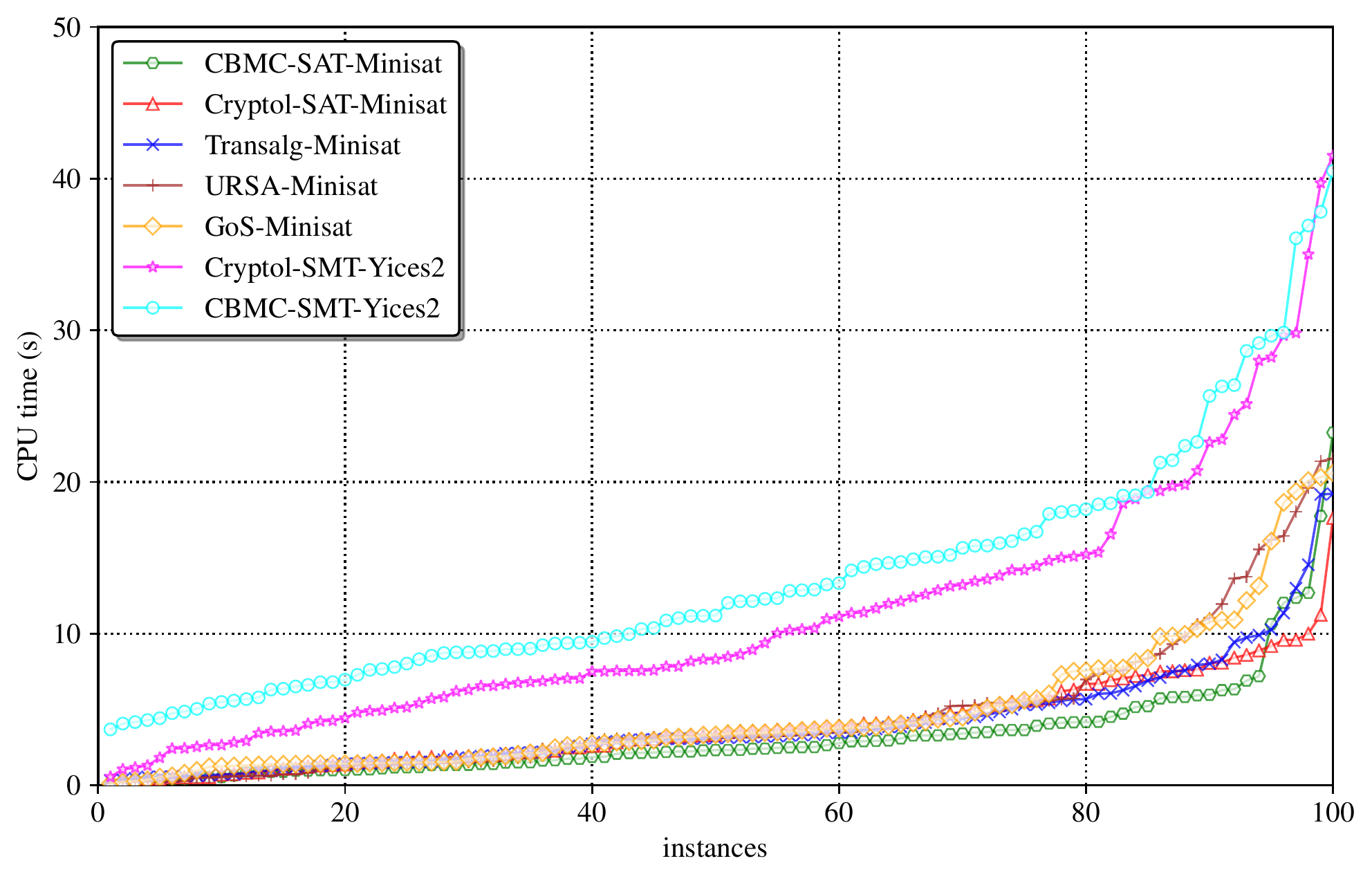}
		\label{fig:geffe-best}
  }
\subfloat[][Wolfram]{
    \includegraphics[width=0.5\textwidth]{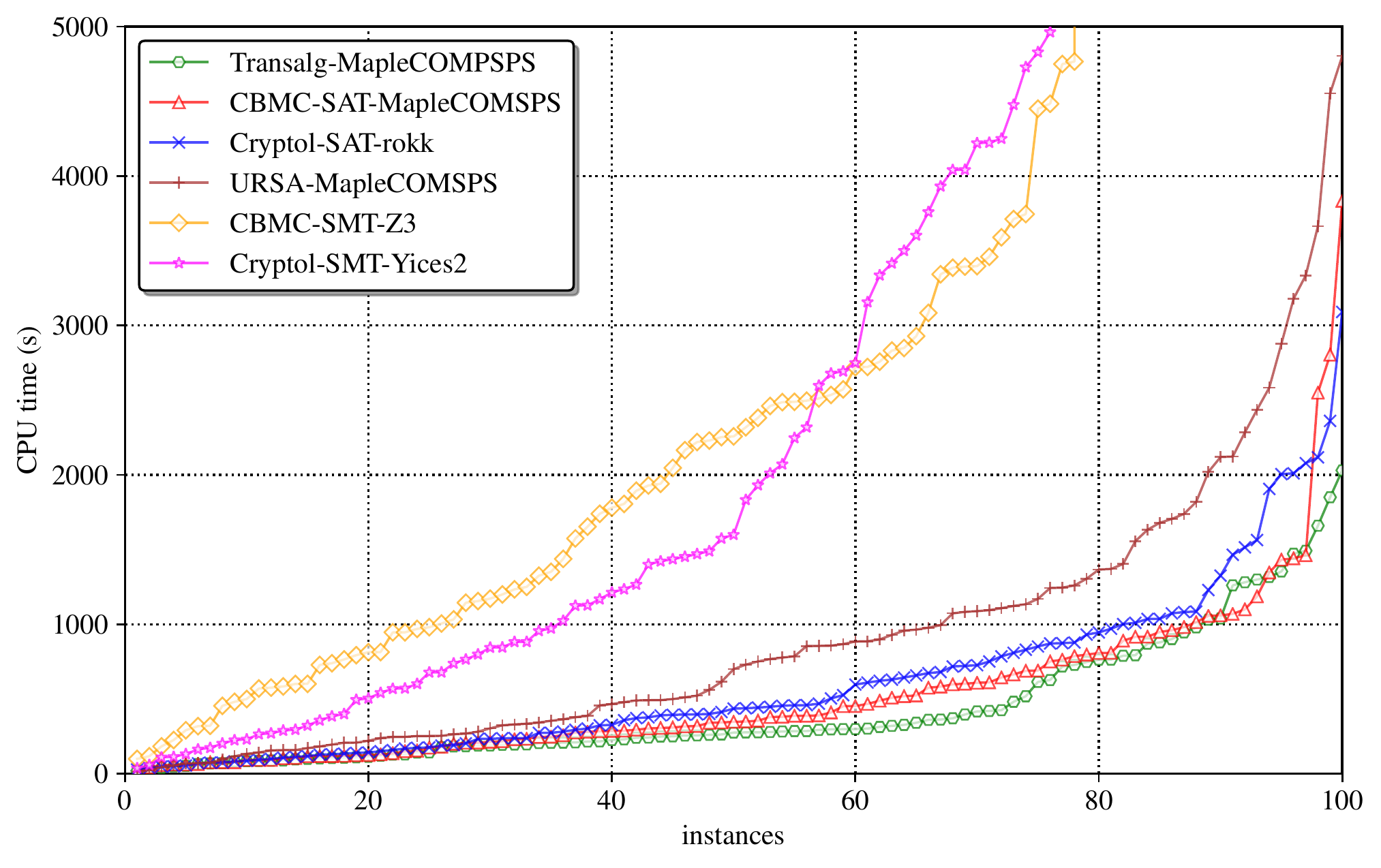}
		\label{fig:wolfram-best}
  }
\vskip\baselineskip
\subfloat[][Bivium30]{
    \includegraphics[width=0.5\textwidth]{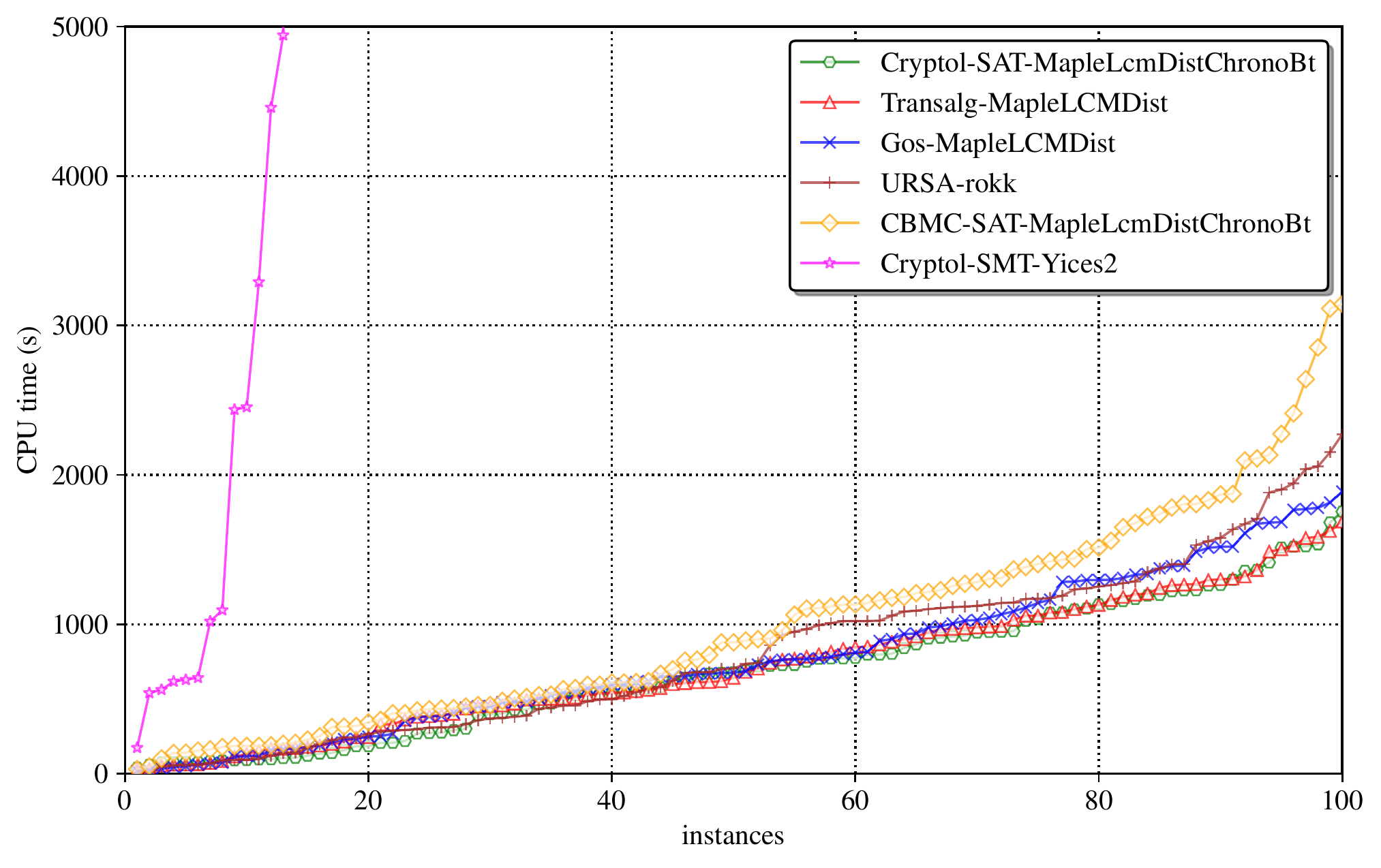}
		\label{fig:bivium-best}
  }
\subfloat[][Grain102]{
    \includegraphics[width=0.5\textwidth]{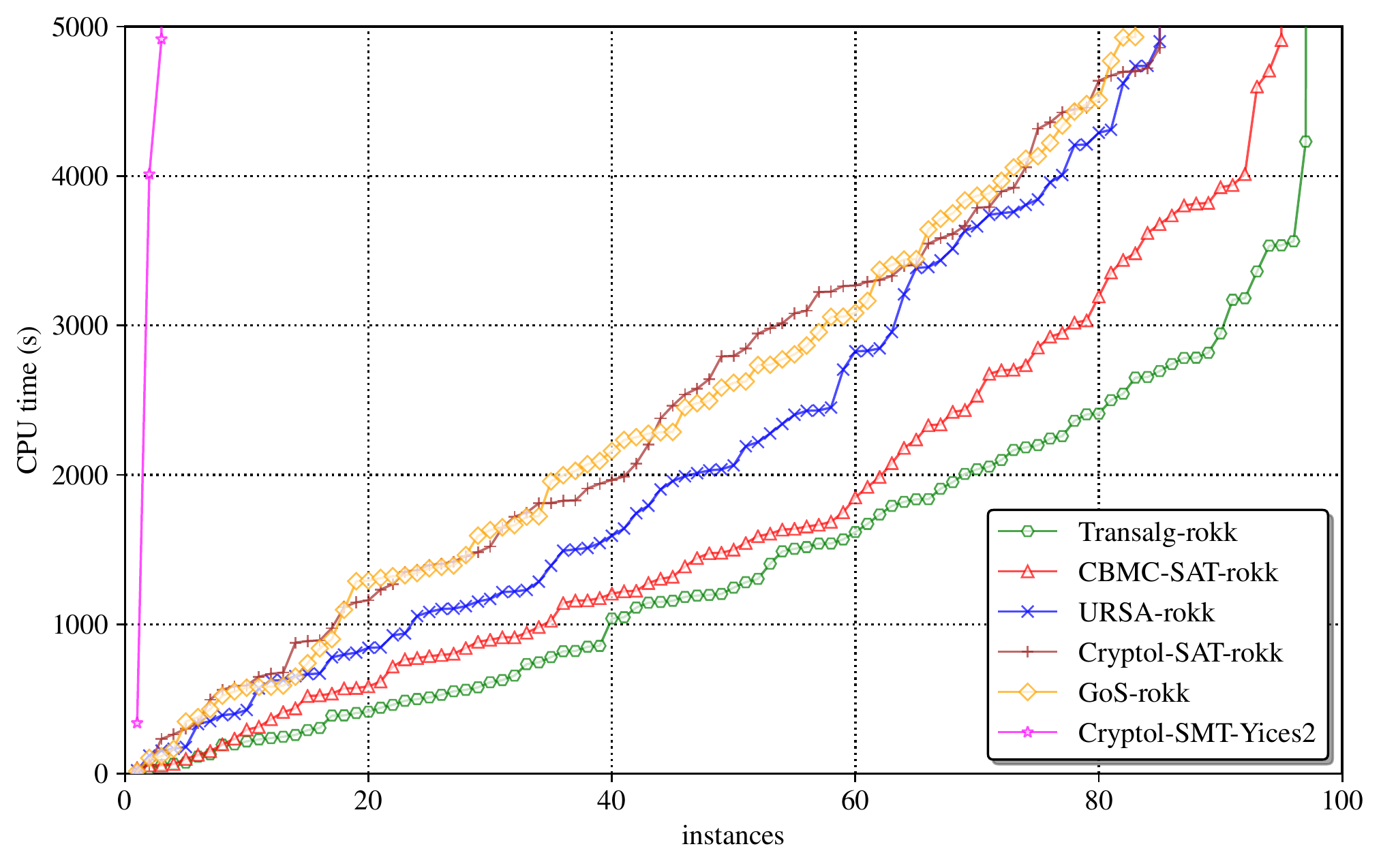}
		\label{fig:grain-best}
  }
  \caption{Comparison of encodings on the best solvers}
  \label{fig:best}
\end{figure}

Let us comment on the obtained results. On the considered problems SAT solvers significantly outperform the SMT solvers. Among the latter, \textsc{Yices} showed the best results. As for SAT, the \textsc{Transalg} encodings showed the best results on Wolfram and the weakened Grain, \textsc{SAW+Cryptol} outperformed the competitors on the weakened Bivium, while \textsc{CBMC} showed the best results on S\_Geffe.

\subsection{Discussion}
The tests we considered above to compare different tools for constructing SAT encodings 
in fact represent one of the most simple cryptographic attacks in SAT form: each a 
single instance with values of variables (which weaken an instance) fixed in 
a straightforward manner by assigning values to variables directly. In practical
cryptographic attacks it is often necessary to introduce more complex constraints.
They may link together several Boolean variables, or require solving sequences of instances, where the next instance is  constructed using the information obtained once the previous is solved, etc.
Thus, in practice the specific abilities of the propositional encoding systems
that make it possible to impose additional constraints and tune a SAT encoding
become exceptionally important.

In this context, we would like to specifically remind the reader about 
several \textsc{Transalg} features described in \autoref{sec:transalg}
that prove to be quite useful when implementing various SAT-based 
cryptographic attacks. 
The distinctive feature is that \textsc{Transalg} can construct and explicitly 
output a template CNF $C(f_n)$ (see Definition \ref{template}) with a number of specific characteristics. 
A template CNF essentially encodes the algorithm in a pure symbolic
execution sense.
It has the clearly outlined input and output variables,
and all the auxiliary and output variables depend on the input variables. 
As a result, in template CNFs constructed by \textsc{Transalg}, 
the variables are represented and numbered in the order in which they are introduced during 
the translation process. It means that the first variables correspond to a 
function's input and the last variables to its output. 
The Boolean variables not relevant
to constructing the output are pruned out during the propositional 
encoding process. 
If necessary, \textsc{Transalg} performs re-numbering
of variables to avoid gaps.

Thanks to the outlined sets of input and output 
variables, it is easy to use template CNFs to test the correctness of constructed SAT 
encodings, as well as to directly interpret satisfying assignments found by a 
SAT solver. The fact that in template CNFs the 
variables corresponding to the function input are outlined, makes it possible to use 
them for implementing the partitioning strategy \cite{Hyvarinen11} in 
a distributed computing environment (see for example \cite{DBLP:conf/pact/SemenovZ15,Semenov2016}) 
without specific preparation. Another important feature of input variables 
is that  they form a SUPBS  \cite{Williams:2003:BTC:1630659.1630827} 
in this template CNF. We will touch these questions in more detail in \autoref{sec:applications}.
Also, using template CNFs we can quickly generate families of instances that encode a cryptanalysis problem: to make a certain SAT instance for function inversion it is sufficient to add to a template CNF the unit clauses encoding the corresponding output. 

When translating an algorithm to SAT, the data structures employed by
\textsc{Transalg} preserve all the connections between the introduced Boolean 
variables and the corresponding cells of the abstract machine's memory. 
This fact allows one to effectively write auxiliary constraints on arbitrary
subsets of program variables. 
For this purpose the corresponding constraints are introduced in a TA program using the \texttt{assert} instruction: they will then be correctly 
transferred to a resulting SAT encoding. This feature is very important 
when implementing the SAT variants of the so-called differential attacks on 
cryptographic hash functions \cite{DBLP:conf/eurocrypt/WangLFCY05,DBLP:conf/eurocrypt/WangY05}. In \textsc{CBMC} such constraints can be imposed by using the \texttt{assume} instruction.

In \autoref{comparison_table} we compare all considered systems with respect to several criteria.
Let us give additional comments to it. Technically, \textsc{CBMC} and \textsc{URSA} can
construct template CNFs. The \textsc{CBMC} system constructs template CNFs directly, 
but in order to do so one has to add to  the corresponding program an empty hypothesis. 
For all practical purposes the obtained SAT encoding is a template CNF (despite having
unit clauses induced by empty hypothesis), but this step does not look entirely natural. \textsc{URSA} adds the values of variables to a constructed SAT encoding
by means of unit clauses. 
It also has an option of outputting the mapping between the variables from a program 
in its domain-specific language and the Boolean variables in a constructed SAT encoding. 
However, to make a template CNF one
has to parse the \textsc{URSA} output and remove the unit clauses corresponding
to function's outputs. To the best of our knowledge, both \textsc{SAW+Cryptol} 
and \textsc{Grain-of-Salt} have no option to output a template CNF with or without additional steps. 

\begin{table}[htbp]
\caption{Main functional abilities of \textsc{Grain-of-Salt}, \textsc{URSA}, \textsc{SAW+Cryptol}, \textsc{CBMC}, and \textsc{Transalg}.}
\label{comparison_table}
	\centering
		\begin{tabular}{|l|c|c|c|c|c|}
		\hline
		System & \textsc{GoS}& \textsc{URSA}& \textsc{SAW+Cryptol} & \textsc{CBMC} & \textsc{Transalg} \\
		\hline
        \hline
        Encodes conditional operators & - & + & + & + & + \\
		\hline
		Supports procedures and       &  &  &  &  &  \\
		functions       & - & - & + & + & + \\
        \hline
        Has embedded solvers          & - & + & + & + & - \\
        \hline
        Constructs SMT encodings      & - & - & + & + & - \\
        \hline
		Outlines sets of input        &  &  &  &  &  \\
        and output variables          & + & + & - & + & + \\
		\hline
        Constructs template CNFs      & - & - & - & +/- & + \\
		\hline
        Adds auxiliary constraints     &  & &  &  & \\
        on variables inside a program                  & - & - & - & + & + \\
        \hline		
		\end{tabular}	
\end{table}

Regarding the sets of input and output variables -- almost all tools can output the corresponding
information. However, in \textsc{Transalg} the input variables are always the first and 
the output variables are always the last, and always in the order specified in a
TA program. The other tools usually provide the mapping of program variables to Boolean
variables, but the ones corresponding to inputs and outputs are usually spread in the set of variables 
and rarely follow any specific order.

As for writing auxiliary constraints on arbitrary subsets of program variables, it appears
that only \textsc{Transalg} and \textsc{CBMC} provide this functionality directly. The mapping
of program variables to encoding ones  (provided by \textsc{URSA}, for example) 
theoretically enables one to add them manually, but this is a very arduous process.

The remaining points in \autoref{comparison_table} reflect the richness of 
a language employed by a propositional encoding tool and its system capabilities
such as whether it has embedded solvers and the ability to work with SMT encodings.

We believe that the lack of embedded solvers in \textsc{Transalg} and other tools
should not be viewed as a drawback, at least in the cryptographic context. Indeed, 
the cryptanalysis instances are typically very hard, and to solve them one usually
has to use the best available tools, such as cutting edge SAT solvers, including 
the parallel ones. Taking into account the fact that almost every year the SAT
solvers become better and better (which can be tracked in annual SAT competitions), 
it is actually quite hard to maintain the list of embedded solvers up to date at 
all times. The existence of various API and Python interfaces for SAT solvers, e.g., \cite{imms-sat18} allows for using external libraries for fast prototyping and testing.

As for the ability to output SMT encodings, our empirical evaluation to date
shows that SMT solvers typically work worse on cryptanalysis instances compared
to cutting edge SAT solvers. We believe that adding this functionality to \textsc{Transalg}
and investigating why SMT solvers often lose in comparisons similar to that summarized in, e.g., \autoref{solving_table} is an interesting direction 
for future development and research.


\section{Inversion of real world cryptographic functions using~their~translation~to~SAT}\label{sec:applications}

In this section, we present our results on SAT-based cryptanalysis of several ciphering systems. Some of these systems were used in practice in recent past, and some of them are still employed at the present moment.
In all cases considered below we used the encodings produced by the \textsc{Transalg} system.

\subsection{Using \textsc{Transalg} to construct SAT-based guess-and-determine attacks on several ciphers}

Guess-and-determine is a general cryptanalysis strategy that can be used to evaluate
the cryptographic weakness of various ciphers. The number of such attacks proposed  in the recent two decades is very large. Here we would like to cite the book 
\cite{Bard:2009:AC:1618541}, the major part of which studies guess-and-determine approach in the context 
of algebraic cryptanalysis.

The basic idea of the guess-and-determine strategy can be described as follows. 
Let $F$ be an arbitrary cipher that works with binary data, and let $E(F)$ be a 
system of Boolean or algebraic equations that corresponds to some cryptanalysis problem for $F$.
For example, for a given known pair $(x,y)$, where $x$ is a plaintext and $y$ is a 
ciphertext, $E(F)$ can be a system from the solution of which one can extract the
key $z$  such that $F(x,z)=y$. Let us denote by $X$ the set of all variables from $E(F)$.
Let $\tilde{X}$, $\tilde{X}\subseteq X$ be such a set of $l$ Boolean variables, that
by assigning values to all variables from $\tilde{X}$ the problem of solving $E(F)$ 
becomes trivial. The simplest example in this context is when $\tilde{X}$ consists of  
variables corresponding to a secret key of a cipher $F$. Also $\tilde{X}$ can be 
formed by variables corresponding to the internal state of a cipher at some time 
moment, for example, to an internal state of keystream generator registers at some
fixed step. In these cases by checking all possible assignments for variables from 
$\tilde{X}$, i.e. by performing exhaustive search over the set $\{0,1\}^l$, we perform 
a \textit{brute force} attack on a cipher $F$. 

For some ciphers it is possible to find a set $B$, $B\subseteq X$, $|B|=s$, which has the following property. Let us consider 
all possible assignments of variables from $B$. For each assignment we set the 
corresponding values to variables from $B$ into a system $E(F)$ 
and spend on solving each constructed weakened system by some fixed algorithm at most 
$t$ seconds (or any other fitting complexity measure). Assume that if we search over the whole 
$\{0,1\}^s$ in such a way, we find the solution for a considered cryptanalysis problem, and spend on this process at most $T=2^s\cdot t$. In this case we can say that 
there is a \textit{guess-and-determine attack} with complexity $T$. The set $B$ is
called a \textit{set of guessed bits}. Guess-and-determine attacks with a complexity significantly 
less than that of brute force attack are of particular interest. Usually, for such attacks 
$s \ll l$. 

In SAT-based cryptanalysis the problems of finding solutions for systems of equations 
of the type $E(F)$ are reduced to SAT. To construct a guess-and-determine attack in this case, one needs to know additional information about variables contained in
the corresponding CNF. In particular, it is very useful to know which CNF variables 
correspond to secret key bits. 
The methodology of constructing template CNFs employed in \textsc{Transalg} allows one to naturally outline sets of guessed bits, and traverse the search space of such sets to construct guess-and-determine attacks with good runtime estimations (see, e.g., \cite{Semenov2016,DBLP:conf/isw/ZaikinK17}).
Below we will briefly describe guess-and-determine attacks on
several ciphers for which SAT encodings were produced by \textsc{Transalg}.

In \cite{DBLP:conf/pact/SemenovZBP11}, a SAT-based guess-and-determine attack on the A5/1 keystream generator was constructed. In that attack the set of guessed bits of size 31 was used. The attack from \cite{DBLP:conf/pact/SemenovZBP11} was later verified in the volunteer computing project SAT@home \cite{csci37} by solving several dozens of the corresponding SAT instances using the technique from \cite{DBLP:conf/pact/SemenovZBP11}. Later in \cite{DBLP:conf/pact/SemenovZ15,Semenov2016} an automatic method for finding sets of guessed bits via optimization of a special function was proposed. A value of the function for a particular set of guessed bits is the runtime estimation of a corresponding guess-and-determine attack. Using this method, in \cite{Semenov2016} a guess-and-determine attack on the Bivium cipher \cite{DBLP:conf/isw/Canniere06} was constructed. The corresponding runtime estimation makes this attack realistic for state-of-the-art distributed computing systems. Using the algorithms from \cite{Semenov2016}, in \cite{DBLP:conf/isw/ZaikinK17} 
guess-and-determine attacks on several variants of the alternating step generator were constructed and implemented on a computing cluster.

In \cite{SemeonovZOKI2018}, a new class of SAT-based guess-and-determine cryptographic attacks was proposed. It is based on the so-called \textit{Inverse Backdoor Sets}
(IBS). IBS is a modification of the notion of Strong Backdoor Set for Constraint
Satisfaction Problems (including SAT), introduced in \cite{Williams:2003:BTC:1630659.1630827}. IBS is oriented
specifically on problems of SAT-based cryptanalysis. In more detail, a \textit{Strong Backdoor Set} for a CNF $C$ with respect to a polynomial algorithm $A$ is such a subset $B$ of
a set of variables $X$ of this CNF that setting values to variables from $B$ in $C$
in any possible way results in a CNF for which SAT is solved by $A$. This definition in its original form does not suit
well to cryptanalysis problems. However, in \cite{SemeonovZOKI2018} we modified it in the context of discrete functions inversion problems. Conceptually, the modification consists
in the following. Instead of demanding that $A$ is polynomial, we limit its runtime by some value $t$. In the role of the set of guessed bits we use an arbitrary $B\subseteq X$ and demand that algorithm $A$ can invert some portion of outputs of function $f_n$ constructed for random inputs from $\{0,1\}^n$ in time $\leq t$ using the assignments of variables from $B$ as hints. The portion of inverted outputs is the probability of a particular random event. If it is relatively large and the power of $B$ is relatively small, then, as it was shown in \cite{SemeonovZOKI2018}, $B$ can
be used to construct on its basis a nontrivial guess-and-determine attack on $f_n$.
The set $B$ defined in such a way is called \textit{Inverse Backdoor Set (IBS)}. The
effectiveness of guess-and-determine attacks based on IBS can be evaluated using
the Monte Carlo method \cite{Metropolis49}. The problem of finding IBS with good runtime estimation of a
corresponding guess-and-determine attack in \cite{SemeonovZOKI2018} was reduced to a problem 
of optimization of a black box function over the Boolean hypercube. Using IBSs, in \cite{SemeonovZOKI2018} the best or close to the best guess-and-determine attacks on several ciphers were constructed. In particular, the runtime estimation of  the constructed attack
using 2 known plaintext (2KP) on the block cipher AES-128 with 2.5 rounds is several dozen times 
better than that of  the attack from \cite{bouillaguet-crypto11} and requires little to no memory,
while the attack from \cite{bouillaguet-crypto11} needs colossal amounts of it. 
In \cite{DBLP:conf/evoW/PavlenkoSU19,DBLP:conf/gecco/PavlenkoBU19} several evolutionary algorithms
were used to minimize black box functions introduced in \cite{SemeonovZOKI2018} resulting in
new guess-and-determine attacks on several keystream generators.

Let us once again emphasize the important features of translating algorithms to SAT,
which are advantageous in the context of construction of guess-and-determine attacks.
Here we first and foremost mean an ability to provide information about the interconnection between the variables in propositional encodings with corresponding elementary operations performed in an original cryptographic algorithm. For example, the ability to outline
the variables that encode an input of a considered function for attacks described in \cite{SemeonovZOKI2018} allows us to use template CNFs for effective generation of large
random samples containing simplified CNFs. Each of them is formed by applying Unit Propagation to a template CNF augmented by the values corresponding to the known function input. When implementing a guess-and-determine attack with realistic
runtime estimation, \textsc{Transalg}'s features make it possible to naturally mount the attack using the incremental SAT technique \cite{DBLP:journals/entcs/EenS03}. In some cases it can
lead to significant performance gains (see, e.g., \cite{DBLP:conf/isw/ZaikinK17}).

\subsection{SAT-based cryptanalysis of hash functions from MD family}
In this subsection, we present examples of application of the \textsc{Transalg} 
system to cryptanalysis of cryptographic hash functions from the MD family.
It should be noted that these functions are still considered to be interesting
among cryptanalysts, and the first successful examples of application of  SAT-based cryptanalysis to real world cryptosystems are related specifically
to hash functions from the MD family \cite{DBLP:conf/SAT/MironovZ06}. The present
subsection is split into several parts: first we consider the problems of 
finding collisions for MD4 and MD5 functions. Then we construct preimage attacks
on truncated variants of the MD4 function.

\subsubsection{Finding Collisions for MD4 and MD5}
Let $f:\{0,1\}^+\rightarrow \{0,1\}^c$ be some cryptographic hash function that works with messages split into blocks of length $n$, $n>c$. It defines
a function of the kind $f_n:\{0,1\}^n\rightarrow \{0,1\}^c$. To produce a SAT encoding for the problem of finding collisions of this function, we essentially
translate the program describing $f_n$ twice using disjoint sets of Boolean variables.
Let $C_1$ and $C_2$ be the corresponding CNFs in which the sets of input and output
variables are denoted by $X^1=\{x_1^1,\ldots,x_n^1\}$, $X^2=\{x_1^2,\ldots,x_n^2\}$ 
and $Y^1=\{y_1^1,\ldots,y_c^1\}$,  $Y^2=\{y_1^2,\ldots,y_c^2\}$, respectively. Then
finding collisions of $f_n$ is reduced to finding an assignment that satisfies the
following Boolean formula:
\begin{equation}
\label{eq13}
C_1\wedge C_2\wedge (y_1^1\equiv y_1^2)\wedge \ldots\wedge(y_c^1\equiv y_c^2 )\wedge 
\left((x_1^1\oplus x_1^2)\vee \ldots\vee(x_n^1\oplus x_n^2)\right).
\end{equation}

Below we consider the problems of constructing collisions for the MD4 and MD5 hash functions, which were actively used up until 2005. Let us first briefly remind the reader about features of the Merkle--Damg{\aa}rd construction \cite{DBLP:conf/crypto/Merkle89,DBLP:conf/crypto/Damgard89a}, which serves as a basis of many cryptographic hash functions. In accordance with this construction, in MD4 and MD5 the process of computing a hash value is considered as a sequence of transformations of data stored in a special 128-bit register, to which we will refer as a \textit{hash register}. The hash register is split into four 32-bit cells.
At the initial stage a message called \textit{Initial Value} (IV), which is specified in  the algorithm's standard, is put into a hash register. Then the contents of the hash register are mixed with a 512-bit block of a hashed message by means of iterative transformations called \textit{steps}. There are 48 and 64 steps in MD4 and MD5, respectively. At each step a 32-bit variable is associated with an arbitrary cell of a hash register. Such variables are called \textit{chaining variables}. The transformations of data in a hash register, which were defined above, specify the so-called \textit{compression function}. We will denote the compression functions used in MD4 and MD5 as $f^{\mathit{MD}4}$ and $f^{\mathit{MD}5}$, respectively. The MD4 and MD5 algorithms can be used to construct hash values for messages of an arbitrary length. 
For this purpose an original message is first padded so that its length becomes a multiple of 512, and then is split into 512-bit blocks. Let $M=M_1,\ldots, M_k$ be a $k$-block message, where $M_i$, $i\in \{1,\ldots,k\}$ are 512-bit blocks. A hash value for message $M$ is constructed iteratively according to the following recurrence relations: $\chi_0=IV$, $\chi_i=f^{\mathit{MD}}(\chi_{i-1}, M_{i})$, $i=1,\ldots,k$ (here MD is either MD4 or MD5). A word $\chi_{k}$ is the resulting hash value of a message $M$. If hash values of two different $k$-block messages coincide, then the corresponding messages form a \textit{$k$-block collision} of the considered hash function.

The MD4 and MD5 algorithms were completely compromised with respect to finding collisions in \cite{DBLP:conf/eurocrypt/WangLFCY05,DBLP:conf/eurocrypt/WangY05}. The cryptanalysis methods used in the mentioned papers belong to a class of the so-called \textit{differential attacks}. In the attacks from \cite{DBLP:conf/eurocrypt/WangLFCY05,DBLP:conf/eurocrypt/WangY05}, the MD4 and MD5 hash functions were applied to two different messages. The processes of constructing hash values for these messages correspond to transformations of the contents of two hash registers. 
The main feature of differential attacks is that additional constraints are imposed on chaining variables associated with the corresponding cells of hash registers in the form of integer differences modulo $2^{32}$.
Also, special constraints on individual bits of these chaining variables can be used. These two types of constraints form the so-called \textit{differential path}. In \cite{DBLP:conf/eurocrypt/WangLFCY05,DBLP:conf/eurocrypt/WangY05},  differential paths were proposed that make it possible to effectively construct single-block collisions for MD4 and two-block collisions for MD5.

As we already mentioned, the first SAT variants of attacks from \cite{DBLP:conf/eurocrypt/WangLFCY05,DBLP:conf/eurocrypt/WangY05} were constructed in \cite{DBLP:conf/SAT/MironovZ06}. To obtain a corresponding propositional encoding, it is first necessary to construct a formula of the kind \eqref{eq13} and then transform it to CNF using the Tseitin transformations. However, the resulting CNF turns out to be extremely hard even for state-of-the-art SAT solvers. A realistically feasible runtime of cryptanalysis is achievable only by adding to a constructed CNF the clauses which encode a differential path. In our experiments, the special \texttt{assert} instruction of the TA language made it possible to implement this step quite easily. We would like to additionally note that the constraints defining a non-zero differential path eliminate the need for constraints of the kind $\left((x_1^1\oplus x_1^2)\vee \ldots\vee(x_n^1\oplus x_n^2)\right)$ in \eqref{eq13}, which indicate the difference between sets of values of the input variables (since only different inputs of a hash function can lead to a non-zero differential path).

In \autoref{tab_hash}, we compare the SAT encodings of differential attacks for finding collisions for MD4 and MD5 used in \cite{DBLP:conf/SAT/MironovZ06} with those constructed by \textsc{Transalg}.

\begin{table}[ht]
\caption{The parameters of SAT encodings for finding collisions of the MD4 and MD5 hash
functions with differential paths from \cite{DBLP:conf/eurocrypt/WangLFCY05,DBLP:conf/eurocrypt/WangY05}.}
\label{tab_hash}
\begin{tabular}{|l|l|r|r|}
\hline
\multicolumn{2}{|l|}{}&SAT encodings from \cite{DBLP:conf/SAT/MironovZ06} & \textsc{Transalg} encodings\\
\hline
MD4& variables&53228&18095\\
\cline{2-4}
&clauses&221440&187033\\
\hline
MD5&variables&89748&34181\\
\cline{2-4}
&clauses&375176&295773 \\
\hline
\end{tabular}
\end{table}

For the problem of finding single block collisions of the MD4 hash function, we managed to find 
about 1000 MD4 collisions within 200 seconds on one core of Intel i7-3770K (16 Gb RAM) using the 
SAT encodings produced by \textsc{Transalg} and the \textsc{Cryptominisat} solver \cite{DBLP:conf/SAT/SoosNC09}.
Note that in \cite{DBLP:conf/SAT/MironovZ06} it took about 500 seconds to construct one single block collision for MD4.

After this we studied the problem of finding two-block collisions of MD5. This process consists of two stages.
At the first stage, we search for two 512-bit blocks $M_1$ and $M'_1$  that satisfy the differential path from \cite{DBLP:conf/eurocrypt/WangY05}.
We denote $\chi_1=f^{\mathit{MD}5}(IV,M_1)$,
$\chi'_1=f^{\mathit{MD}5} (IV,M'_1)$. At the second stage, we look for second 512-bit blocks $M_2$
and $M'_2$ (which also satisfy the differential path from \cite{DBLP:conf/eurocrypt/WangY05}) such that $f^{\mathit{MD}5}(\chi_1,M_2)=f^{\mathit{MD}5}(\chi'_1,M'_2)$.

For the problem of finding a first message blocks pair $(M_1,M'_1)$, which turned out to be quite hard, we used the HPC-cluster ``Academician V.M. Matrosov" \cite{Matrosov-web}. We ran state-of-the-art SAT solvers working in the multi-threaded mode (36 threads) on the cluster.
In particular, we used \textsc{plingeling}, \textsc{treengeling}
(versions from the SAT competition 2014 \cite{lingeling2014}) and \textsc{plingeling}, \textsc{treengeling} \cite{lingeling2017},
\textsc{painless} \cite{painless2017}, \textsc{glucose-syrup} \cite{syrup2017} from the SAT competition 2017. Surprisingly, only \textsc{treengeling} 2014 managed to solve the corresponding SAT instances within a time limit (30 hours).

During these experiments, several message blocks with a lot of zeros in the beginning were found. A more detailed analysis showed that the maximum number of first message bytes that can be set to $0$ simultaneously in $M_{1}$ and $M'_{1}$ is 10 bytes. Assignment of the $11^{\mathrm{th}}$ byte to $0$ in $M_{1}$ and $M'_{1}$ makes the corresponding SAT instance unsatisfiable (which can be proven quickly). Thus we outlined the class of message blocks pairs that satisfy the differential path
from \cite{DBLP:conf/eurocrypt/WangY05} and both blocks have first 10 zero bytes. The problem of 
finding a pair of such blocks is relatively simple and can be
solved using a number of SAT solvers (compared to the situation when the first 10 bytes
are not set to zero). On the corresponding SAT instance we ran four different SAT solvers (\textsc{painless}, \textsc{glucose-syrup}, \textsc{treengeling} 2014, \textsc{treengeling} 2017) each working in multi-threaded mode on one cluster node. In 24 hours each solver managed to find several message blocks pairs, except for \textsc{treengeling} 2014 SAT solver, which found only one. The corresponding results are given in  \autoref{collision_results}.

\begin{table}[ht]
\caption{Finding a pair $M_1$, $M'_1$ that satisfies the differential path from \cite{DBLP:conf/eurocrypt/WangY05}, and both $M_1$ and $M'_1$ have first 10 zero bytes.}
\label{collision_results}
\begin{tabular}{|l|r|r|}
\hline
SAT solver & Solved instances & Avg. time (s)\\
\hline
\textsc{painless} & 3 & 32327\\ \hline
\textsc{glucose-syrup} &3 & 38302\\ \hline
\textsc{treengeling} 2014 &1 &54335\\ \hline
\textsc{treengeling} 2017 & 3 &33357\\ \hline

\end{tabular}
\end{table}

For the obtained pairs of first blocks, the problem of constructing such pairs $(M_2,M'_2)$
that the messages $M_1|M_2$ (the concatenation of two 512-bit blocks $M_1$ and $M_2$) and $M'_1|M'_2$ form a two-block collision for MD5 turned
out to be much simpler: on average one such pair $(M_2,M'_2)$ can be found by
any considered parallel solver in less than 500 seconds on one cluster node. An example of the two-block collision of the described kind is shown in  \autoref{col_exp}.

\begin{table}[ht]
\caption{An example of two-block MD5 collision with first 10 zero bytes.}
\label{col_exp}
	\centering
		\begin{tabular}{|l|l|}
		\hline
		&				\texttt{00 00 00 00 00 00 00 00 00 00 20 74 67 a6 f5 48}\\
		&				\texttt{cb c1 6d \underline{a5} 3e f7 b8 bc 67 a3 8d d9 3c 9b f5 b8}\\
		&				\texttt{55 ed 32 06 06 0a 74 a3 0f b6 84 87 47 \underline{cf} \underline{91} d0}\\
		&				\texttt{db 4c 6f 43 ef 64 f0 8d a4 1d 50 \underline{c6} 26 df 95 fe}\\
		$M_1|M_2$&	\texttt{ff d1 2e c9 a0 90 aa b3 7d e7 e5 bc f2 3a 4e ab}\\
		&				\texttt{24 b8 d4 \underline{13} 4c cc 7b 1b 00 29 eb f5 53 7a 0d d1}\\
		&				\texttt{5d 1f b7 79 af 36 ce 08 1e 44 a2 d0 51 \underline{ec} 91 fb}\\
		&				 \texttt{c5 4c a2 89 75 b3 a3 84 ac 97 7f \underline{f2} 7e 50 d4 56} \\
		\hline
		&				\texttt{00 00 00 00 00 00 00 00 00 00 20 74 67 a6 f5 48}\\
		&				\texttt{cb c1 6d \underline{25} 3e f7 b8 bc 67 a3 8d d9 3c 9b f5 b8} \\
		&				\texttt{55 ed 32 06 06 0a 74 a3 0f b6 84 87 47 \underline{4f} \underline{92} d0}\\
		&				\texttt{db 4c 6f 43 ef 64 f0 8d a4 1d 50 \underline{46} 26 df 95 fe}\\
		$M'_1|M'_2$&	\texttt{ff d1 2e c9 a0 90 aa b3 7d e7 e5 bc f2 3a 4e ab}\\
		&				\texttt{24 b8 d4 \underline{93} 4c cc 7b 1b 00 29 eb f5 53 7a 0d d1} \\
		&				\texttt{5d 1f b7 79 af 36 ce 08 1e 44 a2 d0 51 \underline{6c} 91 fb}\\
		&				\texttt{c5 4c a2 89 75 b3 a3 84 ac 97 7f \underline{72} 7e 50 d4 56} \\
		\hline

		$Hash$&\texttt{c22664780a9766ceb57065eba36af06b}\\
		\hline
		\end{tabular}
\end{table}

In conclusion we would like to once more point out features of the \textsc{Transalg} system
that made it possible to obtain the presented results. It is mainly thanks to the translation
concept of \textsc{Transalg} that allows one to directly work with variables encoding each 
elementary step of a  considered algorithm. That is why we can effectively reflect in SAT
encoding any additional constraints, such as, for example, the ones that specify a differential
path. As far as we know, only \textsc{Transalg} and \textsc{CBMC} allow for adding such constraints, whereas \textsc{SAW+Cryptol}, \textsc{URSA} and \textsc{Grain-of-Salt} do not have this capability.

It should be noted that at the current stage SAT-based cryptanalysis is less effective in application to the
collision search problems for cryptographic hash functions in comparison with specialized methods 
\cite{10.1007/978-3-642-03356-8_4,sasaki2007new}. On the other hand, the use of new SAT encodings and state-of-the-art
SAT solvers makes it possible to find collisions for MD4 hash function about 1000 times faster
than it was done in \cite{DBLP:conf/SAT/MironovZ06}. From our perspective, the potential 
for further improvements in this direction is far from being exhausted. It should be also 
mentioned that SAT-based cryptanalysis is, apparently, the most effective for preimage attacks on cryptographic hash functions. Below we build a new preimage attack on the 39-step version of the MD4 hash function using \textsc{Transalg}.

\subsubsection{Preimage attacks on truncated variants of MD4}
Despite the fact that MD4 is compromised with respect to collision finding, the problem of 
finding preimages for this function is still considered to be extremely hard. While it is believed that MD4 is not highly resistant to preimage attacks, all the
arguments of this kind are mostly theoretical \cite{Leurent08}. We are not aware of  papers in which the inversion problem of full-round MD4 would be solved in reasonable time. To the best of our knowledge, until recently, the paper \cite{DBLP:conf/SAT/DeKV07} was considered to be the best practical attack on 
MD4 since it made it possible to invert a truncated variant of MD4 with 39 (out of 
48) steps. The attack from \cite{DBLP:conf/SAT/DeKV07} is a SAT-based variant of the attack proposed by
H.~Dobbertin in \cite{DBLP:conf/fse/Dobbertin98}. Let us briefly review the results from these papers.

In fact in \cite{DBLP:conf/fse/Dobbertin98} it was shown that the problem of inverting the first two rounds of 
MD4 (i.e. 32 steps) is not computationally hard. The main idea of that paper was to
add some additional constraints on several chaining variables. These constraints
significantly weaken the system of equations corresponding to the process of filling
the hash register of MD4 during the first two rounds. In more detail, H. Dobbertin proposed to fix with constant $K$ the values of 12 chaining variables and showed that choosing the value of $K$ at random with high probability leads to a consistent system which can be easily solved. 

In \cite{DBLP:conf/SAT/DeKV07} a SAT-based attack on MD4 was proposed that used ideas from \cite{DBLP:conf/fse/Dobbertin98}. More precisely,
in \cite{DBLP:conf/SAT/DeKV07} the constant $K$ was fixed to $0$. Also, the authors of \cite{DBLP:conf/SAT/DeKV07} rejected one of the constraints from \cite{DBLP:conf/fse/Dobbertin98}. Thus, 
in \cite{DBLP:conf/SAT/DeKV07} 11 constraints were used instead of 12. The constraints of ``Dobbertin'' type were added to the propositional encoding 
of the MD4 algorithm in the form of unit clauses. Hereinafter we refer to additional
constraints of the ``Dobbertin'' type on chaining variables as \textit{relaxation constraints}.
In some cases the application of such constraints leads to propagation of the values of a large number of other variables. The variables that represent the unknown preimage of a known hash pose the main interest in this context.

In \cite{DBLP:conf/SAT/DeKV07}, apart from the two-round variant of MD4, there were considered preimage attacks on
truncated MD4 variants with $k$ steps, up to and including $k=39$. For an arbitrary $k<48$
we will refer to a corresponding truncated variant of MD4 compression function as to MD4-$k$. The best result
presented in \cite{DBLP:conf/SAT/DeKV07} was a successful inversion of MD4-39 for several hash values
of a special kind. To solve each of such problems it took about 8 hours of the \textsc{MiniSat}
SAT solver. It is surprising that the computational results achieved in \cite{DBLP:conf/SAT/DeKV07} remained state-of-the art for 10 years. In \cite{GribanovaS18} we significantly improved them. It was the result of 
using a special technique which reduced the problem of finding \textit{promising} relaxation 
constraints to the problem of optimization of a black box function over the Boolean hypercube.
In all experiments in \cite{GribanovaS18} we used the SAT encodings constructed by the \textsc{Transalg} system. Below let us briefly review results obtained in \cite{GribanovaS18}.

To automatically take into account information about relaxation constraints, special variables called \textit{switching variables} \cite{mit2016} were added to the corresponding TA programs. 
The main idea of
this approach consists in the following. Let $C$ be a CNF that encodes the inversion of some
function and $X$ be a set of Boolean variables from $C$. Assume that we need to add to $C$ 
new constraints that specify some predicate over variables from a set $\tilde{X}$, $\tilde{X}
\subseteq X$. Let $R(\tilde{X})$ be a formula specifying this predicate. Now let us introduce
new Boolean variable $u$, $u\notin X$. Consider the formula $C'= C\wedge (\neg u\vee R(\tilde{
X}))$. It is clear that the constraint $R(\tilde{X})$ will be inactive when $u=0$ and 
active when $u=1$. Let us refer to variables similar to $u$ as {switching variables}. 

In application to preimage attacks on MD4-$k$, switching variables make it possible to reduce the problem of finding effective relaxation constraints (of the ``Dobbertin'' type)
to the optimization problem over the Boolean hypercube. With that purpose in \cite{GribanovaS18} we introduce a special measure $\mu$ that heuristically evaluates the effectiveness of a considered set of 
relaxation constraints. Each particular set of relaxation constraints is defined by an
assignment of switching variables. The measure $\mu$ is a black box function. The
relaxation constraints for which the value of $\mu$ lies in a particular range are considered
to be promising. Thus, the arguments of the considered function are switching variables and its values are the values of $\mu$ on the corresponding sets of relaxation constraints.
The function defined that way is maximized over the Boolean hypercube, each point of which 
represents an assignment of switching variables. Since the constructed function does not have an
analytical representation, it is sensible to use metaheuristic methods for its maximization.
In particular in \cite{GribanovaS18} we used an algorithm from the \textit{tabu search} \cite{Glover:1997:TS:549765} class.
We view as promising such sets of relaxation
constraints whose activation results in derivation by the Unit Propagation rule
of a relatively large number of variables corresponding to the hashed message in a SAT encoding (the number of such variables gives the value of function $\mu$). Similar to \cite{DBLP:conf/SAT/DeKV07} in the role of relaxation constraints we used the constraints meaning that the  corresponding chaining variables should take the value $K=0$. 
As a result, we managed to find new
relaxation constraints that make it possible to invert the MD4-39 hash function much
faster than in \cite{DBLP:conf/SAT/DeKV07}. Let us briefly mention the results of computational experiments from \cite{GribanovaS18}.

Let us note here that 
based on the features of the MD4 algorithm \cite{DBLP:conf/fse/Dobbertin98}
it is impossible to 
impose constraints on the first four and the last (preceding the calculation of the final
hash value) four steps of the MD4-39 algorithm. 
According to this, the sets of new relaxation
constraints were selected (using the values of the corresponding switching variables) from 
the set of cardinality $31$. Thus, the problem of maximization of a function described above over the
Boolean hypercube $\{0,1\}^{31}$  was considered. The details of experiments can be found in
\cite{GribanovaS18}. As a result we found two new sets of relaxation constraints. We denote them as $\rho_1$ and
$\rho_2$ and they represent the following
assignments of corresponding switching variables:
\begin{equation*}
\begin{array}{ll}
\rho_1 : &0000000001101110111011101000000\\
\rho_2 : &0000000000101110111011101100000
\end{array}
\end{equation*}
For example, vector $\rho_1$ specifies the set of $12$ relaxation constraints: chaining variables 
on steps $14, 15, 17, 18, 19, 21, 22, 23, 25, 26, 27, 29$ are assigned with the value $K=0$. 
The application of the relaxation constraints specified by $\rho_1$ and $\rho_2$ allows one to
find preimages of the MD4-39 hash function for known hash values $0^{128}$ and $1^{128}$ within 
one minute of \textsc{MiniSat} 2.2 runtime. Note, that using constraints from \cite{DBLP:conf/SAT/DeKV07} the solution of the preimage finding problem for $1^{128}$ requires about 2 hours, and the preimage finding 
problem for $0^{128}$ cannot be solved in 8 hours. The corresponding results are presented in \autoref{preimage_search_1}, where $\rho_{De}$ denotes the set of relaxation constraints described in \cite{DBLP:conf/SAT/DeKV07} and 
$\rho_{Dobbertin}$ denotes the variant of Dobbertin's constraints from \cite{DBLP:conf/fse/Dobbertin98} with constant
$K=0$. 
\begin{table}[ht]
\caption{Finding the MD4-39 preimages for hash values $0^{128}$ and $1^{128}$.}
\label{preimage_search_1}
\centering
\begin{tabular}{|l|r|r|}
\hline
Relaxation & \multicolumn{2}{c|}{Result / Solving time (s)} \\
\cline{2-3}
constraints &\multirow{2}{*}{$\chi = 0^{128}$}& \multirow{2}{*}{$\chi = 1^{128}$}\\  
 & & \\
\hline
$\rho_1$ & SAT / 20 & SAT / 10 \\
\hline
$\rho_2$ & SAT / 60 & UNSAT / $<$ 1\\
\hline
$\rho_{Dobbertin}$& SAT / 20 & Unknown / $>$ 30000 \\
\hline
$\rho_{De}$ & Unknown / $>$ 30000 & SAT / 7000 \\
\hline
\end{tabular}
\end{table}
Below, these relaxation constraints are specified by the vectors of values of switching
variables from $\{0,1\}^{31}$ (in the notation similar to that of $\rho_1$ and $\rho_2$):
\begin{equation*}
\begin{array}{ll}
\rho_{Dobbertin}: &0000000011101110111011100000000\\
\rho_{De} :          &0000000001101110111011100000000
\end{array}
\end{equation*}
What is particularly interesting is that the application of new sets of relaxation constraints 
$\rho_1$ and $\rho_2$ also allows one to find preimages of MD4-39 for randomly generated 128-bit
Boolean vectors persistently. To obtain this result, we considered a test set of 500 randomly 
generated vectors from $\{0,1\}^{128}$. Regarding each of these vectors we assumed that it is a hash value of MD4-39. After that the preimage finding problem for this value was solved 
using constraints specified by vectors $\rho_1$ and $\rho_2$. For the prevailing part of the
tasks (65--75\%) the solutions were successfully found using \textsc{MiniSat} 2.2.
The average time of finding one preimage was less than 1 minute. The rest (25--35\% of the tasks)
corresponded to 128-bit vectors for which there were no MD4-39 preimages under constraints 
specified by $\rho_1$ and $\rho_2$ (this fact was proven by the SAT solver in under 1 minute on average). These results are presented in  \autoref{preimage_search_2}. Note that even in 
a few hours we did not manage to solve the preimage finding problem for any vector from the test
set using constraints from \cite{DBLP:conf/SAT/DeKV07} or \cite{DBLP:conf/fse/Dobbertin98}.

\begin{table}[ht]
\caption{Finding the MD4-39 preimages for 500 randomly generated 128-bit Boolean vectors. Instances with preimages are satisfiable, while those with no preimages are unsatisfiable.}
\label{preimage_search_2}
\centering
\begin{tabular}{|l|r|r|r|r|}
\hline
Relaxation&Avg.& Max.& \multicolumn{2}{c|}{Solved instances (in \% of total} \\ 
constraints &solving&solving& \multicolumn{2}{c|}{number of instances)} \\
\cline{4-5}
&time (s)&time (s)&with preimages&with no preimages\\
\hline
$\rho_1$& 12 & 80 & 65 & 35 \\ \hline
$\rho_2$& 46 & 250 & 75 & 25 \\ \hline
\end{tabular}
\end{table}

Let us summarize the results of the present section. From our point of view, we convincingly demonstrated the power of SAT-based cryptanalysis methods. 
We believe that future ideas both in the area of reduction to SAT and in
algorithms of state-of-the-art SAT solvers will make it possible to increase the effectiveness of 
SAT-based cryptanalysis and extend the spectrum of its applications.

\section{Related works}\label{sec:related}

The ideas of using general purpose combinatorial algorithms to solve cryptanalysis problems can
be found in many papers starting from 90-th years of XX-th century. Apparently, S.A. Cook and D.G. Mitchell 
were the first to propose using SAT solving algorithms in cryptanalysis 
in \cite{Cook97findinghard}. The first example of propositional encoding  of a cryptographic problem
(in particular of DES cryptanalysis) was given in \cite{Massacci:1999:UWR:1624218.1624261,DBLP:journals/jar/MassacciM00}. In \cite{DBLP:conf/frocos/JovanovicJ05} the
problems of finding collisions of a number of cryptographic hash functions were reduced to
SAT. In \cite{DBLP:conf/SAT/MironovZ06} propositional encodings of hash functions from the MD family were also constructed. The authors of \cite{DBLP:conf/SAT/MironovZ06} added to these encodings the
constraints that encode the differential paths introduced in \cite{DBLP:conf/eurocrypt/WangLFCY05,DBLP:conf/eurocrypt/WangY05}. This addition made it possible to 
persistently construct single-block collisions for MD4 (it took about 10 minutes per collision).
Thus, \cite{DBLP:conf/SAT/MironovZ06} can be considered to be the first paper in which SAT-based
cryptanalysis was successfully applied to relevant cryptographic algorithms. In the book \cite{Bard:2009:AC:1618541} SAT solvers are considered to be the primary tool for solving problems of 
algebraic cryptanalysis. It should be noted that in all the mentioned papers no automated system was used to construct
propositional encodings of the considered functions. 

In the present paper we described in detail the principles of constructing propositional encodings of discrete functions with the focus on functions employed in cryptography. We also compared several different systems that can produce such SAT encodings in an automatic mode. First it is the well-known \textsc{CBMC} system for symbolic verification \cite{ckl2004,DBLP:series/faia/Kroening09}, which have been developed for more than 15 years. \textsc{CBMC} is a generic system and it is not designed with cryptanalysis problems in mind. However, as we show in Section 4, it allows one to perform the majority of actions available to a limited number of considered domain-specific systems. Here we mean \textsc{SAW+Cryptol}, \textsc{Grain-of-Salt}, \textsc{URSA} and \textsc{Transalg}.

The first version of the \textsc{Cryptol} language was published in 2003 \cite{LewisM03}. Its second version \cite{ErkokPLPV09,DBLP:conf/fmcad/ErkokCW09} was later augmented by SAW \cite{Carter2013} that allowed producing SAT encodings. In 2010, the \textsc{Grain-of-Salt} system was proposed  \cite{DBLP:conf/tools/Soos10}. Approximately at this time we started the development of the \textsc{Transalg} software tool, which we describe in the present paper (\textsc{Transalg} was first mentioned in papers in Russian in 2011 \cite{OtpSem11}). In 2012, the \textsc{URSA} system, aimed at reducing to SAT various constraint programming problems, was published \cite{journals/lmcs/predrag}. It can be applied to construct encodings of cryptographic functions as well. Note that \textsc{SAW+Cryptol}, \textsc{URSA} and \textsc{Transalg} can encode to SAT algorithmic
descriptions of a very wide class of functions working with binary data. Meanwhile,
\textsc{Grain-of-Salt} is designed to work only with keystream generators based on shift
registers. We considered the pros and cons of all mentioned systems in detail in Section~\ref{sec:compar}.

\textsc{CBMC}, \textsc{Transalg} and other similar systems are based on symbolic execution of a program specifying a considered function. The idea to transform programs
to Boolean formulas was first proposed by S.A. Cook in his paper \cite{DBLP:conf/stoc/Cook71} which led to the creation and development of the theory of NP-completeness. The notion ``Symbolic Execution'' first
appeared in the paper \cite{King:1976:SEP:360248.360252} by J.C.~King, where it is defined as a process of 
interpretation of a program in a special extended semantics, within the context of 
which it is allowed to take symbols as input and put formulas as output. 
Currently, symbolic execution combined with Bounded Model Checking is actively used in software verification (see, e.g., \cite{DBLP:series/faia/Kroening09}).

As we mentioned above, SAT-based cryptanalysis is still actively developing. The 
cryptographic attacks that employ SAT solvers show very good results for a number of 
keystream generators: Geffe, Wolfram (present paper), Crypto-1, Hitag2 (see \cite{DBLP:conf/SAT/SoosNC09}). In
\cite{DBLP:conf/pact/SemenovZBP11}, a successful SAT-based attack on the widely known A5/1 cryptographic keystream generator was described.
Later several dozen SAT instances that encode the cryptanalysis problem for  A5/1 were solved in the SAT@home
volunteer computing project \cite{csci37}. This result together with other attacks on A5/1
(see \cite{OE_a51,Guneysu2008,Nohl2010}) provides an exhaustive argument towards
not using A5/1 any more. The Bivium stream cipher \cite{DBLP:conf/isw/Canniere06} is a popular object of
algebraic and SAT-based cryptanalysis \cite{Mcdonald_attackingbivium,DBLP:conf/SAT/EibachPV08,DBLP:conf/SAT/SoosNC09,Eibach2010}. 
In \cite{Semenov2016} a SAT-based guess-and-determine attack on Bivium was proposed. The corresponding runtime estimation turned out to be realistic for modern distributed computing systems. The LTE stream cipher ZUC was analyzed by SAT in \cite{Lafitte15}. In \cite{DBLP:conf/isw/ZaikinK17}, SAT-based guess-and-determine attacks on several variants of the alternating step generator were described. SAT-based cryptanalysis of stream ciphers from the CAESAR competition was described in \cite{DBLP:conf/secrypt/DwivediKMNPW17}.

In \cite{SemeonovZOKI2018} a new class of SAT-based guess-and-determine attacks was described, in which the notion of Inverse Backdoor Set (IBS) is used. IBS is a modification of a well-known notion of Strong Backdoor Set \cite{Williams:2003:BTC:1630659.1630827}. It made it possible to construct the best or close to the best guess-and-determine attacks on several ciphers. For example, the attack on 2.5-round
AES-128 with 2 Known Plaintexts, presented in \cite{SemeonovZOKI2018}, is significantly better than the previously best known attack on this cipher proposed in \cite{bouillaguet-crypto11}. In \cite{DBLP:conf/evoW/PavlenkoSU19,DBLP:conf/gecco/PavlenkoBU19},  evolutionary and genetic algorithms were used to minimize the objective function from \cite{SemeonovZOKI2018} in application to SAT-based  guess-and-determine attacks on weakened variants of the Trivium stream cipher \cite{DBLP:conf/isw/Canniere06}. Note that the function introduced in \cite{SemeonovZOKI2018} to associate with a particular IBS the estimation of 
effectiveness of a corresponding guess-and-determine attack is a concretization of the
notion of SAT-immunity, introduced by N.~Courtois in \cite{Courtois-eprint,Courtois-tatra,Courtois-cryptologia}. 

As we already noted, \cite{DBLP:conf/SAT/MironovZ06} was the first paper to demonstrate the applicability of SAT-based
cryptanalysis to relevant cryptographic algorithms. In that paper, using the \textsc{MiniSat}
solver \cite{DBLP:conf/sat/EenS03} it was possible to quite effectively find single-block collisions for MD4.
Using new propositional encoding methods (in particular, the \textsc{Transalg} system) and
state-of-the-art SAT solvers one can find preimages for MD4 and MD5 several hundred
times faster than it was done in \cite{DBLP:conf/SAT/MironovZ06}. Nevertheless, on the current stage SAT-based 
cryptanalysis is less effective than specialized methods (see, for example \cite{sasaki2007new,DBLP:journals/iacr/Stevens12,StevensKP16}) on problems of finding collisions of cryptographic hash functions. However, as far as we know, it is the SAT-based approach that yields best known preimage attacks on  truncated variants 
of hash functions \cite{DBLP:conf/SAT/DeKV07,Nossum2012,10.1007/978-3-319-66263-3_16}. In application to MD4-39, for a long time the SAT-based preimage attack from \cite{DBLP:conf/SAT/DeKV07} was considered to be the best. In \cite{GribanovaS18} we significantly improved the results from \cite{DBLP:conf/SAT/DeKV07}. It was possible for the large part thanks to functional capabilities of the \textsc{Transalg} system.

\section{Conclusion and future work}

In this paper, we study the principles of encoding the problems of inversion
of discrete functions from a wide class to the Boolean satisfiability
problem with a focus on cryptographic applications. We provide
the theoretical basis of SAT-based cryptanalysis and use it to
design the domain-specific system called \textsc{Transalg} for use in 
algebraic cryptanalysis.

In the comparison with relevant software tools for constructing
propositional encodings, such as \textsc{CBMC}, \textsc{SAW+Cryptol},
\textsc{URSA}, and \textsc{Grain-of-Salt}, we showed that 
the \textsc{Transalg} encoding concepts often make it possible 
to build SAT encodings of cryptanalysis problems which are on par
or better than that constructed by competitors. 
From the results of our study, it follows that in the cryptographic
context the overall functional capabilities of \textsc{Transalg}  
match that of the \textsc{CBMC} system which is the recognized 
leader in SAT-based Bounded Model Checking.
We also show how the distinctive features of \textsc{Transalg} can 
be useful in algebraic cryptanalysis on the example of the
applications described in the final part of the paper. 

In our opinion, \textsc{Transalg} often allows one to make better encodings than competition because it
uses a number of techniques to reduce the redundancy of the encodings
and also employs Boolean minimization (in form of the \textsc{Espresso} software tool) 
to make subformulas' representation in CNFs more compact. 
However, investigating this phenomenon in detail will take 
a large amount of computational experiments and time to comprehend their results. 
We are going to study these issues in the nearest future.

As a final comment, we would like to once more emphasize the theoretical 
and practical importance of SAT-based cryptanalysis and note that the
corresponding problems can be viewed as interesting challenges for researchers.
Therefore, they may stimulate the development of new algorithms and SAT solving techniques. 
We believe that the results presented in this paper will be useful in that context.


\section*{Acknowledgments}
\noindent
Authors express deep gratitude to professors Daniel Kroening, Predrag Jani\v ci\'c, and Aaron Tomb for feedback regarding several important details of \textsc{CBMC}, \textsc{URSA} and \textsc{Cryptol}, respectively. We are grateful to anonymous reviewers for their valuable comments that made it possible to significantly improve the quality of the present paper.

The research was funded by Russian Science Foundation (project No. 16-11-10046). Irina Gribanova and Stepan Kochemazov are additionally supported by the Council for Grants of the President of the Russian Federation (stipends SP-3545.2019.5 and SP-2017.2019.5, respectively).

\bibliographystyle{alpha}

\begin{thebibliography}{BHvMW09}

\bibitem[ABE00]{10.1007/3-540-46419-0_28}
Parosh~Aziz Abdulla, Per Bjesse, and Niklas E{\'e}n.
\newblock Symbolic reachability analysis based on {SAT}-solvers.
\newblock In {\em Tools and Algorithms for the Construction and Analysis of
  Systems ({TACAS})}, volume 1785 of {\em Lecture Notes in Computer Science}, pages
  411--425, 2000.

\bibitem[AS17]{syrup2017}
Gilles Audemard and Laurent Simon.
\newblock Glucose and {Syrup} in the {SAT'17}.
\newblock In Tom{\'{a}}s Balyo, Marijn J.~H. Heule, and Matti J{\"{a}}rvisalo,
  editors, {\em {SAT Competition} 2017}, volume B-2017-1, pages 16--17, 2017.

\bibitem[Bar09]{Bard:2009:AC:1618541}
Gregory~V. Bard.
\newblock {\em Algebraic Cryptanalysis}.
\newblock Springer Publishing Company, Incorporated, 1st edition, 2009.

\bibitem[BCC{\etalchar{+}}99]{Biere:1999:SMC:309847.309942}
Armin Biere, Alessandro Cimatti, Edmund~M. Clarke, Masahiro Fujita, and Yunshan
  Zhu.
\newblock Symbolic model checking using {SAT} procedures instead of {BDDs}.
\newblock In {\em Proc. of the 36th Annual ACM/IEEE Design Automation
  Conference ({DAC})}, pages 317--320, 1999.

\bibitem[BCC{\etalchar{+}}03]{Biere2003BoundedMC}
Armin Biere, Alessandro Cimatti, Edmund~M. Clarke, Ofer Strichman, and Yunshan
  Zhu.
\newblock {Bounded Model Checking}.
\newblock {\em Advances in Computers}, 58:117--148, 2003.

\bibitem[BCCZ99]{Biere:1999:SMC:646483.691738}
Armin Biere, Alessandro Cimatti, Edmund~M. Clarke, and Yunshan Zhu.
\newblock Symbolic model checking without {BDDs}.
\newblock In {\em Proc. of the 5th International Conference on Tools and
  Algorithms for Construction and Analysis of Systems ({TACAS})}, volume 1579
  of {\em Lecture Notes in Computer Science}, pages 193--207, 1999.

\bibitem[BCD{\etalchar{+}}11]{CVC4}
Clark Barrett, Christopher~L. Conway, Morgan Deters, Liana Hadarean, Dejan
  Jovanovi{\'c}, Tim King, Andrew Reynolds, and Cesare Tinelli.
\newblock {CVC4}.
\newblock In {\em Proc. of the 23rd International Conference on Computer
  Aided Verification ({CAV})}, volume 6806 of {\em Lecture Notes in Computer
  Science}, pages 171--177, 2011.
\newblock Snowbird, Utah.

\bibitem[BDF11]{bouillaguet-crypto11}
Charles Bouillaguet, Patrick Derbez, and Pierre-Alain Fouque.
\newblock Automatic search of attacks on round-reduced {AES} and applications.
\newblock In {\em Proc. of Advances in Cryptology ({CRYPTO})}, volume 6841 of {\em
  Lecture Notes in Computer Science}, pages 169--187, 2011.

\bibitem[BHvMW09]{DBLP:series/faia/2009-185}
Armin Biere, Marijn Heule, Hans van Maaren, and Toby Walsh, editors.
\newblock {\em Handbook of Satisfiability}, volume 185 of {\em Frontiers in
  Artificial Intelligence and Applications}. IOS Press, 2009.

\bibitem[Bie07]{Biere-FMV-TR-07-1}
Armin Biere.
\newblock The {AIGER And-Inverter Graph (AIG)} format version 20071012.
\newblock Technical Report 07/1, Institute for Formal Models and Verification,
  Johannes Kepler University, Altenbergerstr. 69, 4040 Linz, Austria, 2007.

\bibitem[Bie14]{lingeling2014}
Armin Biere.
\newblock Yet another local search solver and lingeling and friends entering
  the {SAT Competition} 2014.
\newblock In Anton Belov, Daniel Diepold, Marijn Heule, and Matti
  J{\"{a}}rvisalo, editors, {\em {SAT Competition} 2014}, volume B-2014-2,
  pages 39--40, 2014.

\bibitem[Bie17]{lingeling2017}
Armin Biere.
\newblock {CaDiCaL}, {Lingeling}, {Plingeling}, {Treengeling}, {YalSAT}
  entering the {SAT} competition 2017.
\newblock In Tom{\'{a}}s Balyo, Marijn J.~H. Heule, and Matti J{\"{a}}rvisalo,
  editors, {\em {SAT Competition} 2017}, volume B-2017-1, pages 14--15, 2017.

\bibitem[BKNW09]{DBLP:conf/ijcai/BessiereKNW09}
Christian Bessiere, George Katsirelos, Nina Narodytska, and Toby Walsh.
\newblock Circuit complexity and decompositions of global constraints.
\newblock In {\em Proc. of the 21st International Joint
  Conference on Artificial Intelligence ({IJCAI})}, pages 412--418, 2009.

\bibitem[Bru84]{Bruer}
J.~Bruer.
\newblock On pseudo-random sequences as crypto generators.
\newblock In {\em Proc. of the Int. Zurich Seminar on Digital Comm.},
  volume~46, pages 157--161, 1984.

\bibitem[BSVMH84]{Brayton:1984:LMA:577427}
Robert~King Brayton, Alberto~L. Sangiovanni-Vincentelli, Curtis~T. McMullen,
  and Gary~D. Hachtel.
\newblock {\em Logic Minimization Algorithms for VLSI Synthesis}.
\newblock Kluwer Academic Publishers, Norwell, MA, USA, 1984.

\bibitem[BSZK18]{OE_a51}
Vadim Bulavintsev, Alexander Semenov, Oleg Zaikin, and Stepan Kochemazov.
\newblock A bitslice implementation of {Anderson's} attack on {A5/1}.
\newblock {\em Open Engineering}, 8(1):7--16, 2018.

\bibitem[Can06]{DBLP:conf/isw/Canniere06}
Christophe~De Canni{\`{e}}re.
\newblock Trivium: {A} stream cipher construction inspired by block cipher
  design principles.
\newblock In {\em Proc. of the 9th International Conference on Information Security ({ISC})}, volume 4176 of {\em Lecture Notes in Computer Science}, pages
  171--186, 2006.

\bibitem[CB07]{CourtoisB2007}
Nicolas~T. Courtois and Gregory~V. Bard.
\newblock Algebraic cryptanalysis of the data encryption standard.
\newblock In {\em Proc. of Cryptography and Coding}, volume 4887 of {\em Lecture Notes
  in Computer Science}, pages 152--169, 2007.

\bibitem[CFH{\etalchar{+}}13]{Carter2013}
Kyle Carter, Adam Foltzer, Joe Hendrix, Brian Huffman, and Aaron Tomb.
\newblock {SAW}: The software analysis workbench.
\newblock In {\em Proc. of the 2013 ACM SIGAda Annual Conference on High
  Integrity Language Technology ({HILT})}, pages 15--18, 2013.

\bibitem[CGS12]{Courtois-tatra}
Nicolas~T. Courtois, Jerzy~A. Gawinecki, and Guangyan Song.
\newblock Contradiction immunity and guess-then-determine attacks on {GOST}.
\newblock {\em Tatra Mountains Mathematical Publications}, 53(1):2--13, 2012.

\bibitem[CKL04]{ckl2004}
Edmund Clarke, Daniel Kroening, and Flavio Lerda.
\newblock A tool for checking {ANSI-C} programs.
\newblock In {\em Proc. of Tools and Algorithms for the Construction and Analysis of
  Systems ({TACAS})}, volume 2988 of {\em Lecture Notes in Computer
  Science}, pages 168--176, 2004.

\bibitem[CM97]{Cook97findinghard}
Stephen~A. Cook and David~G. Mitchell.
\newblock Finding hard instances of the satisfiability problem: A survey.
\newblock In {\em DIMACS Series in Discr. Math. and Theoretical Comp. Sci.},
  volume~35, pages 1--17. American Mathematical Society, 1997.

\bibitem[Coo71]{DBLP:conf/stoc/Cook71}
Stephen~A. Cook.
\newblock The complexity of theorem-proving procedures.
\newblock In {\em Proc. of the 3rd Annual {ACM} Symposium on Theory of
  Computing}, pages 151--158, 1971.

\bibitem[Cou13]{Courtois-cryptologia}
Nicolas~T. Courtois.
\newblock Low-complexity key recovery attacks on {GOST} block cipher.
\newblock {\em Cryptologia}, 37(1):1--10, 2013.

\bibitem[Cou15]{Courtois-eprint}
Nicolas~T. Courtois.
\newblock Algebraic complexity reduction and cryptanalysis of {GOST}.
\newblock Cryptology ePrint Archive, Report 2011/626, 2011--2015.
\newblock \url{http://eprint.iacr.org/2011/626}.

\bibitem[Dam89]{DBLP:conf/crypto/Damgard89a}
Ivan Damg{\aa}rd.
\newblock A design principle for hash functions.
\newblock In {\em Proc. of Advances in Cryptology ({CRYPTO})}, volume 435 of {\em
  Lecture Notes in Computer Science}, pages 416--427, 1989.

\bibitem[DG84]{dowling-jlogp84}
William~F. Dowling and Jean~H. Gallier.
\newblock Linear-time algorithms for testing the satisfiability of
  propositional horn formulae.
\newblock {\em J. Log. Program.}, 1(3):267--284, 1984.

\bibitem[DKM{\etalchar{+}}17]{DBLP:conf/secrypt/DwivediKMNPW17}
Ashutosh~Dhar Dwivedi, Milos Kloucek, Pawel Morawiecki, Ivica Nikolic, Josef
  Pieprzyk, and Sebastian W{\'{o}}jtowicz.
\newblock {SAT}-based cryptanalysis of authenticated ciphers from the {CAESAR}
  competition.
\newblock In Pierangela Samarati, Mohammad~S. Obaidat, and Enrique Cabello,
  editors, {\em Proc. of the 14th International Joint Conference on
  e-Business and Telecommunications {(ICETE} 2017) - Volume 4: SECRYPT}, pages 237--246. SciTePress, 2017.

\bibitem[DKV07]{DBLP:conf/SAT/DeKV07}
Debapratim De, Abishek Kumarasubramanian, and Ramarathnam Venkatesan.
\newblock Inversion attacks on secure hash functions using {SAT} solvers.
\newblock In {\em Proc. of Theory and Applications of Satisfiability Testing ({SAT})}, volume 4501 of {\em Lecture Notes in Computer Science}, pages
  377--382, 2007.

\bibitem[DLL62]{Davis:1962}
Martin Davis, George Logemann, and Donald Loveland.
\newblock A machine program for theorem-proving.
\newblock {\em Commun. ACM}, 5(7):394--397, 1962.

\bibitem[dMB08]{Z3}
Leonardo de~Moura and Nikolaj Bj{\o}rner.
\newblock Z3: An efficient {SMT} solver.
\newblock In {\em Proc. of Tools and Algorithms for the Construction and Analysis of Systems ({TACAS})}, volume 4963 of {\em Lecture Notes in Computer Science}, pages 337--340, 2008.

\bibitem[Dob98]{DBLP:conf/fse/Dobbertin98}
Hans Dobbertin.
\newblock The first two rounds of {MD4} are not one-way.
\newblock In {\em Proc. of Fast Software Encryption ({FSE})}, volume 1372 of {\em Lecture Notes in Computer Science}, pages 284--292. Springer, 1998.

\bibitem[DP60]{Davis:1960:CPQ:321033.321034}
Martin Davis and Hilary Putnam.
\newblock A computing procedure for quantification theory.
\newblock {\em J. ACM}, 7(3):201--215, 1960.

\bibitem[Dut14]{YICES}
Bruno Dutertre.
\newblock Yices 2.2.
\newblock In {\em Proc. of Computer Aided Verification ({CAV})}, volume 8559 of 
{\em Lecture Notes in Computer Science}, pages 737--744, 2014.

\bibitem[ECW09]{DBLP:conf/fmcad/ErkokCW09}
Levent Erk{\"{o}}k, Magnus Carlsson, and Adam Wick.
\newblock Hardware/software co-verification of cryptographic algorithms using
  {Cryptol}.
\newblock In {\em Proc. of 9th International Conference on Formal Methods
  in Computer-Aided Design ({FMCAD})}, pages 188--191. {IEEE}, 2009.

\bibitem[EM09]{ErkokPLPV09}
Levent Erk\"{o}k and John Matthews.
\newblock Pragmatic equivalence and safety checking in {Cryptol}.
\newblock In {\em Proc. of the 3rd Workshop on Programming Languages
  Meets Program Verification ({PLPV})}, pages 73--82, New York, NY, USA, 2009.
  ACM.

\bibitem[EPV08]{DBLP:conf/SAT/EibachPV08}
Tobias Eibach, Enrico Pilz, and Gunnar V{\"o}lkel.
\newblock Attacking {Bivium} using {SAT} solvers.
\newblock In {\em Theory and Applications of Satisfiability Testing ({SAT})}, volume 4996 of {\em Lecture Notes in Computer Science}, pages 63--76,
  2008.

\bibitem[ES03]{DBLP:journals/entcs/EenS03}
Niklas E{\'{e}}n and Niklas S{\"{o}}rensson.
\newblock Temporal induction by incremental {SAT} solving.
\newblock {\em Electr. Notes Theor. Comput. Sci.}, 89(4):543--560, 2003.

\bibitem[ES04]{DBLP:conf/sat/EenS03}
Niklas E{\'{e}}n and Niklas S{\"{o}}rensson.
\newblock An extensible {SAT-solver}.
\newblock In {\em Proc. of the 6th International Conference on Theory and Applications of Satisfiability Testing ({SAT})}, volume 2919 of
  {\em Lecture Notes in Computer Science}, pages 502--518, 2004.

\bibitem[EVP10]{Eibach2010}
Tobias Eibach, Gunnar V{\"o}lkel, and Enrico Pilz.
\newblock Optimising {Gr{\"o}bner} bases on {Bivium}.
\newblock {\em Mathematics in Computer Science}, 3(2):159--172, Apr 2010.

\bibitem[FBSK17]{painless2017}
Ludovic~Le Frioux, Souheib Baarir, Julien Sopena, and Fabrice Kordon.
\newblock painless-maplecomsps.
\newblock In Tom{\'{a}}s Balyo, Marijn J.~H. Heule, and Matti J{\"{a}}rvisalo,
  editors, {\em {SAT Competition} 2017}, volume B-2017-1, pages 26--27, 2017.

\bibitem[Gef73]{Geffe}
P.~Geffe.
\newblock How to protect data with ciphers that are really hard to break.
\newblock {\em Electronics}, 46(1):99--101, 1973.

\bibitem[GJ79]{Garey}
Michael~R. Garey and David~S. Johnson.
\newblock {\em Computers and Intractability: A Guide to the Theory of
  NP-Completeness}.
\newblock W. H. Freeman \& Co., New York, NY, USA, 1979.

\bibitem[GKN{\etalchar{+}}08]{Guneysu2008}
Tim G\"{u}neysu, Timo Kasper, Martin Novotn\'{y}, Christof Paar, and Andy Rupp.
\newblock {Cryptanalysis with COPACOBANA}.
\newblock {\em IEEE Trans. Comput.}, 57(11):1498--1513, 2008.

\bibitem[GKNS07]{Gebser07}
Martin Gebser, Benjamin Kaufmann, Andr{\'e} Neumann, and Torsten Schaub.
\newblock clasp: A conflict-driven answer set solver.
\newblock In {\em Proc. of Logic Programming and Nonmonotonic Reasoning}, volume 4483 of {\em Lecture Notes in Computer Science}, pages 260--265, 2007.

\bibitem[GL97]{Glover:1997:TS:549765}
Fred Glover and Manuel Laguna.
\newblock {\em Tabu Search}.
\newblock Kluwer Academic Publishers, Norwell, MA, USA, 1997.

\bibitem[Gol08]{Goldreich:2008:CCC:1373317}
Oded Goldreich.
\newblock {\em Computational Complexity: A Conceptual Perspective}.
\newblock Cambridge University Press, New York, NY, USA, 1 edition, 2008.

\bibitem[GS18]{GribanovaS18}
Irina Gribanova and Alexander Semenov.
\newblock Using automatic generation of relaxation constraints to improve the
  preimage attack on 39-step {MD4}.
\newblock In {\em Proc. of the 41st International Convention on Information and
  Communication Technology, Electronics and Microelectronics ({MIPRO})},
  pages 1174--1179. IEEE, 2018.

\bibitem[GZK{\etalchar{+}}17]{mit2016}
Irina Gribanova, Oleg Zaikin, Stepan Kochemazov, Ilya Otpuschennikov, and
  Alexander Semenov.
\newblock The study of inversion problems of cryptographic hash functions from
  {MD} family using algorithms for solving {Boolean} satisfiability problem.
\newblock In {\em Proc. of International Conference Mathematical and Information
  Technologies}, volume 1839, pages 98--113. CEUR-WS, 2017.

\bibitem[HJM07]{Hell:2007:GSC:1358393.1358401}
Martin Hell, Thomas Johansson, and Willi Meier.
\newblock Grain: a stream cipher for constrained environments.
\newblock {\em Int. J. Wire. Mob. Comput.}, 2(1):86--93, May 2007.

\bibitem[Hyv11]{Hyvarinen11}
Antti E.~J. Hyv{\"{a}}rinen.
\newblock {\em {Grid Based Propositional Satisfiability Solving}}.
\newblock PhD thesis, Aalto University, 2011.

\bibitem[Ign]{web:mkplot}
Alexey Ignatiev.
\newblock mkplot: a {Python} script to create cactus and scatter plots based on
  matplotlib. {URL}: https://github.com/alexeyignatiev/mkplot.

\bibitem[IMMS18]{imms-sat18}
Alexey Ignatiev, Antonio Morgado, and Jo{\~a}o~P. Marques-Silva.
\newblock {PySAT:} {A} {Python} toolkit for prototyping with {SAT} oracles.
\newblock In {\em Proc. of Theory and Applications of Satisfiability Testing  ({SAT})}, volume 10929 of {\em Lecture Notes in Computer Science}, pages
  428--437, 2018.

\bibitem[Jan12]{journals/lmcs/predrag}
Predrag Janicic.
\newblock {URSA}: a system for uniform reduction to {SAT}.
\newblock {\em Logical Methods in Computer Science}, 8(3):1--39, 2012.

\bibitem[JBH12]{DBLP:journals/jar/JarvisaloBH12}
Matti J{\"{a}}rvisalo, Armin Biere, and Marijn Heule.
\newblock Simulating circuit-level simplifications on {CNF}.
\newblock {\em J. Autom. Reasoning}, 49(4):583--619, 2012.

\bibitem[JJ05]{DBLP:conf/frocos/JovanovicJ05}
Dejan Jovanovic and Predrag Janicic.
\newblock Logical analysis of hash functions.
\newblock In {\em Proc. of the 5th International Workshop on Frontiers of Combining Systems ({FroCoS})}, volume 3717 of {\em Lecture Notes in Computer Science}, pages 200--215, 2005.

\bibitem[JJ09]{Jarvisalo2009}
Matti J{\"a}rvisalo and Tommi Junttila.
\newblock Limitations of restricted branching in clause learning.
\newblock {\em Constraints}, 14(3):325--356, Sep 2009.

\bibitem[Kin76]{King:1976:SEP:360248.360252}
James~C. King.
\newblock Symbolic execution and program testing.
\newblock {\em Commun. ACM}, 19(7):385--394, July 1976.

\bibitem[Kro09]{DBLP:series/faia/Kroening09}
Daniel Kroening.
\newblock Software verification.
\newblock In Biere et~al. \cite{DBLP:series/faia/2009-185}, pages 505--532.

\bibitem[Leu08]{Leurent08}
Ga{\"e}tan Leurent.
\newblock {MD4} is not one-way.
\newblock In {\em Proc. of Fast Software Encryption (FSE)}, volume 5086 of {\em Lecture Notes in Computer Science}, pages 412--428, 2008.

\bibitem[LGPC16]{DBLP:conf/sat/LiangGPC16}
Jia~Hui Liang, Vijay Ganesh, Pascal Poupart, and Krzysztof Czarnecki.
\newblock Learning rate based branching heuristic for {SAT} solvers.
\newblock In {\em Proc. of Theory and Applications of Satisfiability Testing ({SAT})}, volume 9710 of {\em Lecture Notes in Computer Science}, pages 123--140. Springer, 2016.

\bibitem[LM03]{LewisM03}
J.~R. {Lewis} and B.~{Martin}.
\newblock Cryptol: high assurance, retargetable crypto development and
  validation.
\newblock In {\em IEEE Military Communications Conference, 2003. MILCOM 2003.},
  volume~2, pages 820--825, 2003.

\bibitem[LMH15]{Lafitte15}
Fr{\'e}d{\'e}ric Lafitte, Olivier Markowitch, and Dirk~Van Heule.
\newblock {SAT} based analysis of {LTE} stream cipher {ZUC}.
\newblock {\em Journal of Information Security and Applications}, 22:54 -- 65,
  2015.
\newblock Special Issue on Security of Information and Networks.

\bibitem[LOG{\etalchar{+}}17]{maplecomsps2017}
Jia~Hui Liang, Chanseok Oh, Vijay Ganesh, Krzysztof Czarnecki, and Pascal
  Poupart.
\newblock {MapleCOMSPS\_LRB\_VSIDS and MapleCOMSPS\_CHB\_VSIDS}.
\newblock In Tom{\'{a}}s Balyo, Marijn J.~H. Heule, and Matti J{\"{a}}rvisalo,
  editors, {\em {SAT Competition} 2017}, volume B-2017-1, pages 20--21, 2017.

\bibitem[Mar09]{Maric09}
Filip Mari\'{c}.
\newblock Formalization and implementation of modern {SAT} solvers.
\newblock {\em J. Autom. Reason.}, 43(1):81--119, 2009.

\bibitem[Mas99]{Massacci:1999:UWR:1624218.1624261}
Fabio Massacci.
\newblock Using {Walk-SAT} and {Rel-SAT} for cryptographic key search.
\newblock In {\em Proc. of the 16th International Joint Conference on Artifical Intelligence (IJCAI)}, pages 290--295. Morgan Kaufmann Publishers Inc., 1999.

\bibitem[mat]{Matrosov-web}
{Irkutsk Supercomputer Center of SB RAS. URL}: \url{http://hpc.icc.ru}.

\bibitem[MCP07]{Mcdonald_attackingbivium}
Cameron Mcdonald, Chris Charnes, and Josef Pieprzyk.
\newblock {Attacking Bivium with MiniSat}.
\newblock Technical Report 2007/040, ECRYPT Stream Cipher Project, 2007.

\bibitem[Mer89]{DBLP:conf/crypto/Merkle89}
Ralph~C. Merkle.
\newblock A certified digital signature.
\newblock In {\em Proc. of Advances in Cryptology ({CRYPTO})}, volume 435 of {\em Lecture Notes in Computer Science}, pages 218--238, 1989.

\bibitem[MM00]{DBLP:journals/jar/MassacciM00}
Fabio Massacci and Laura Marraro.
\newblock Logical cryptanalysis as a {SAT} problem.
\newblock {\em J. Autom. Reasoning}, 24(1/2):165--203, 2000.

\bibitem[MMZ{\etalchar{+}}01]{Moskewicz:2001}
Matthew~W. Moskewicz, Conor~F. Madigan, Ying Zhao, Lintao Zhang, and Sharad
  Malik.
\newblock Chaff: Engineering an efficient {SAT} solver.
\newblock In {\em Proc. of the 38th Annual Design Automation Conference ({DAC})}, pages 530--535, 2001.

\bibitem[MS91]{MeierS91}
Willi Meier and Othmar Staffelbach.
\newblock Analysis of pseudo random sequences generated by cellular automata.
\newblock In {\em Proc. of Advances in Cryptology ({EUROCRYPT})}, volume 547 of {\em Lecture Notes in Computer Science}, pages 186--199, 1991.

\bibitem[MS99]{DBLP:journals/tc/Marques-SilvaS99}
Jo{\~{a}}o~P. Marques{-}Silva and Karem~A. Sakallah.
\newblock {GRASP}: a search algorithm for propositional satisfiability.
\newblock {\em {IEEE} Trans. Computers}, 48(5):506--521, 1999.

\bibitem[MS08]{JPMS2008}
Jo{\~a}o~P. Marques-Silva.
\newblock Practical applications of {Boolean satisfiability}.
\newblock In {\em Proc. of the 9th International Workshop on Discrete Event Systems}, pages 74--80, 2008.

\bibitem[MSLM09]{MSLM09}
Jo{\~a}o~P. Marques-Silva, In{\^e}s Lynce, and Sharad Malik.
\newblock Conflict-driven clause learning {SAT} solvers.
\newblock In Biere et~al. \cite{DBLP:series/faia/2009-185}, pages 131--153.

\bibitem[MSS96]{Marques-Silva:1997}
Jo{\~a}o~P. Marques-Silva and Karem~A. Sakallah.
\newblock {GRASP}: a new search algorithm for satisfiability.
\newblock In {\em Proc. of the 1996 IEEE/ACM International Conference on
  Computer-aided Design ({ICCAD})}, pages 220--227, 1996.

\bibitem[MU49]{Metropolis49}
N~Metropolis and S~Ulam.
\newblock The {Monte Carlo} method.
\newblock {\em J. Amer. statistical assoc.}, 44(247):335--341, 1949.

\bibitem[MVO96]{Menezes:1996:HAC:548089}
Alfred~J. Menezes, Scott~A. Vanstone, and Paul C.~Van Oorschot.
\newblock {\em Handbook of Applied Cryptography}.
\newblock CRC Press, Inc., Boca Raton, FL, USA, 1st edition, 1996.

\bibitem[MZ06]{DBLP:conf/SAT/MironovZ06}
Ilya Mironov and Lintao Zhang.
\newblock Applications of {SAT} solvers to cryptanalysis of hash functions.
\newblock In {\em Proc. of Theory and Applications of Satisfiability Testing  ({SAT})}, volume 4121 of {\em Lecture Notes in Computer Science}, pages 102--115, 2006.

\bibitem[NLG{\etalchar{+}}17]{NejatiLGCG17}
Saeed Nejati, Jia~Hui Liang, Catherine Gebotys, Krzysztof Czarnecki, and Vijay
  Ganesh.
\newblock Adaptive restart and {CEGAR}-based solver for inverting cryptographic
  hash functions.
\newblock In {\em Proc. of Verified Software. Theories, Tools, and Experiments}, volume
  10712 of {\em Lecture Notes in Computer Science}, pages 120--131, 2017.

\bibitem[NNS{\etalchar{+}}17]{10.1007/978-3-319-66263-3_16}
Saeed Nejati, Zack Newsham, Joseph Scott, Jia~Hui Liang, Catherine Gebotys,
  Pascal Poupart, and Vijay Ganesh.
\newblock A propagation rate based splitting heuristic for divide-and-conquer
  solvers.
\newblock In {\em Proc. of Theory and Applications of Satisfiability Testing ({SAT})}, volume 10491 of {\em Lecture Notes in Computer Science}, pages 251--260, 2017.

\bibitem[Noh10]{Nohl2010}
Karsten Nohl.
\newblock Attacking phone privacy.
\newblock In {\em Proc. of {BlackHat 2010 Lecture Notes, Las-Vegas, USA}}, pages 1--6, 2010.

\bibitem[Nos12]{Nossum2012}
Vegard Nossum.
\newblock {SAT-based preimage attacks on SHA-1}.
\newblock Master's thesis, University of Oslo, Department of Informatics, 2012.

\bibitem[NR18]{ChronoBT}
Alexander Nadel and Vadim Ryvchin.
\newblock Chronological backtracking.
\newblock In {\em Proc. Theory and Applications of Satisfiability Testing ({SAT})}, volume 10929 of {\em Lecture Notes in Computer Science}, pages 111--121, 2018.

\bibitem[OGS]{transalg-repository}
Ilya Otpuschennikov, Irina Gribanova, and Alexander Semenov.
\newblock {Transalg} repository, {URL}: https://gitlab.com/transalg/transalg.

\bibitem[OGZ{\etalchar{+}}]{instances-repository}
Ilya Otpuschennikov, Irina Gribanova, Oleg Zaikin, Stepan Kochemazov, and
  Alexander Semenov.
\newblock {SAT} and {SMT} encodings of cryptograhic problems made by the
  {Transalg} tool and its competitors, {URL}:
  https://github.com/{Nauchnik}/transalg-comparison.

\bibitem[OGZS]{satencodings-repository}
Ilya Otpuschennikov, Irina Gribanova, Oleg Zaikin, and Alexander Semenov.
\newblock The repository of {SAT} encodings for various cryptographic
  primitives constructed by {Transalg}, {URL}:
  https://gitlab.com/satencodings/satencodings.

\bibitem[Oh17]{pulsar2017}
Chanseok Oh.
\newblock {COMiniSatPS Pulsar and GHackCOMSPS}.
\newblock In Tom{\'{a}}s Balyo, Marijn J.~H. Heule, and Matti J{\"{a}}rvisalo,
  editors, {\em {SAT Competition} 2017}, volume B-2017-1, pages 12--13, 2017.

\bibitem[OS11]{OtpSem11}
Ilya Otpuschennikov and Alexander Semenov.
\newblock Technology for translating combinatorial problems into boolean
  equations.
\newblock {\em Prikl. Diskr. Mat.}, 1(11):96--115, 2011.

\bibitem[OSG{\etalchar{+}}16]{DBLP:conf/ecai/OtpuschennikovS16}
Ilya Otpuschennikov, Alexander Semenov, Irina Gribanova, Oleg Zaikin, and
  Stepan Kochemazov.
\newblock Encoding cryptographic functions to {SAT} using {TRANSALG} system.
\newblock In {\em Proc. of the 22nd European Conference on Artificial Intelligence ({ECAI})}, volume 285 of {\em Frontiers in Artificial Intelligence and Applications}, pages 1594--1595. {IOS} Press, 2016.

\bibitem[PBG05]{Prasad2005}
Mukul~R. Prasad, Armin Biere, and Aarti Gupta.
\newblock A survey of recent advances in {SAT-based} formal verification.
\newblock {\em International Journal on Software Tools for Technology
  Transfer}, 7(2):156--173, 2005.

\bibitem[PBU19]{DBLP:conf/gecco/PavlenkoBU19}
Artem Pavlenko, Maxim Buzdalov, and Vladimir Ulyantsev.
\newblock Fitness comparison by statistical testing in construction of
  {SAT}-based guess-and-determine cryptographic attacks.
\newblock In {\em Proc. of the Genetic and Evolutionary Computation
  Conference ({GECCO})}, pages 312--320. {ACM}, 2019.

\bibitem[PSU19]{DBLP:conf/evoW/PavlenkoSU19}
Artem Pavlenko, Alexander Semenov, and Vladimir Ulyantsev.
\newblock Evolutionary computation techniques for constructing {SAT}-based
  attacks in algebraic cryptanalysis.
\newblock In {\em Proc. of the 22nd International Conference Applications of Evolutionary Computation}, volume 11454 of {\em Lecture Notes in Computer Science}, pages 237--253, 2019.

\bibitem[PSZ12]{csci37}
Mikhail Posypkin, Alexander Semenov, and Oleg Zaikin.
\newblock Using {BOINC} desktop grid to solve large scale {SAT} problems.
\newblock {\em Computer Science}, 13(1):25, 2012.

\bibitem[Sie84]{Siegenthaler84}
Thomas Siegenthaler.
\newblock Correlation-immunity of nonlinear combining functions for
  cryptographic applications.
\newblock {\em {IEEE} Trans. Information Theory}, 30(5):776--780, 1984.

\bibitem[SKP16]{StevensKP16}
Marc Stevens, Pierre Karpman, and Thomas Peyrin.
\newblock Freestart collision for full {SHA-1}.
\newblock In {\em Proc. of Advances in Cryptology ({EUROCRYPT})}, volume 9665 of {\em Lecture Notes in Computer Science}, pages 459--483. Springer Berlin Heidelberg, 2016.

\bibitem[SMF12]{LLBMC}
Carsten Sinz, Florian Merz, and Stephan Falke.
\newblock {LLBMC}: A bounded model checker for {LLVM's} intermediate
  representation.
\newblock In {\em Proc. of Tools and Algorithms for the Construction and Analysis of Systems ({TACAS})}, volume 7214 of {\em Lecture Notes in Computer Science}, pages 542--544, 2012.

\bibitem[SNC09]{DBLP:conf/SAT/SoosNC09}
Mate Soos, Karsten Nohl, and Claude Castelluccia.
\newblock Extending {SAT} solvers to cryptographic problems.
\newblock In {\em Proc. of Theory and Applications of Satisfiability Testing ({SAT})}, volume 5584 of {\em Lecture Notes in Computer Science}, pages 244--257, 2009.

\bibitem[Soo10]{DBLP:conf/tools/Soos10}
Mate Soos.
\newblock {Grain of Salt} - an automated way to test stream ciphers through
  {SAT} solvers.
\newblock In {\em Proc. of Workshop on Tools for Cryptanalysis ({Tools})}, pages 131--144, 2010.

\bibitem[SSA{\etalchar{+}}09]{10.1007/978-3-642-03356-8_4}
Marc Stevens, Alexander Sotirov, Jacob Appelbaum, Arjen Lenstra, David Molnar,
  Dag~Arne Osvik, and Benne de~Weger.
\newblock Short chosen-prefix collisions for {MD5} and the creation of a rogue
  {CA} certificate.
\newblock In {\em Proc. of Advances in Cryptology ({CRYPTO})}, volume 5677 of {\em Lecture Notes in Computer Science}, pages 55--69, 2009.

\bibitem[Ste12]{DBLP:journals/iacr/Stevens12}
Marc Stevens.
\newblock Single-block collision attack on {MD5}.
\newblock {\em {IACR} Cryptology ePrint Archive}, 2012:40, 2012.

\bibitem[Sto16]{Stoffelen:2016:OSI:3081631.3081642}
Ko~Stoffelen.
\newblock Optimizing {S}-box implementations for several criteria using {SAT}
  solvers.
\newblock In {\em Proc. of the 23rd International Conference on Fast Software Encryption ({FSE})}, volume 9783 of {\em Lecture Notes in Computer Science}, pages 140--160. Springer Berlin Heidelberg, 2016.

\bibitem[SWOK07]{sasaki2007new}
Yu~Sasaki, Lei Wang, Kazuo Ohta, and Noboru Kunihiro.
\newblock New message difference for {MD4}.
\newblock In {\em Proc. of Fast Software Encryption ({FSE})}, volume 4593 of {\em Lecture Notes in Computer Science}, pages 329--348, 2007.

\bibitem[SZ15]{DBLP:conf/pact/SemenovZ15}
Alexander Semenov and Oleg Zaikin.
\newblock Using {Monte Carlo} method for searching partitionings of hard
  variants of {Boolean} satisfiability problem.
\newblock In {\em Proc. of the 13th International Conference on Parallel Computing Technologies ({PaCT})}, volume 9251 of {\em Lecture Notes in Computer
Science}, pages 222--230, 2015.

\bibitem[SZ16]{Semenov2016}
Alexander Semenov and Oleg Zaikin.
\newblock Algorithm for finding partitionings of hard variants of {Boolean}
  satisfiability problem with application to inversion of some cryptographic
  functions.
\newblock {\em SpringerPlus}, 5(1):1--16, 2016.

\bibitem[SZBP11]{DBLP:conf/pact/SemenovZBP11}
Alexander Semenov, Oleg Zaikin, Dmitry Bespalov, and Mikhail Posypkin.
\newblock Parallel logical cryptanalysis of the generator {A5/1} in {BNB}-grid
  system.
\newblock In {\em Proc. of the 11th International Conference on Parallel Computing Technologies ({PaCT})}, volume 6873 of {\em Lecture Notes in Computer Science}, pages 473--483, 2011.

\bibitem[SZO{\etalchar{+}}18]{SemeonovZOKI2018}
Alexander Semenov, Oleg Zaikin, Ilya Otpuschennikov, Stepan Kochemazov, and
  Alexey Ignatiev.
\newblock On cryptographic attacks using backdoors for {SAT}.
\newblock In {\em Proc. of the Thirty-Second {AAAI} Conference on Artificial
  Intelligence ({AAAI})}, pages 6641--6648, 2018.

\bibitem[Tse70]{Tseitin70}
Grigori~S Tseitin.
\newblock On the complexity of derivation in propositional calculus.
\newblock {\em Studies in constructive mathematics and mathematical logic, part
  II, Seminars in mathematics}, pages 115--125, 1970.

\bibitem[{von}51]{vonN51}
John {von Neumann}.
\newblock {The general and logical theory of automata}.
\newblock In L.~A. Jeffress, editor, {\em Cerebral Mechanisms in Behavior --
  The Hixon Symposium}, pages 1--31. John Wiley \& Sons, 1951.

\bibitem[WGS03]{Williams:2003:BTC:1630659.1630827}
Ryan Williams, Carla~P. Gomes, and Bart Selman.
\newblock Backdoors to typical case complexity.
\newblock In {\em Proc. of the 18th International Joint Conference on Artificial
  Intelligence ({IJCAI})}, pages 1173--1178, 2003.

\bibitem[WLF{\etalchar{+}}05]{DBLP:conf/eurocrypt/WangLFCY05}
Xiaoyun Wang, Xuejia Lai, Dengguo Feng, Hui Chen, and Xiuyuan Yu.
\newblock Cryptanalysis of the hash functions {MD4} and {RIPEMD}.
\newblock In {\em Proc. of Advances in Cryptology ({EUROCRYPT})}, volume 3494 of
  {\em Lecture Notes in Computer Science}, pages 1--18, 2005.

\bibitem[Wol86]{Wolfram}
Stephen Wolfram.
\newblock Random sequence generation by cellular automata.
\newblock {\em Advances in Applied Mathematics}, 7(2):123 -- 169, 1986.

\bibitem[WY05]{DBLP:conf/eurocrypt/WangY05}
Xiaoyun Wang and Hongbo Yu.
\newblock How to break {MD5} and other hash functions.
\newblock In {\em Proc. of Advances in Cryptology ({EUROCRYPT})}, volume 3494 of
  {\em Lecture Notes in Computer Science}, pages 19--35, 2005.

\bibitem[XLL{\etalchar{+}}17]{maplelcm2017}
Fan Xiao, Mao Luo, Chu-Min Li, Felip Many{\`{a}}, and Zhipeng L{\"{u}}.
\newblock {MapleLRB\_LCM, Maple\_LCM, Maple\_Dist, MapleLRB\_LCMoccRestart and
  Glucose-3.0+width in SAT Competition 2017}.
\newblock In Tom{\'{a}}s Balyo, Marijn J.~H. Heule, and Matti J{\"{a}}rvisalo,
  editors, {\em {SAT Competition} 2017}, volume B-2017-1, pages 22--23, 2017.

\bibitem[XWCC18]{Scavel2018}
Yang Xu, Guanfeng Wu, Qingshan Chen, and Shuwei Chen.
\newblock {Maple\_LCM\_Scavel} and {Maple\_LCM\_Scavel\_200}.
\newblock In Marijn Heule, Matti J{\"a}rvisalo, and Martin Suda, editors, {\em
  Proc.~of {SAT Competition} 2018 -- Solver and Benchmark Descriptions}, volume
  B-2018-1 of {\em Department of Computer Science Series of Publications B},
  page~30. University of Helsinki, 2018.

\bibitem[YO14]{yasumoto-rokk}
Takeru Yasumoto and Takumi Okuwaga.
\newblock Rokk 1.0.1.
\newblock In Anton Belov, Daniel Diepold, Marijn Heule, and Matti
  J{\"{a}}rvisalo, editors, {\em SAT Competition 2014}, page~70, 2014.

\bibitem[ZK17]{DBLP:conf/isw/ZaikinK17}
Oleg Zaikin and Stepan Kochemazov.
\newblock An improved {SAT-based} guess-and-determine attack on the alternating
  step generator.
\newblock In {\em Proc. of the 20th Information Security Conference ({ISC})}, volume 10599 of {\em Lecture Notes in Computer Science}, pages 21--38, 2017.

\bibitem[ZMMM01]{Zhang:2001:ECD:603095.603153}
Lintao Zhang, Conor~F. Madigan, Matthew~H. Moskewicz, and Sharad Malik.
\newblock Efficient conflict driven learning in a {Boolean} satisfiability
  solver.
\newblock In {\em Proc. of IEEE/ACM International Conference on Computer-aided
  Design ({ICCAD})}, pages 279--285, 2001.

\end{thebibliography}
\newcommand{\etalchar}[1]{$^{#1}$}

\begin{appendices}
\newpage

\section{Detailed comparison of all considered solvers}\label{appA}

\begin{figure}[ht]
	\centering
	\subfloat[][\textsc{URSA}]{
    \includegraphics[width=0.5\textwidth]{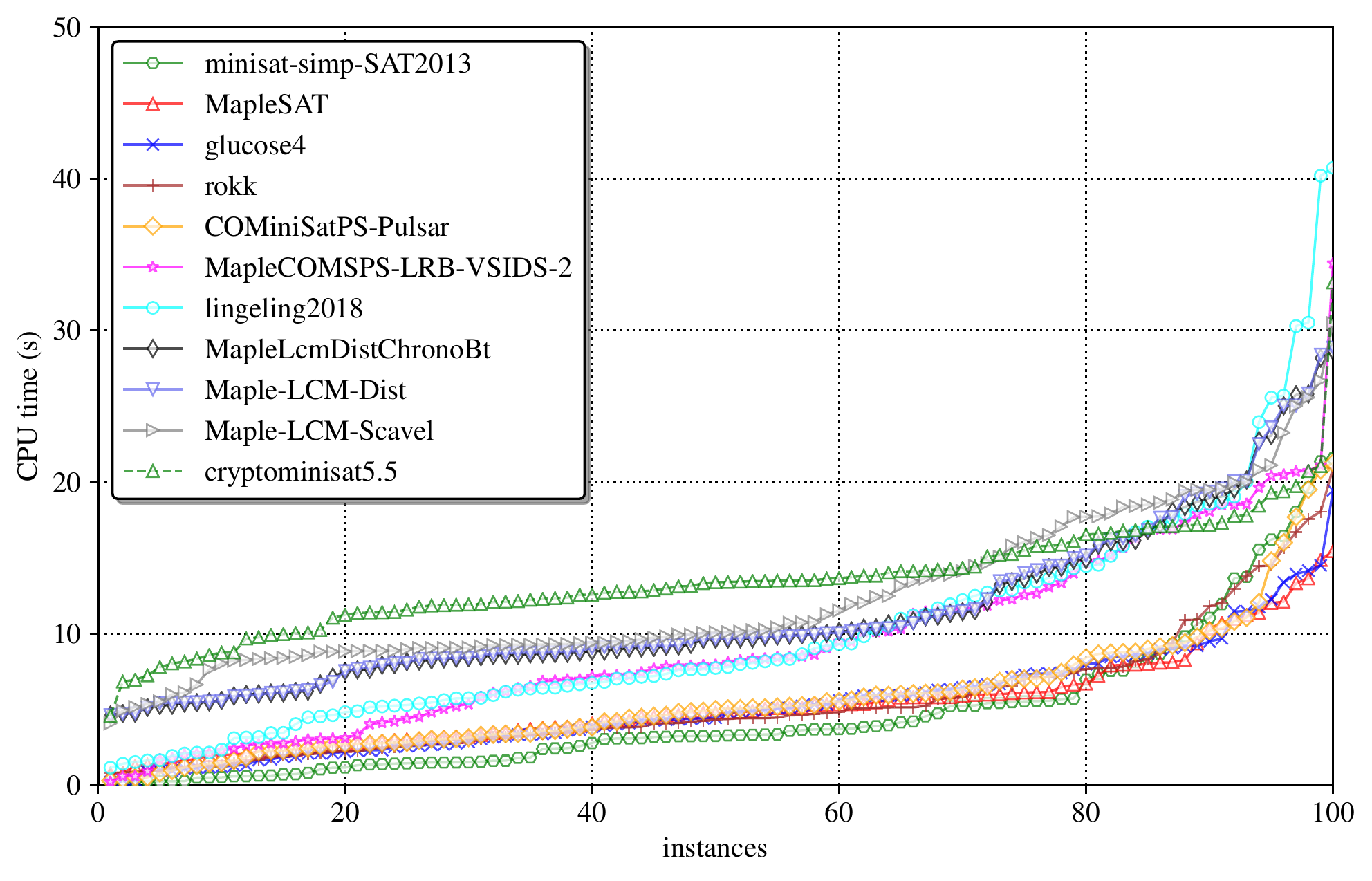}
		\label{fig:geffe-ursa}
  }
 	\subfloat[][\textsc{Transalg}]{
    \includegraphics[width=0.5\textwidth]{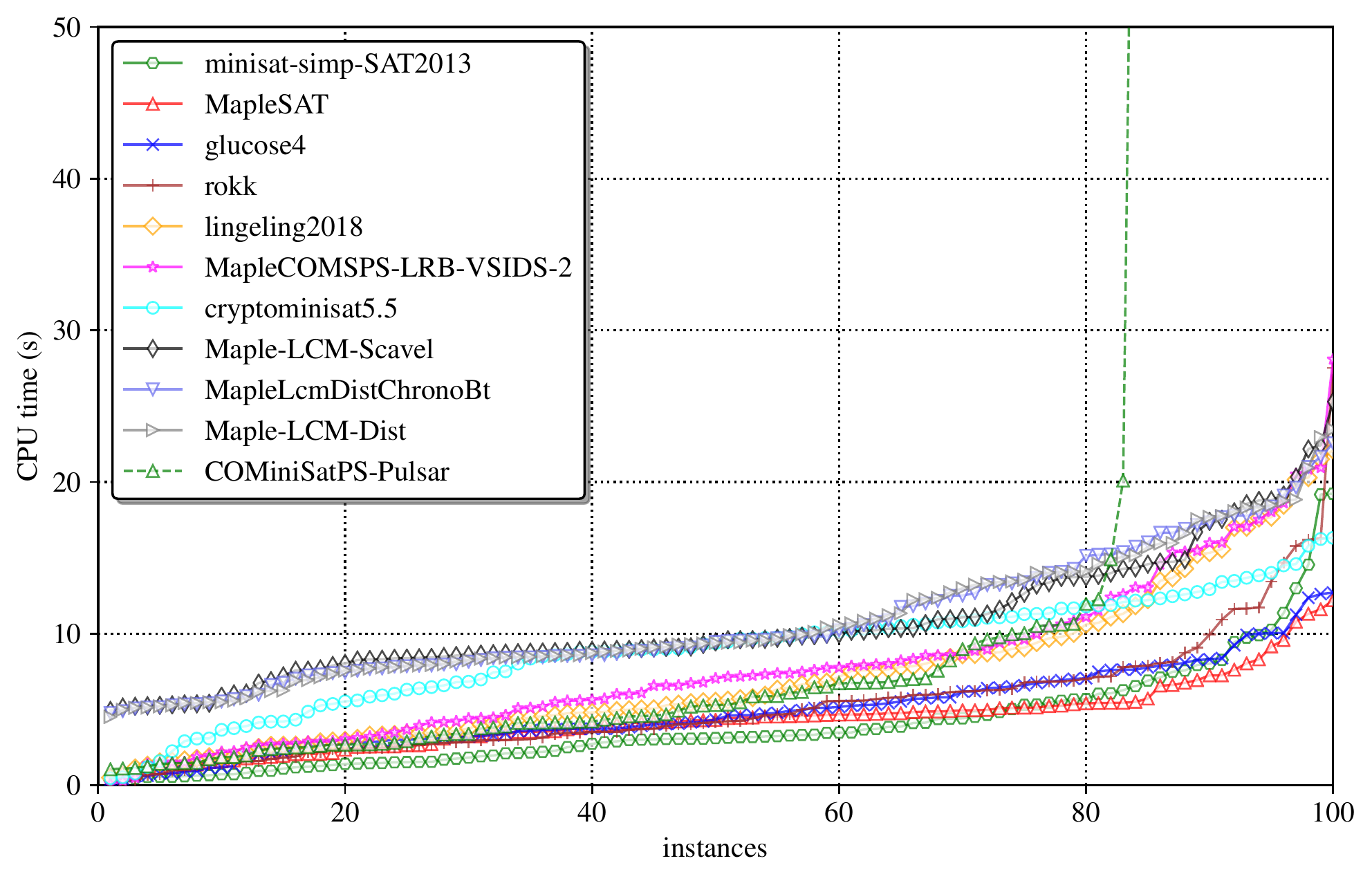}
		\label{fig:geffe-transalg}
  }
   \vskip\baselineskip
	\subfloat[][\textsc{Cryptol-SAT}]{
    \includegraphics[width=0.5\textwidth]{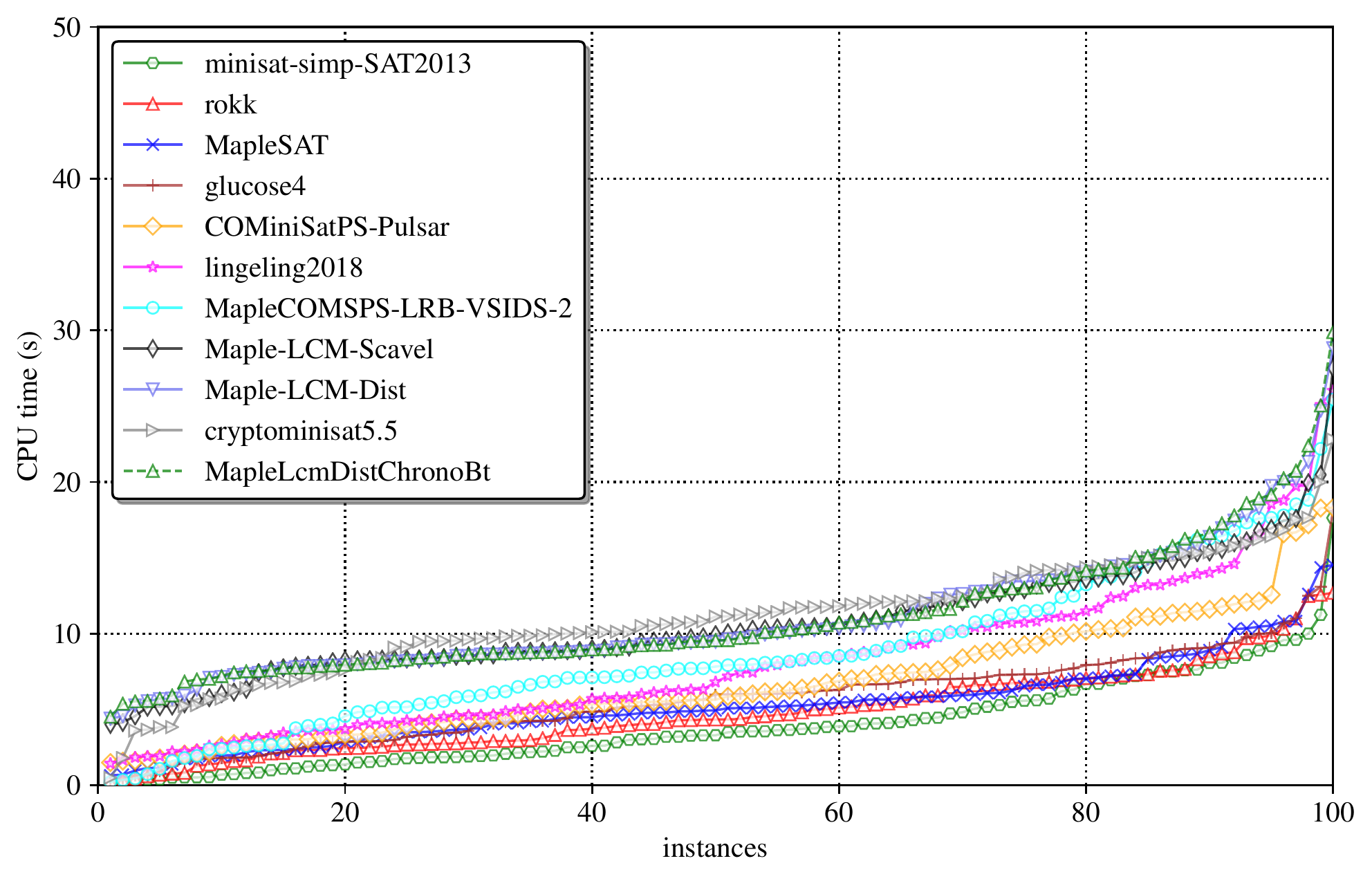}
		\label{fig:geffe-cryptol}
  }
	\subfloat[][\textsc{Grain-of-Salt}]{
    \includegraphics[width=0.5\textwidth]{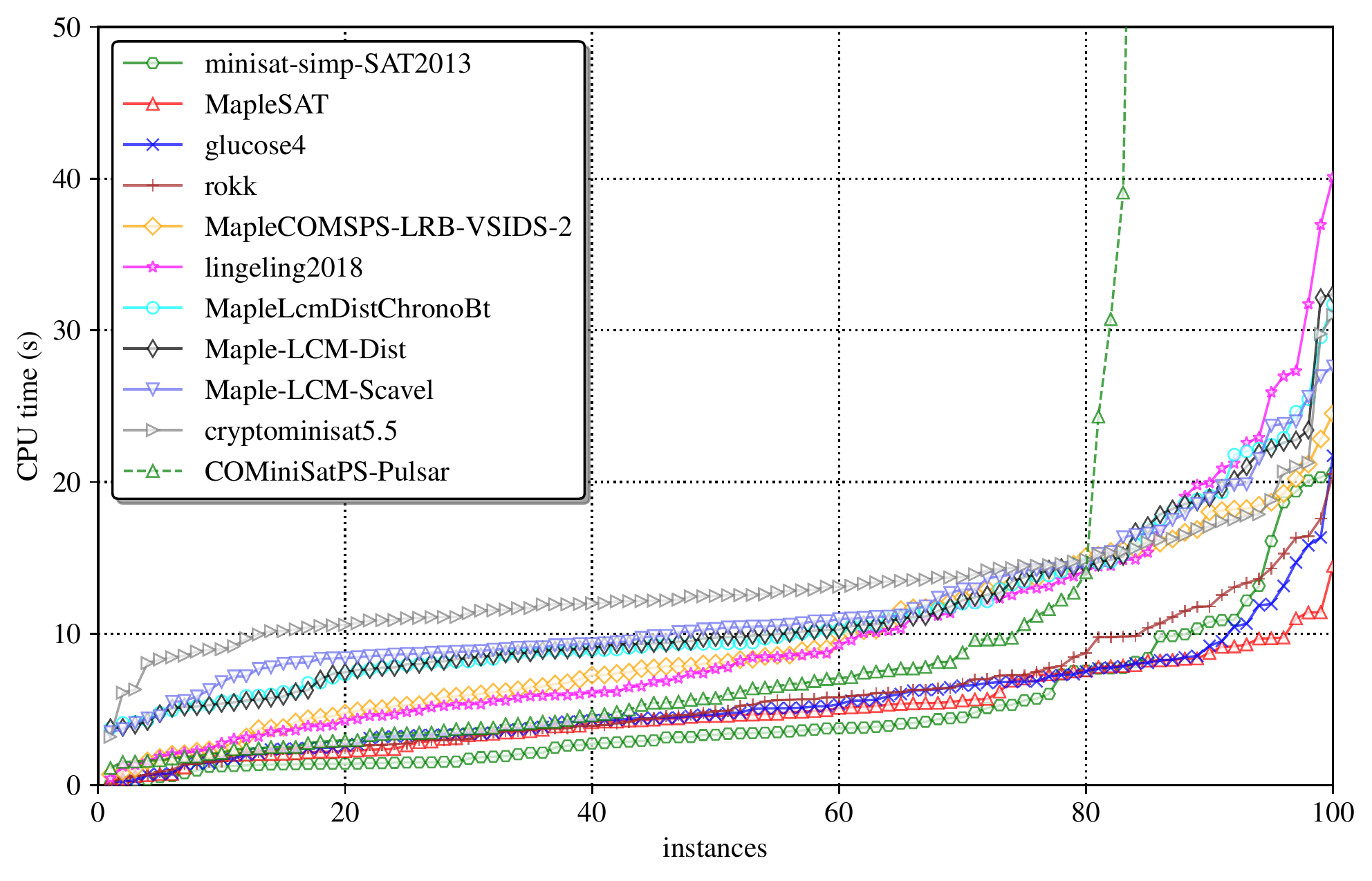}
	\label{fig:geffe-gos}
  }
  	\caption{Solving cryptanalysis problem for S\_Geffe via different SAT encodings (see Section \ref{sec:compar}). Part 1.}
	\label{fig:geffe-detailed}
\end{figure}

\begin{figure}[ht]
	\centering
  	\subfloat[][\textsc{CBMC-SAT}]{
    \includegraphics[width=0.5\textwidth]{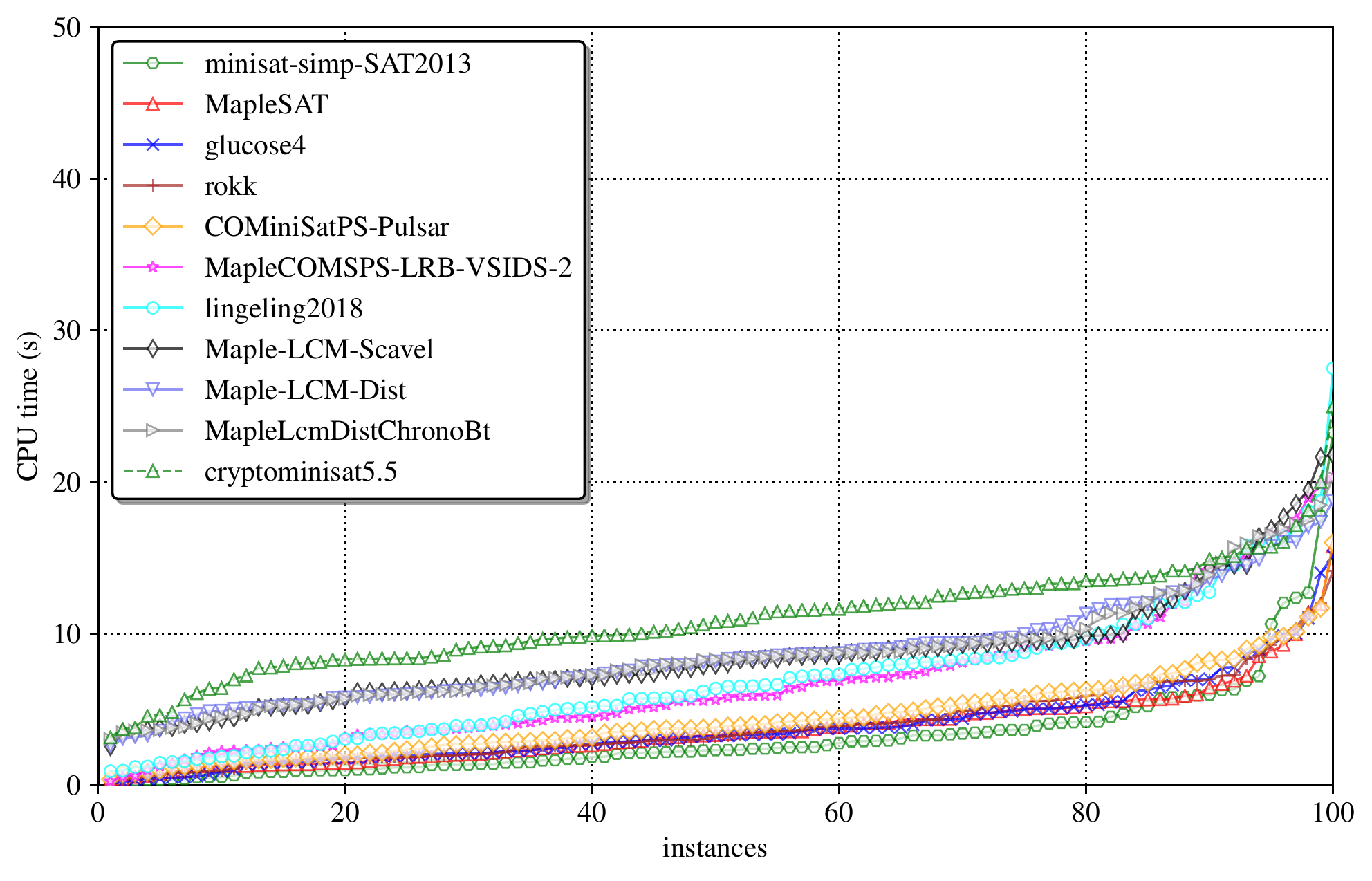}
	\label{fig:geffe-cbmc}
  }
  \subfloat[][\textsc{CBMC-SMT}]{
    \includegraphics[width=0.5\textwidth]{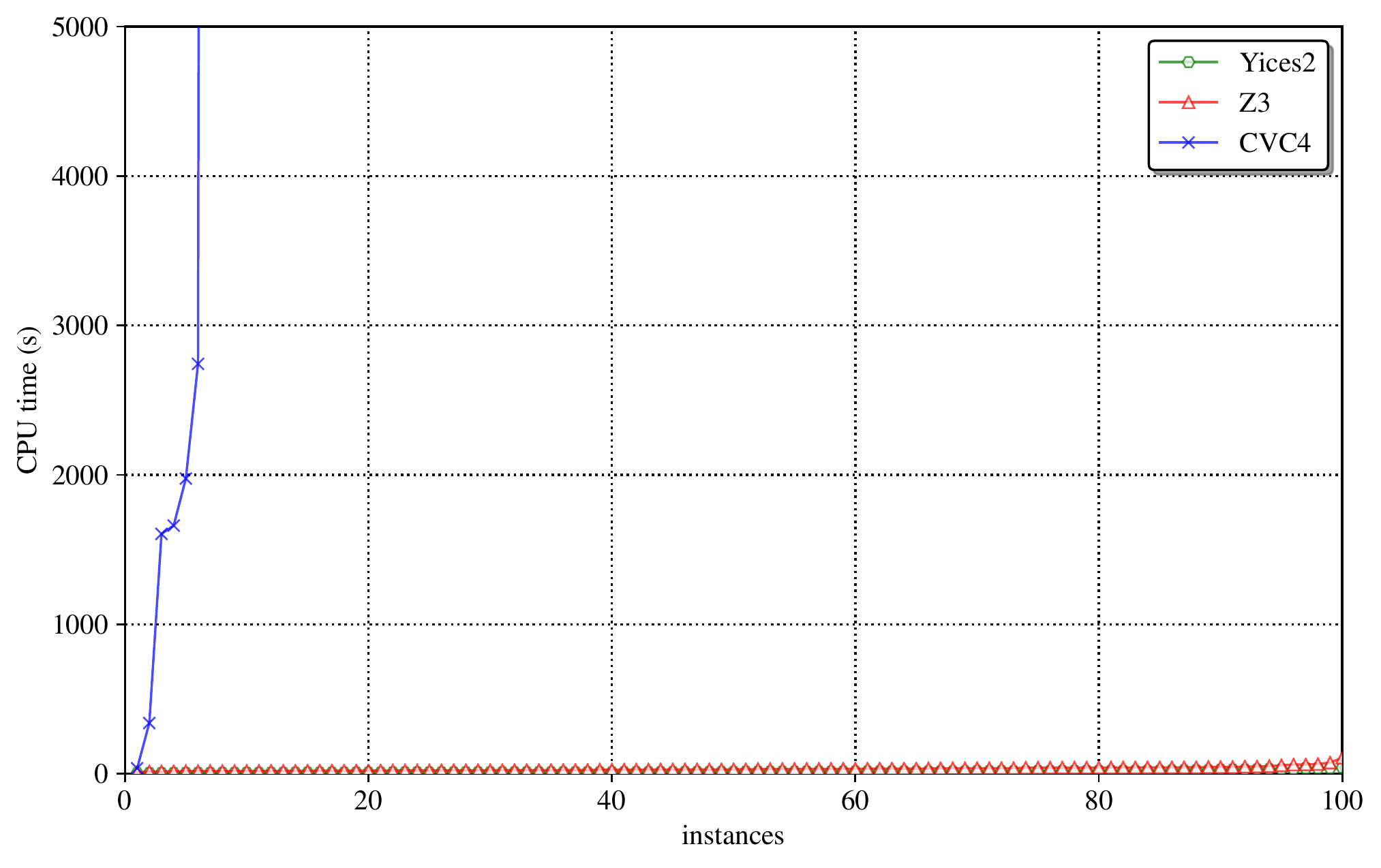}
	\label{fig:geffe-smt-cbmc}
  }
  \vskip\baselineskip
    \subfloat[][\textsc{Cryptol-SMT}]{
    \includegraphics[width=0.5\textwidth]{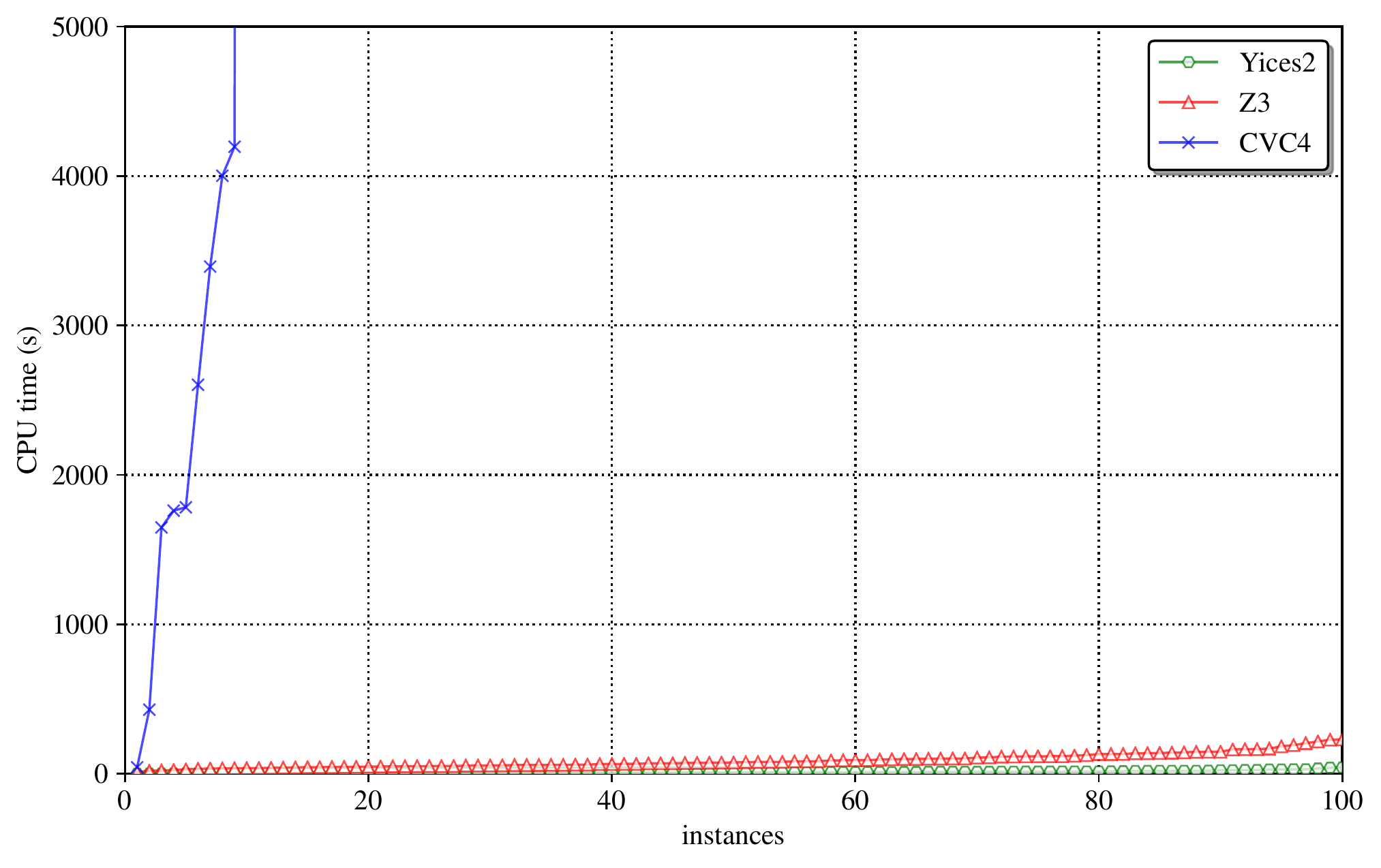}
	\label{fig:geffe-smt-cryptol}
  }
  	\caption{Solving S\_Geffe cryptanalysis instances using different encodings (see Section \ref{sec:compar}). Part 2.}
	\label{fig:geffe-detailed_part2}
\end{figure}

\begin{figure}[ht]
	\centering
	\subfloat[][\textsc{URSA}]{
    \includegraphics[width=0.5\textwidth]{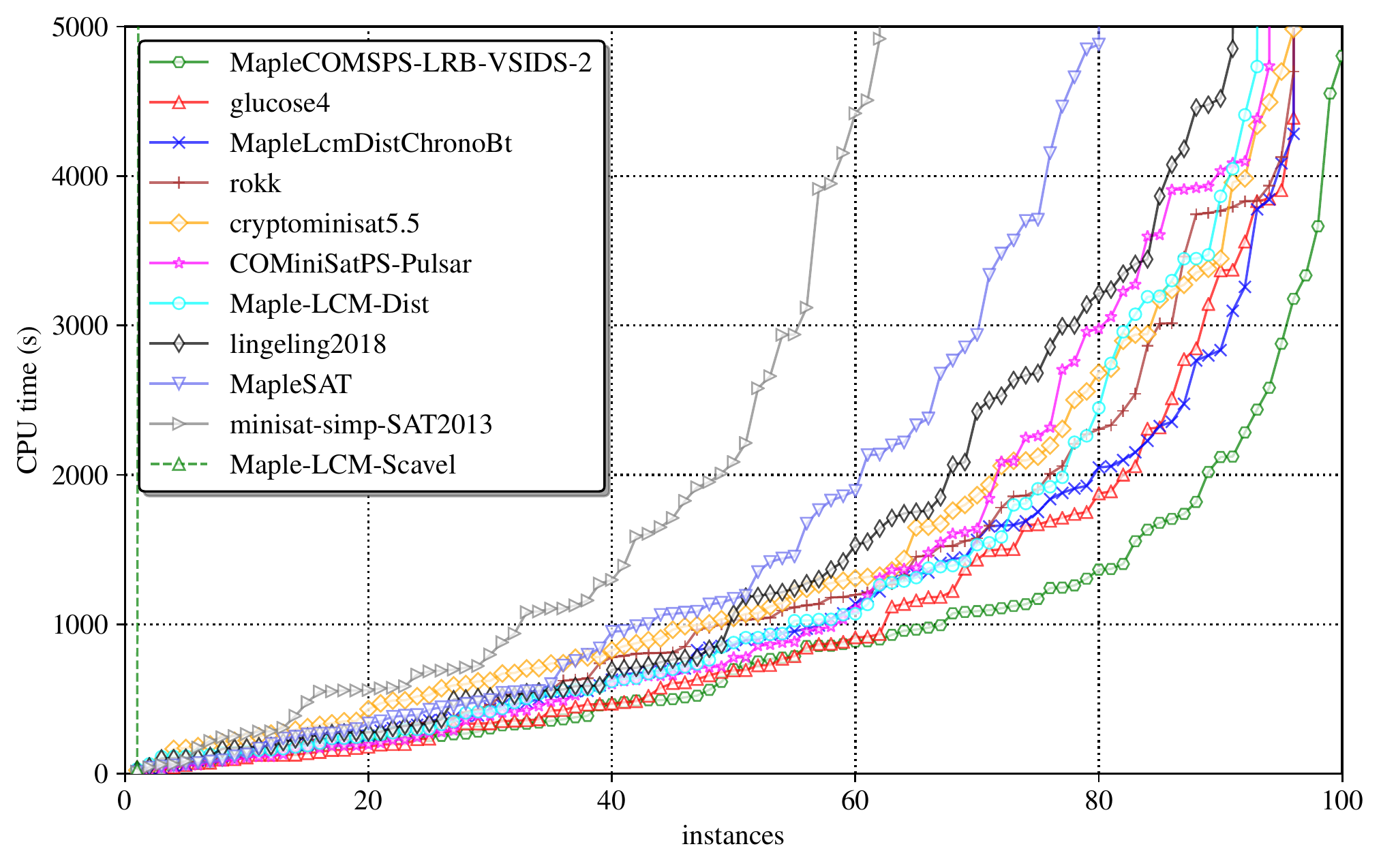}
		\label{fig:wolfram-ursa}
  }
 	\subfloat[][\textsc{Transalg}]{
    \includegraphics[width=0.5\textwidth]{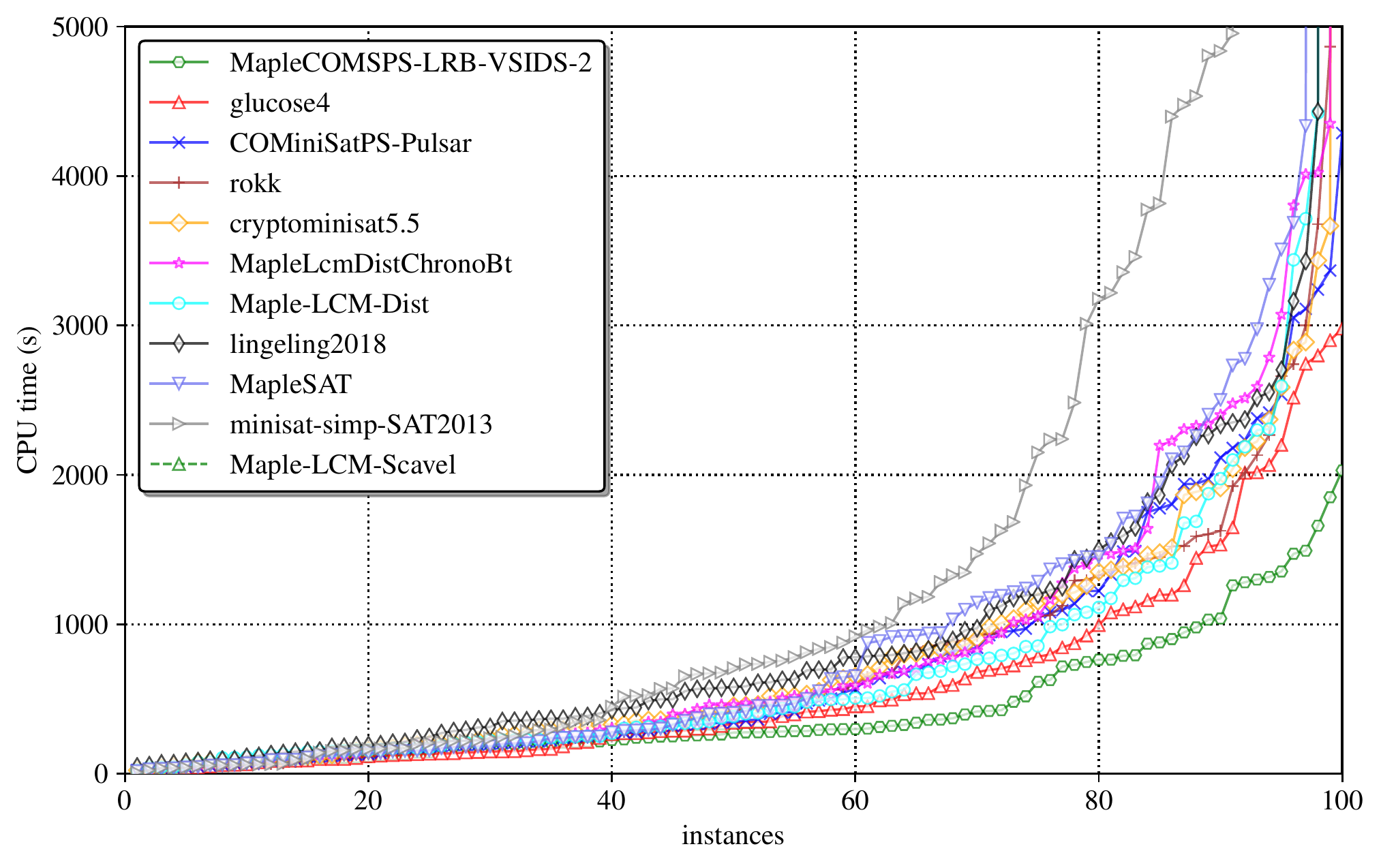}
		\label{fig:wolfram-transalg}
  }
   \vskip\baselineskip
    \subfloat[][\textsc{Cryptol-SAT}]{
    \includegraphics[width=0.5\textwidth]{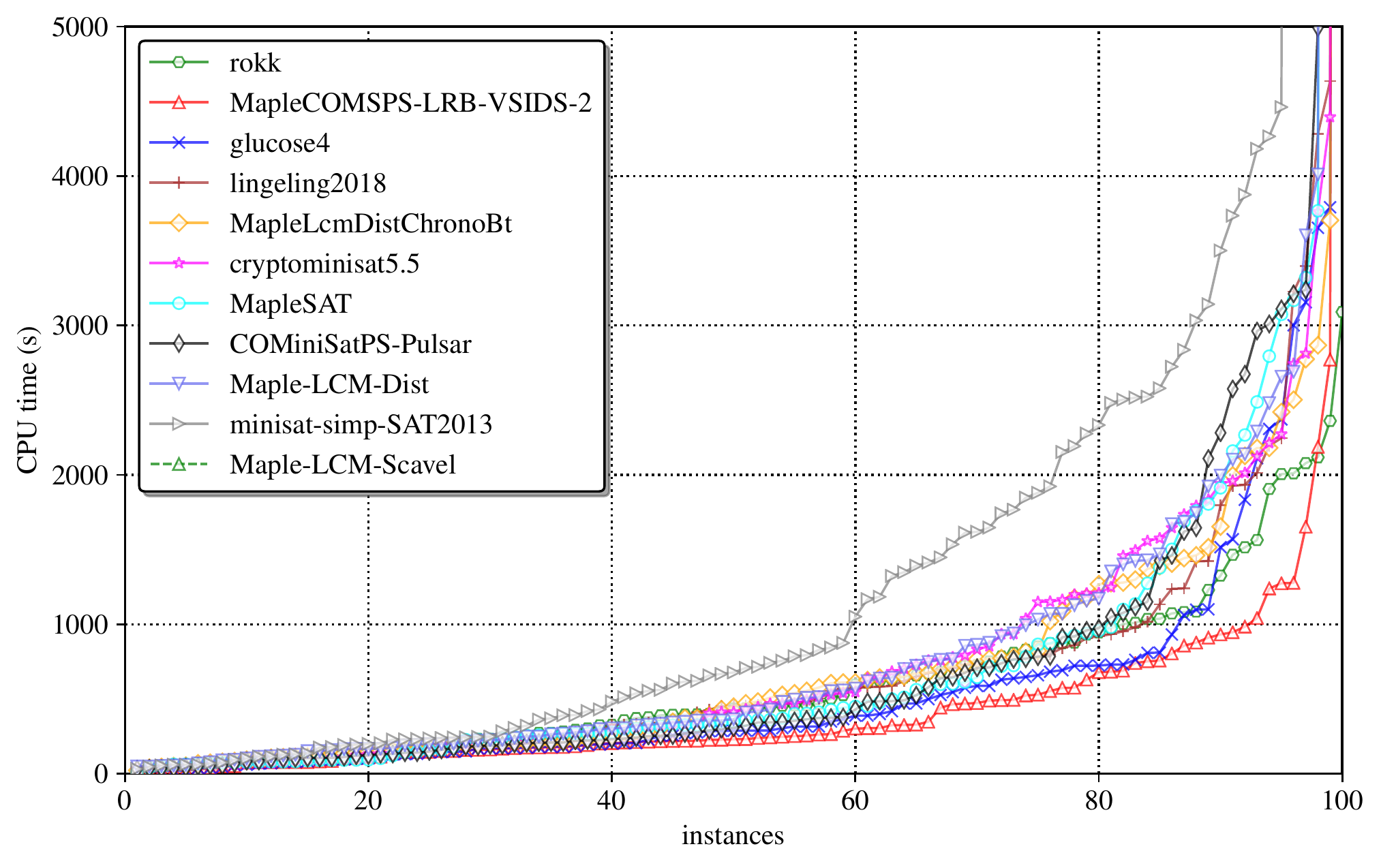}
		\label{fig:wolfram-cryptol-sat}
  }
      \subfloat[][\textsc{CBMC-SAT}]{
    \includegraphics[width=0.5\textwidth]{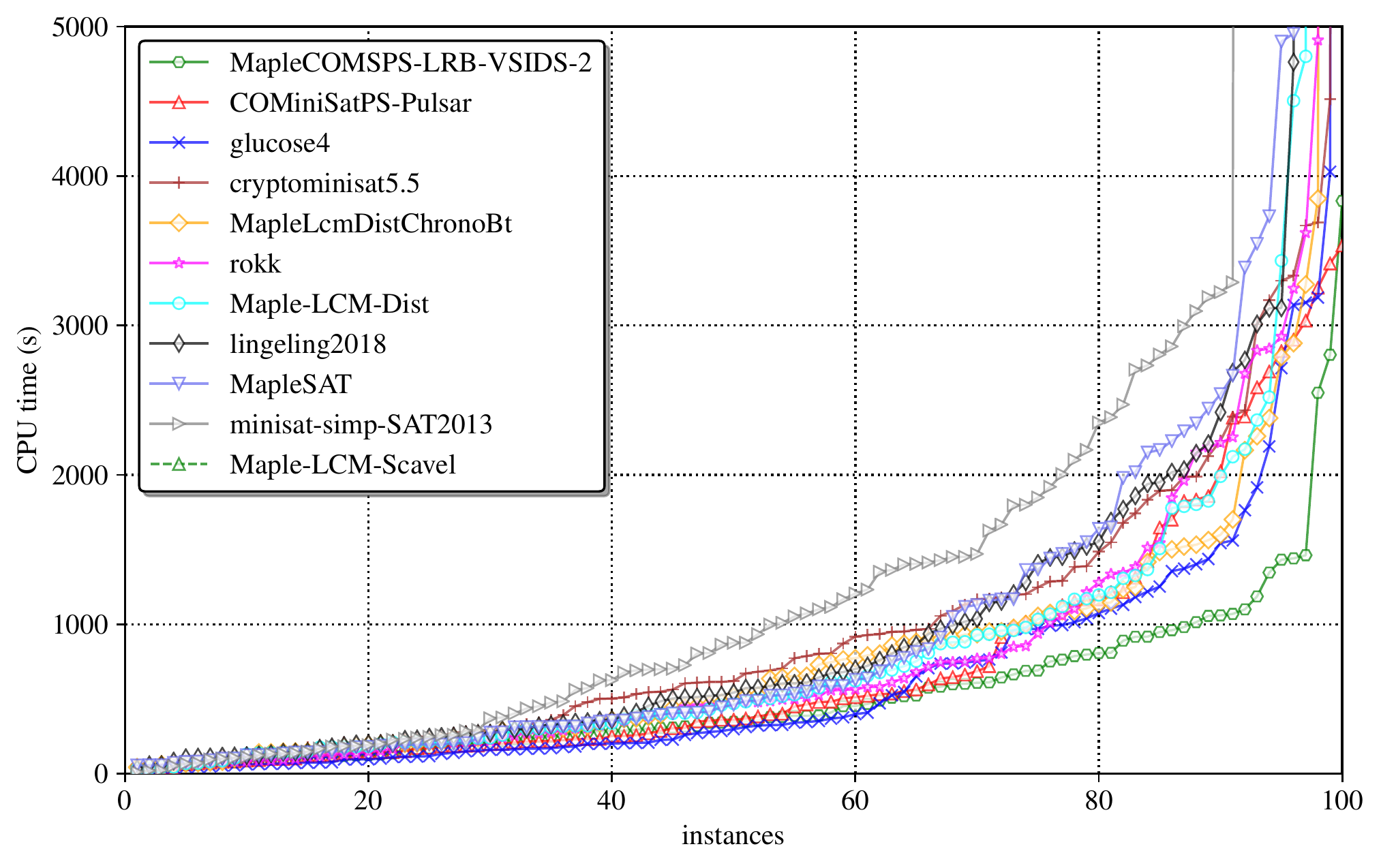}
		\label{fig:wolfram-cmbc-sat}
  }
   \vskip\baselineskip
  	\subfloat[][\textsc{Cryptol-SMT}]{
    \includegraphics[width=0.5\textwidth]{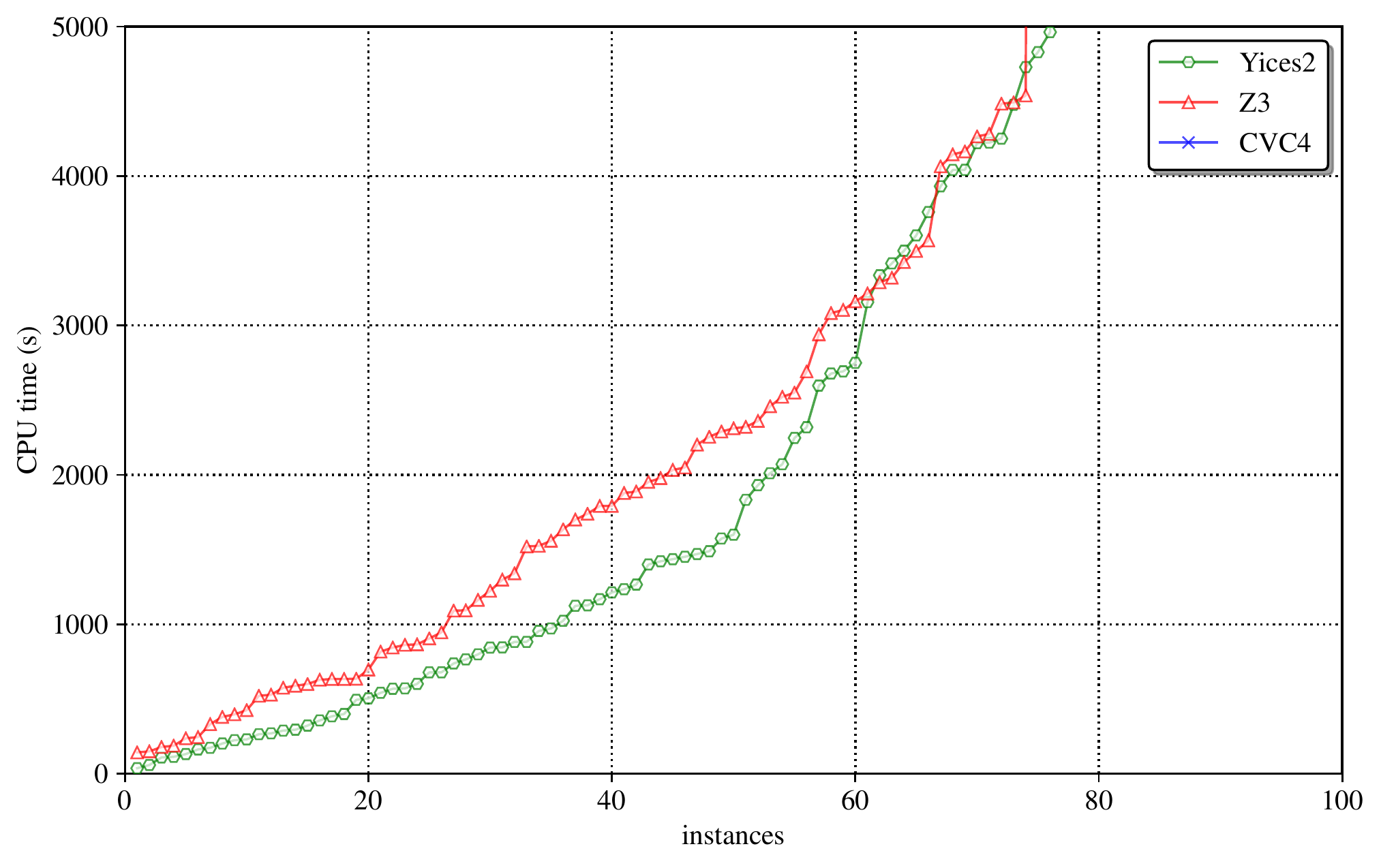}
		\label{fig:wolfram-smt-cryptol}
  }
 	\subfloat[][\textsc{CBMC-SMT}]{
    \includegraphics[width=0.5\textwidth]{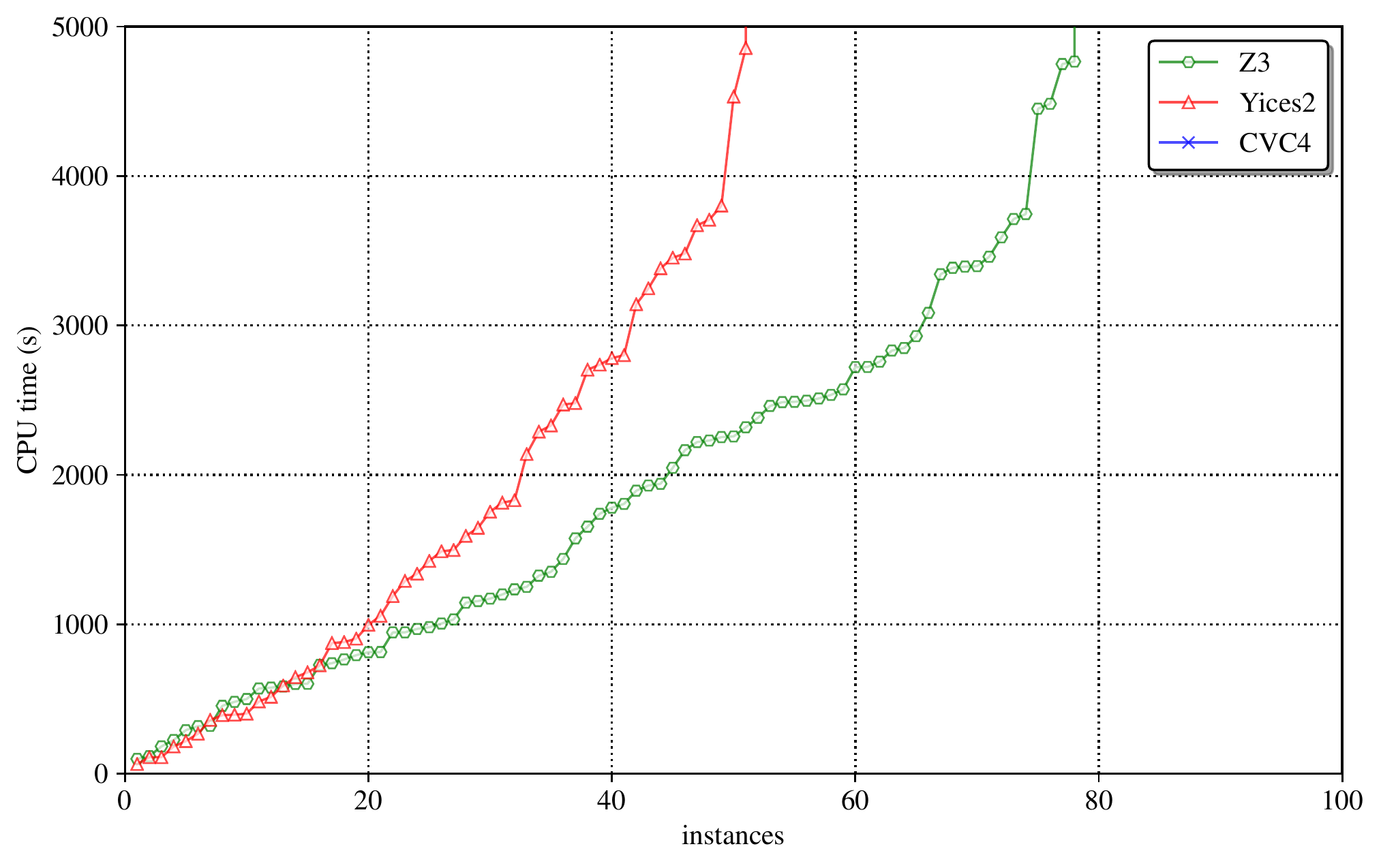}
		\label{fig:wolfram-smt-cbmc}
  }
  	\caption{Solving Wolfram cryptanalysis problem via different SAT encodings (see Section \ref{sec:compar})}
	\label{fig:wolfram-detailed}
\end{figure}

\begin{figure}[ht]
	\centering
	\subfloat[][\textsc{URSA}]{
    \includegraphics[width=0.5\textwidth]{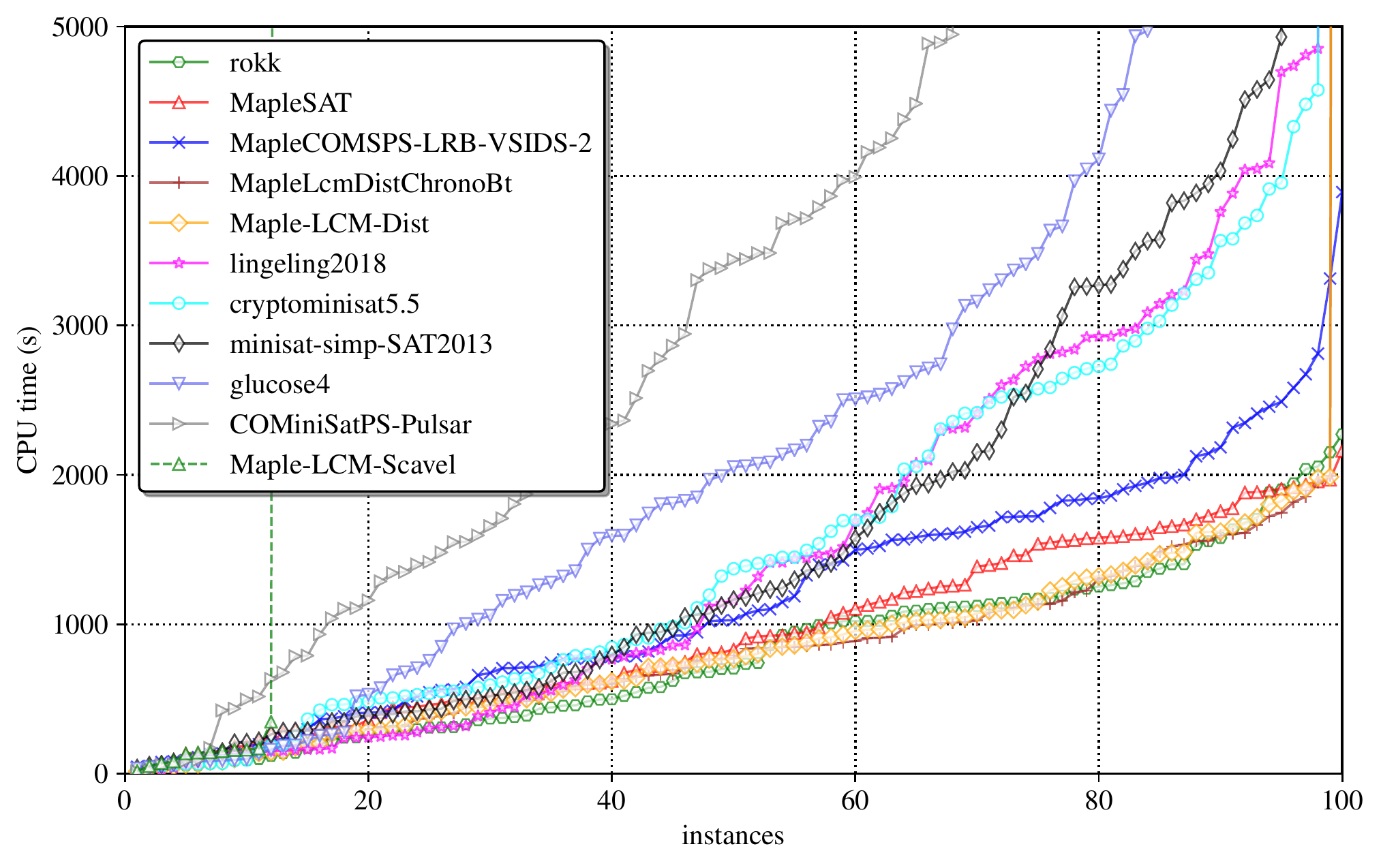}
		\label{fig:bivium-ursa}
  }
 	\subfloat[][\textsc{Transalg}]{
    \includegraphics[width=0.5\textwidth]{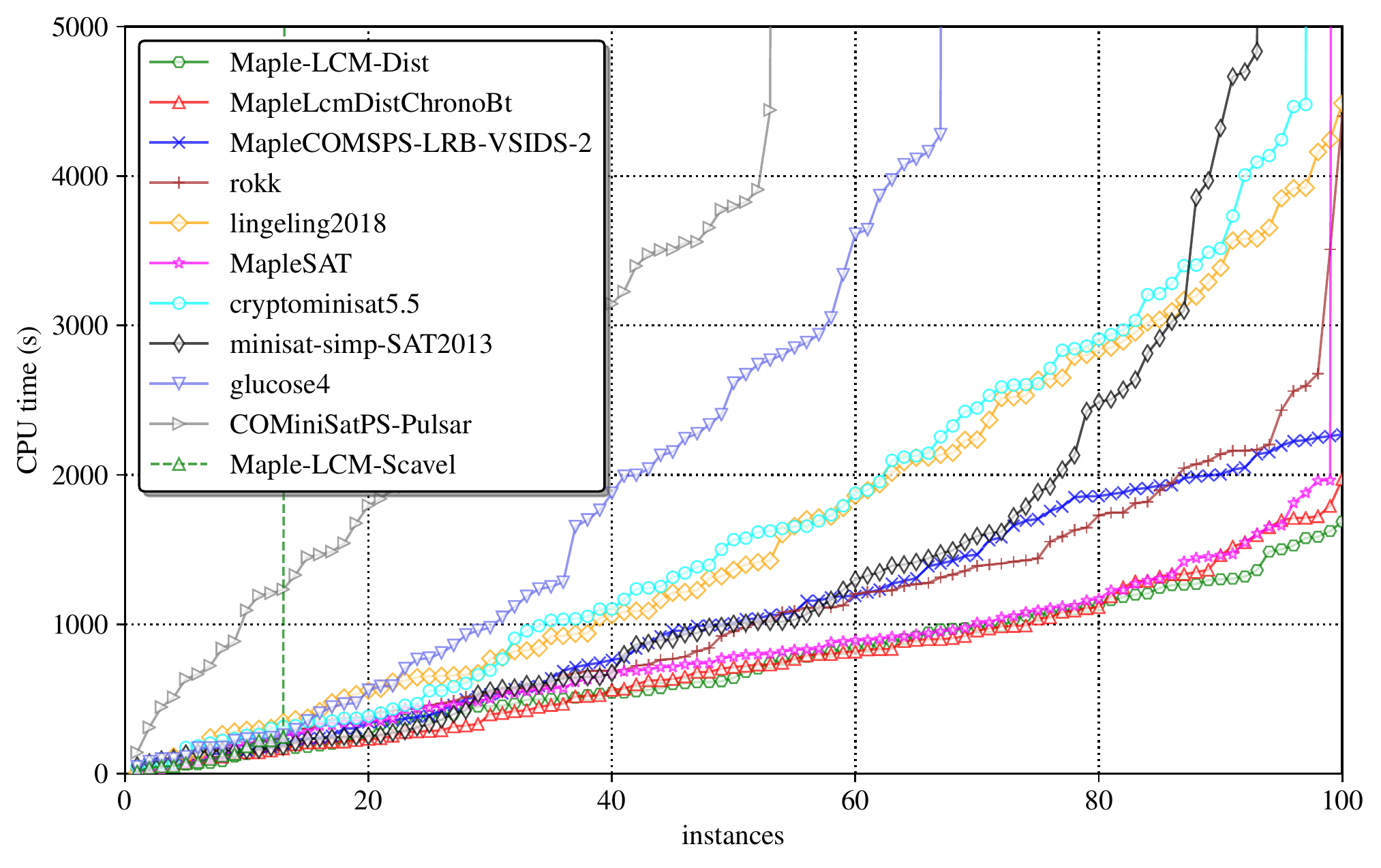}
		\label{fig:bivium-transalg}
  }
   \vskip\baselineskip
	\subfloat[][\textsc{Cryptol-SAT}]{
    \includegraphics[width=0.5\textwidth]{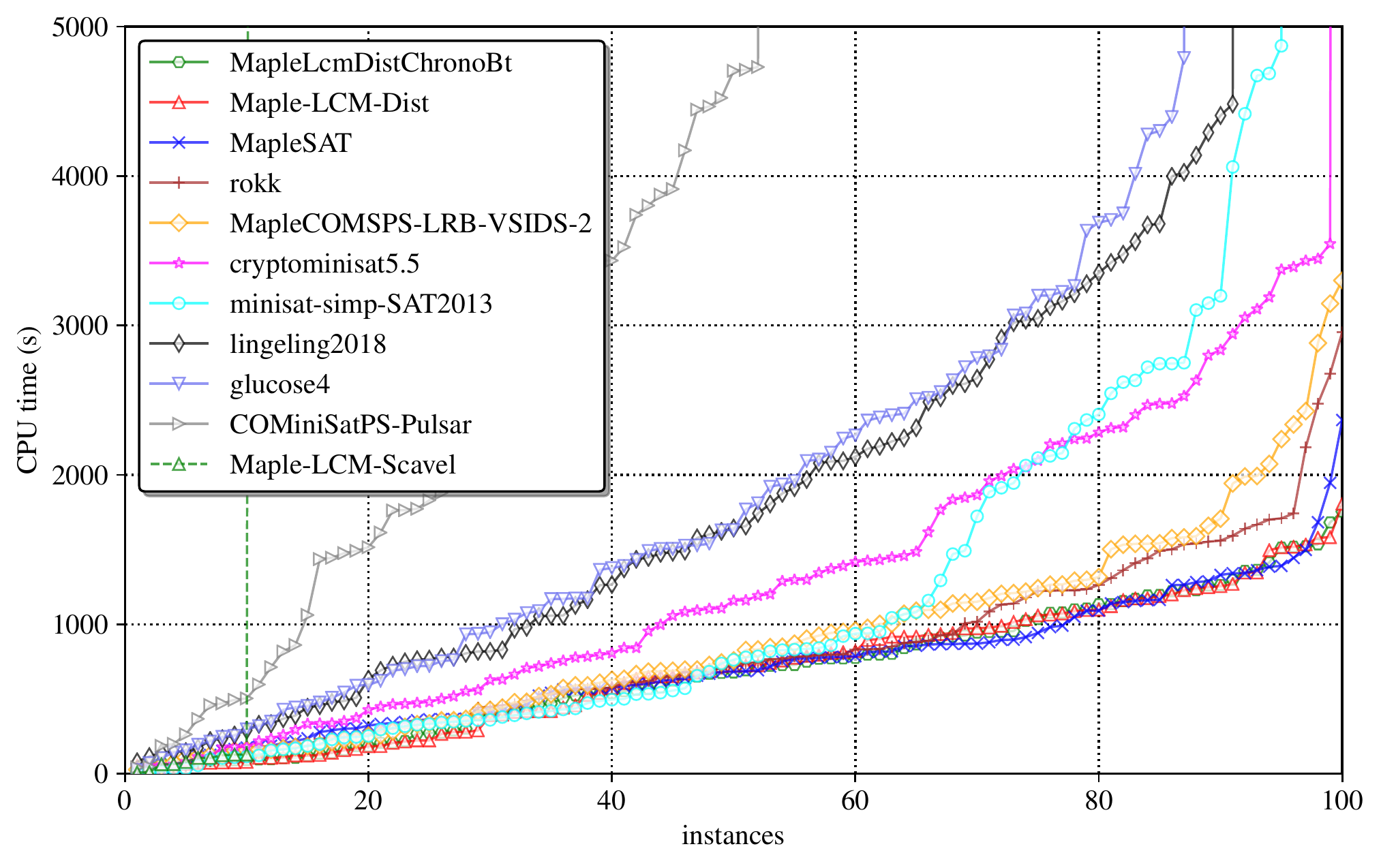}
		\label{fig:bivium-cryptol}
  }
	\subfloat[][\textsc{Grain-of-Salt}]{
    \includegraphics[width=0.5\textwidth]{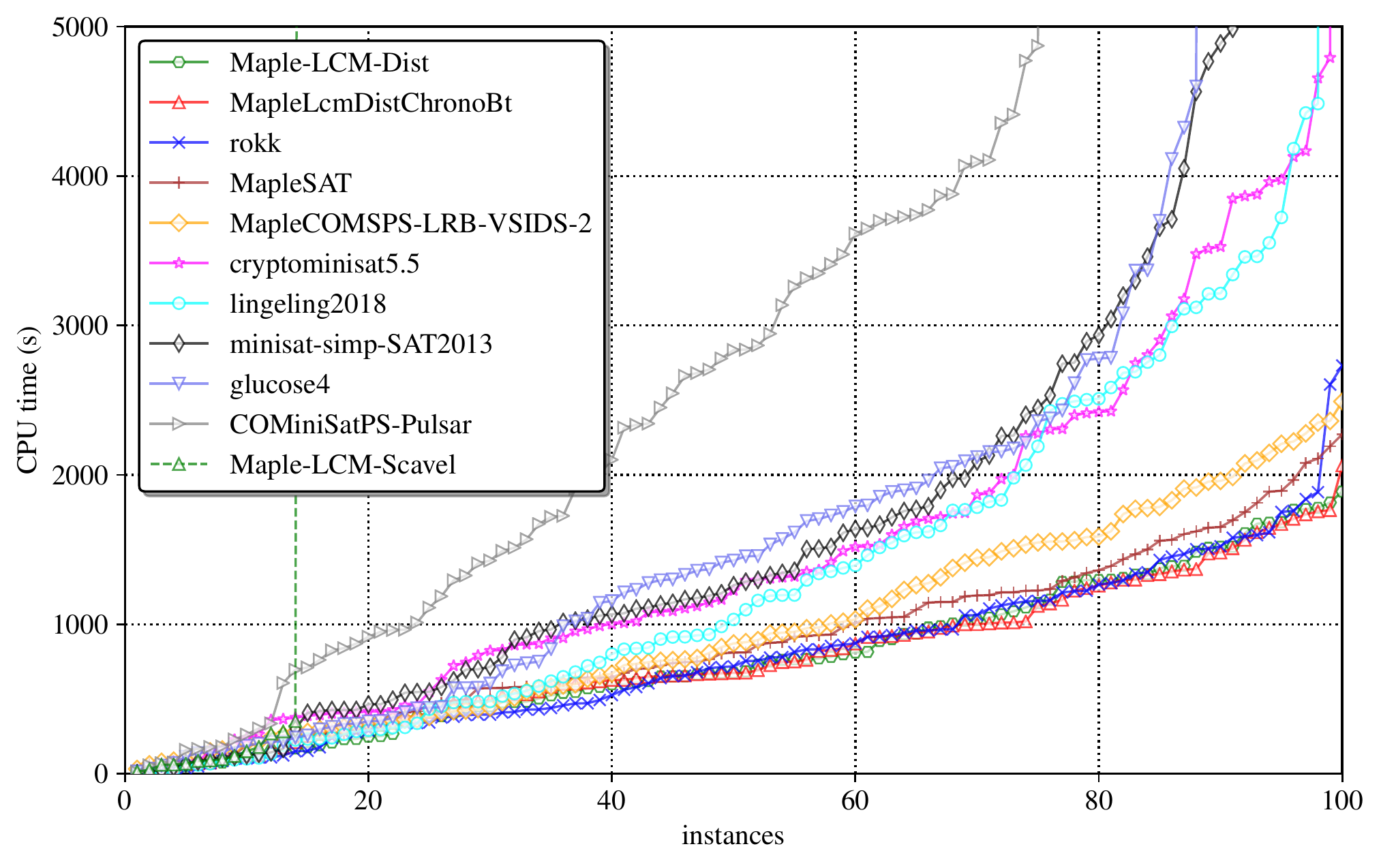}
	\label{fig:bivium-gos}
  }
  \vskip\baselineskip
  \subfloat[][\textsc{CBMC-SAT}]{
    \includegraphics[width=0.5\textwidth]{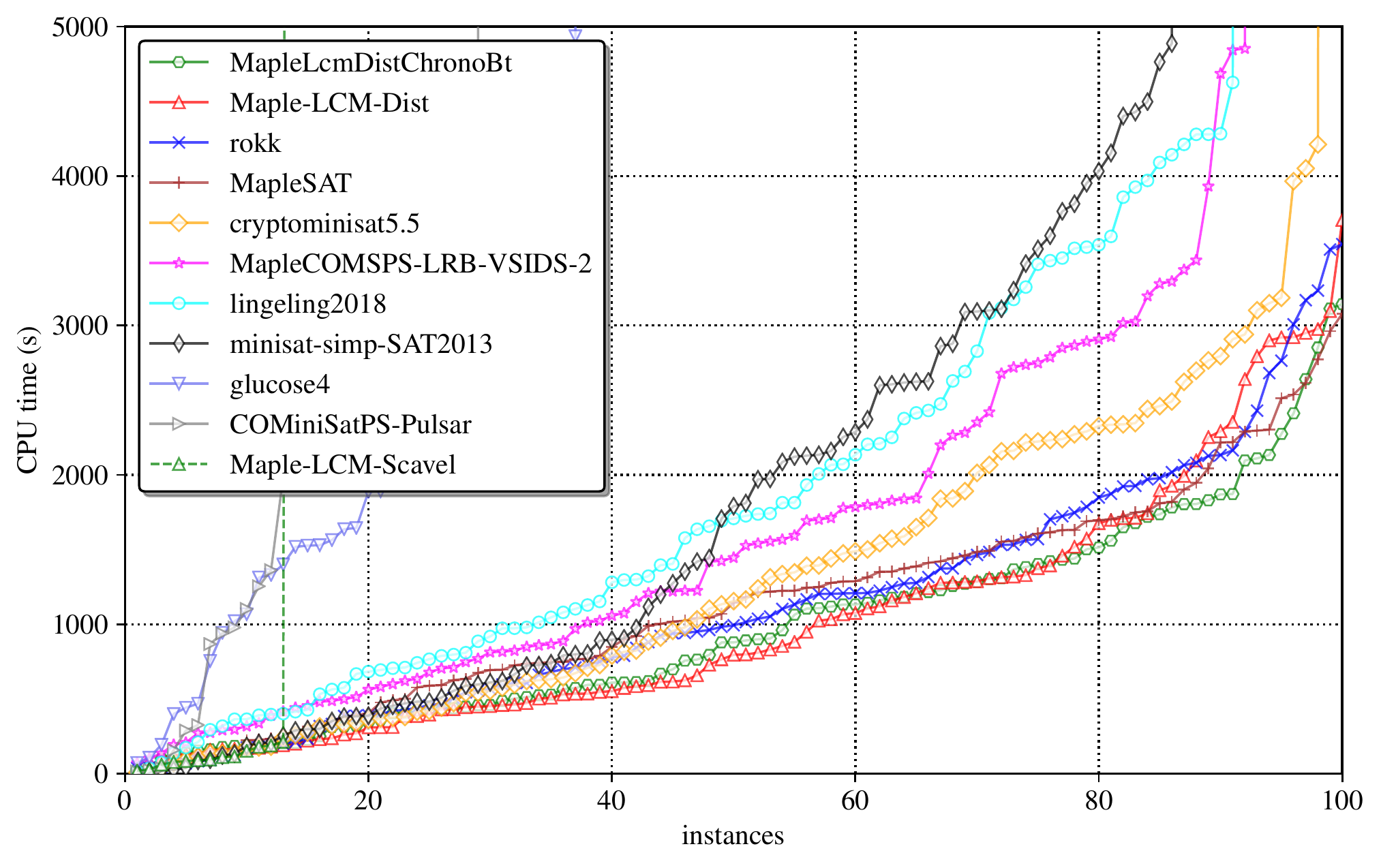}
		\label{fig:bivium-cbmc-sat}
  }
 	\subfloat[][\textsc{Cryptol-SMT}]{
    \includegraphics[width=0.5\textwidth]{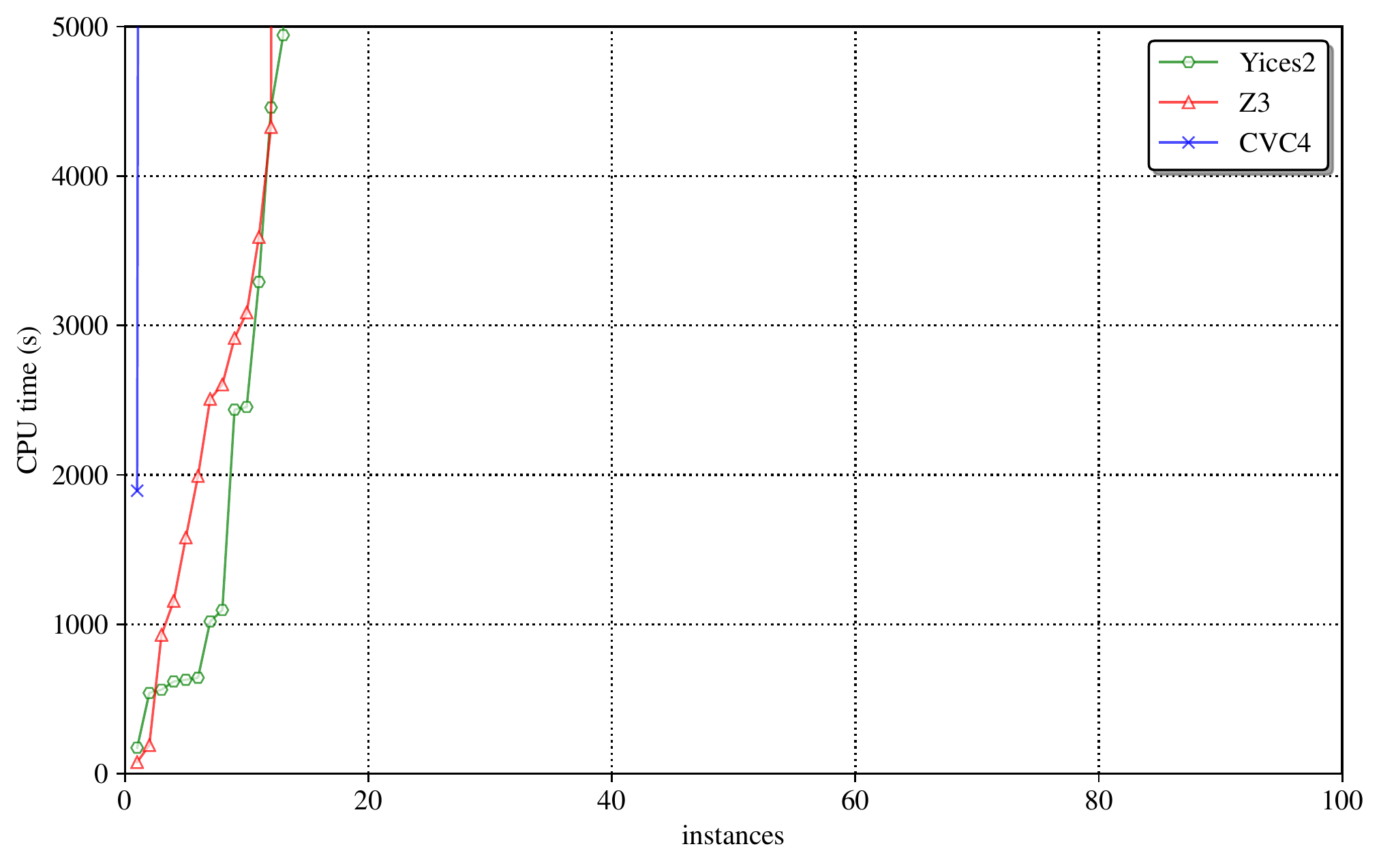}
		\label{fig:bivium-cryptol-smt}
  }
  	\caption{Solving Bivium30 cryptanalysis problem via different SAT encodings (see Section \ref{sec:compar}).}
	\label{fig:bivium-detailed}
\end{figure}

\begin{figure}[ht]
	\centering
	\subfloat[][\textsc{URSA}]{
    \includegraphics[width=0.5\textwidth]{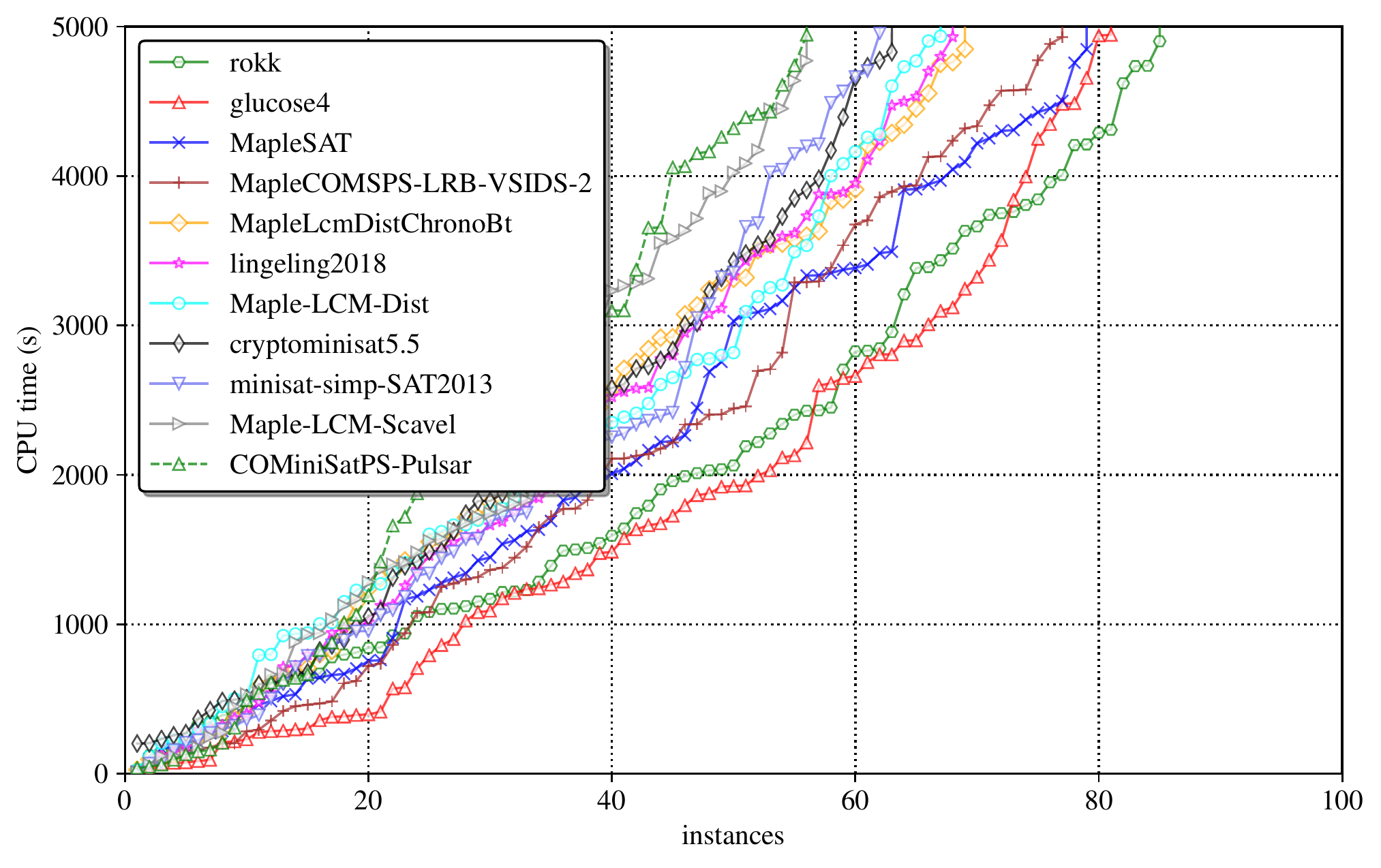}
		\label{fig:grain-ursa}
  }
 	\subfloat[][\textsc{Transalg}]{
    \includegraphics[width=0.5\textwidth]{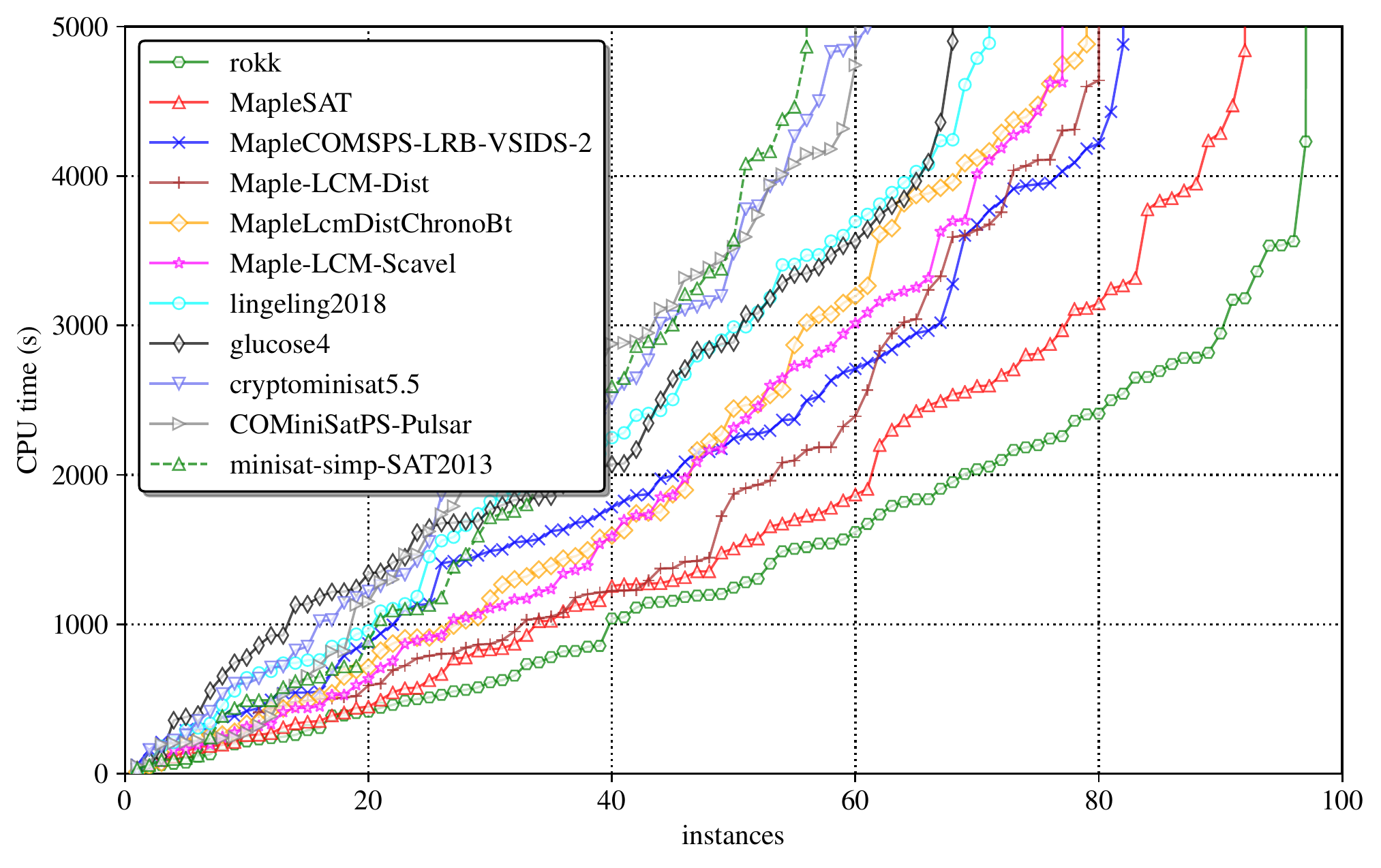}
		\label{fig:grain-transalg}
  }
   \vskip\baselineskip
	\subfloat[][\textsc{Cryptol-SAT}]{
    \includegraphics[width=0.5\textwidth]{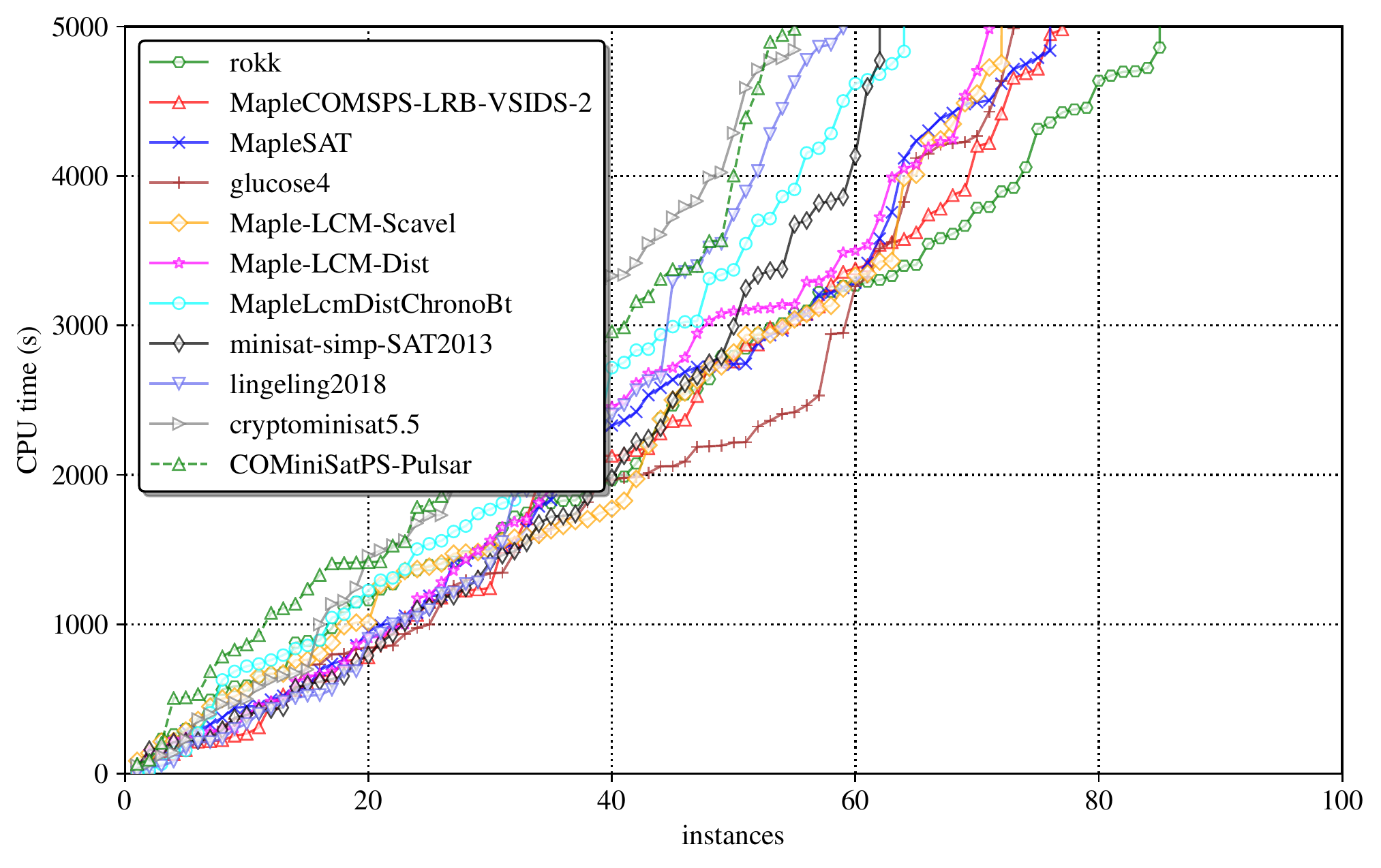}
		\label{fig:grain-cryptol-sat}
  }
	\subfloat[][\textsc{CBMC-SAT}]{
    \includegraphics[width=0.5\textwidth]{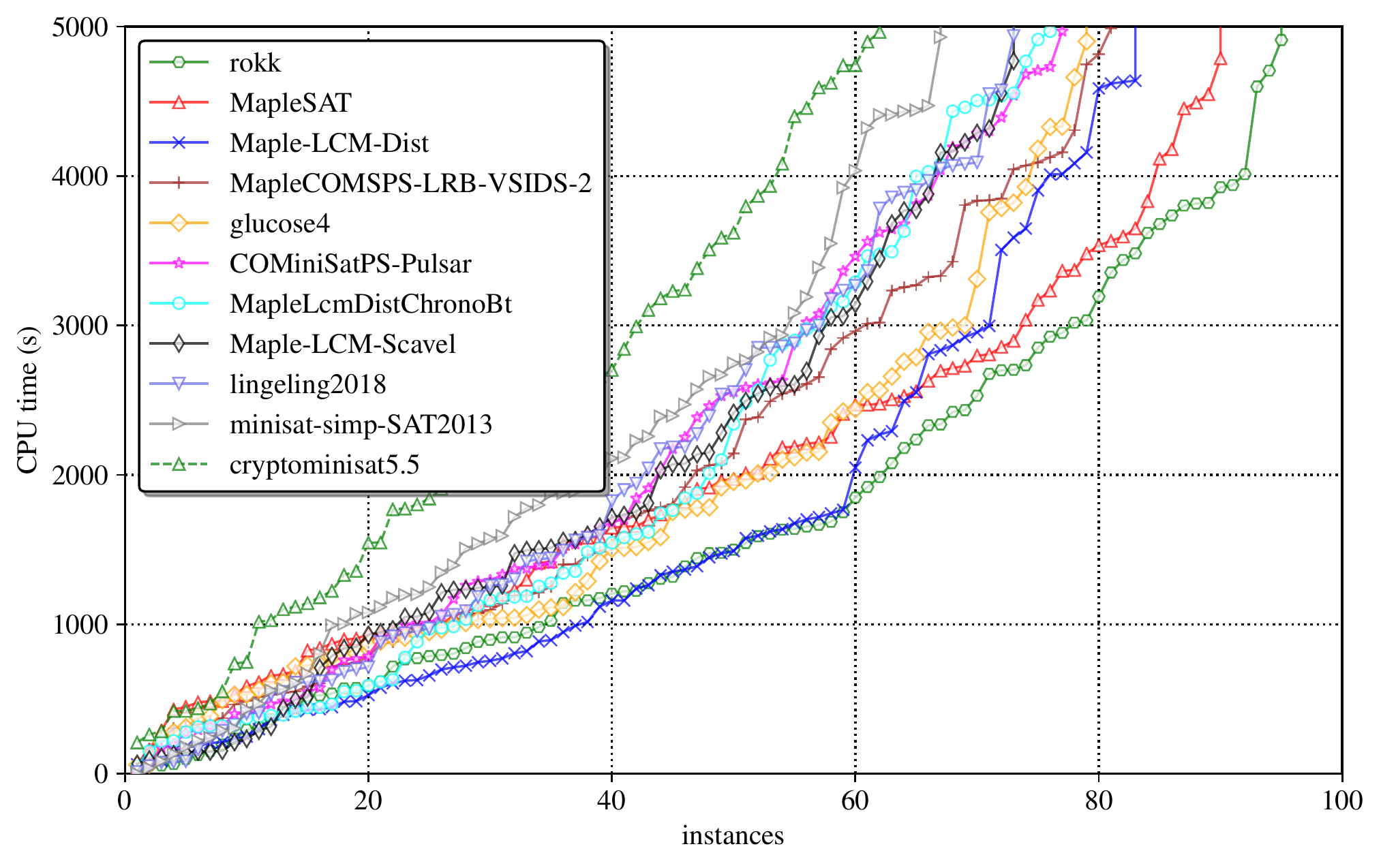}
	\label{fig:grain-cbmc-sat}
  }
  \vskip\baselineskip
	\subfloat[][\textsc{Cryptol-SMT}]{
    \includegraphics[width=0.5\textwidth]{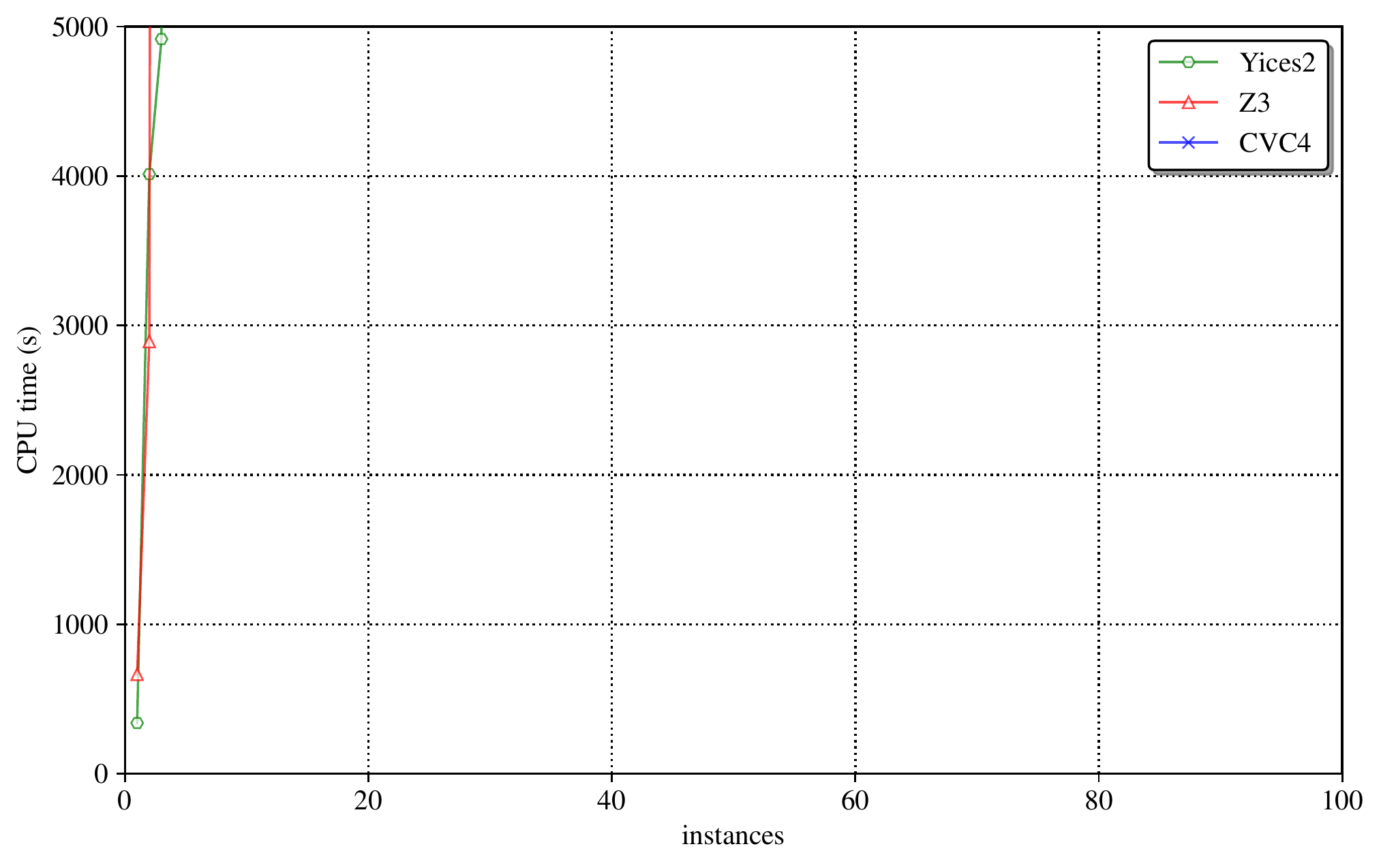}
		\label{fig:grain-cryptol-smt}
  }
	\subfloat[][\textsc{Grain-of-Salt}]{
    \includegraphics[width=0.5\textwidth]{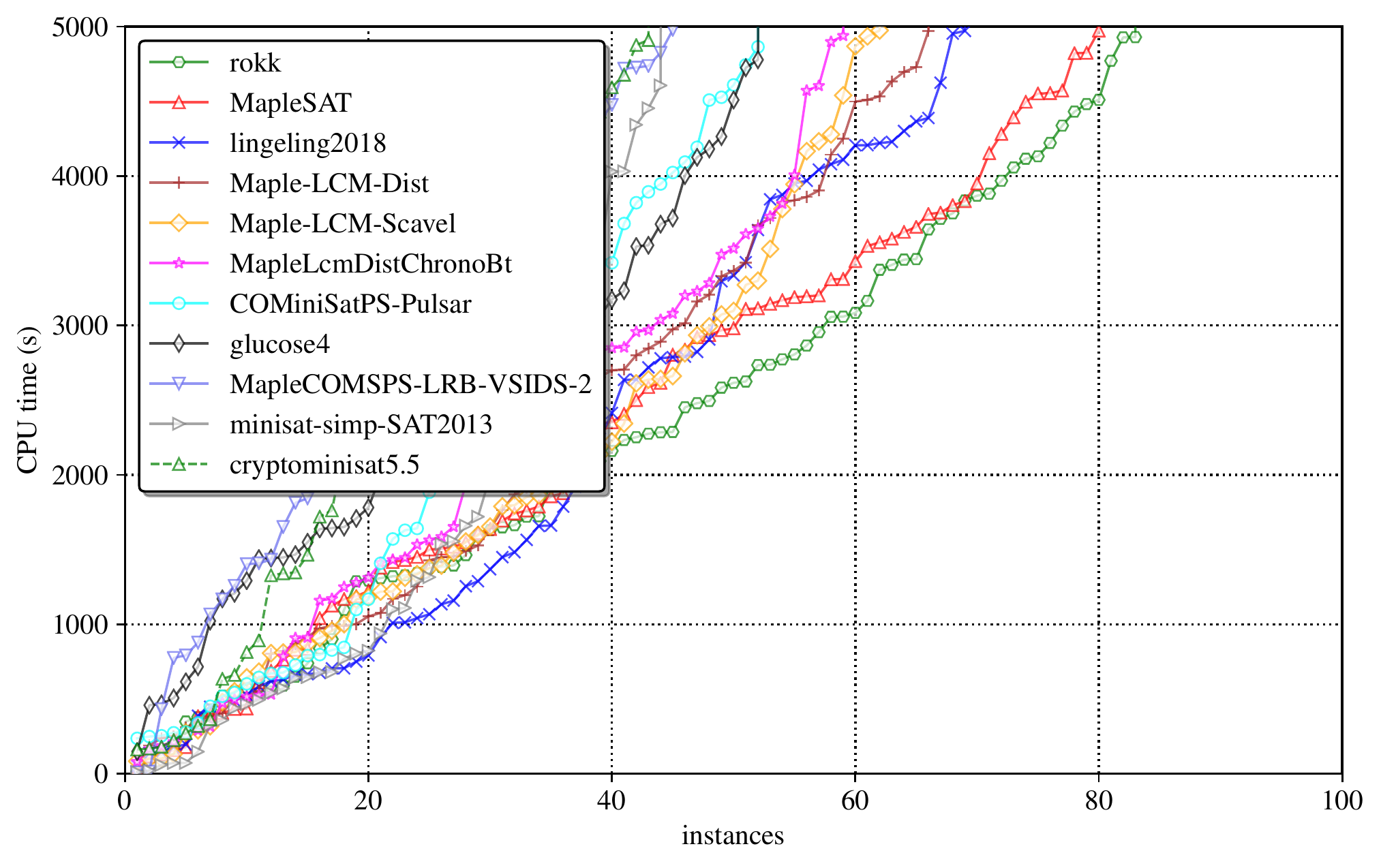}
	\label{fig:grain-gos}
  }
  	\caption{Solving Grain102 cryptanalysis problem via different SAT encodings (see Section \ref{sec:compar})}
	\label{fig:grain-detailed}
\end{figure}

\end{appendices}

\end{document}